\documentclass[sigconf,screen]{acmart} 

\newif\iftrack
\trackfalse 

\newif\ifcameratrack
\cameratrackfalse 

\usepackage{framed} 

\usepackage{fontawesome5}
\usepackage{xspace}
\usepackage{colortbl} 

\usepackage{color} 

\definecolor{OldGray}{RGB}{193,193,193} 

\definecolor{forestgreen}{RGB}{34, 139, 34}

\newcommand{\etc}{{etc.}\xspace}
\newcommand{\etal}{et~al.\@\xspace}
\newcommand{\ie}{i.e.,\@\xspace}
\newcommand{\eg}{e.g.,\@\xspace}

\usepackage{soul}
\iftrack{}
\else 
  \makeatletter
  \renewcommand\st[1]{\@bsphack\@esphack}%
  \makeatother
\fi

\let\stC=\textst
\ifcameratrack{}
\else 
  \makeatletter
  \renewcommand\stC[1]{\@bsphack\@esphack}%
  \makeatother
\fi

\definecolor{revision}{RGB}{255,127,14}

\definecolor{camerarevision}{RGB}{255,127,14}

\newcommand{\camerarevision}[2]{\ifcameratrack{\textcolor{red}{\stC{#1}}\textcolor{forestgreen}{#2}}\else{#2}\fi}

\definecolor{execlitcol}{RGB}{33, 150, 243} 
\definecolor{evallitcol}{RGB}{76, 175, 80} 
\definecolor{goallitcol}{RGB}{255, 152, 0} 
\definecolor{foundlitcol}{RGB}{156, 39, 176} 
\definecolor{criteriatcol}{HTML}{8e219c} 
\definecolor{shadecolor}{HTML}{F4ECF7}

\newcommand{\literacychip}[2]{%
  \setlength{\fboxsep}{0pt}%
  \colorbox{#1!20}{%
      \sffamily#2\strut
  }%
}

\newcommand{\execlit}[1]{\literacychip{execlitcol}{#1}}
\newcommand{\evallit}[1]{\literacychip{evallitcol}{#1}}
\newcommand{\goallit}[1]{\literacychip{goallitcol}{#1}}
\newcommand{\foundlit}[1]{\literacychip{foundlitcol}{#1}}

\newcommand{\gulfs}{\textsc{Gulfs}\xspace}
\newcommand{\abstractions}{\textsc{Abstractions}\xspace}

\newcommand{\intergulfs}[1]{interaction gulfs}
\newcommand{\intragulfs}[1]{visualization interaction gulfs}

\newcommand{\analysislit}{\goallit{\faLightbulb~Analysis literacy*}\xspace}
\newcommand{\tasklit}{\goallit{\faTasks~Task literacy*}\xspace}
\newcommand{\toollit}{\goallit{\faTools~Tool literacy*}\xspace}
\newcommand{\orchlit}{\execlit{\faCogs~Workflow literacy*}\xspace}
\newcommand{\intlit}{\execlit{\faHandPointer~Interaction literacy}\xspace}
\newcommand{\devlit}{\execlit{\faMouse~Device literacy}\xspace}
\newcommand{\evaluationlit}{\evallit{\faSearchPlus~Insight literacy*}\xspace}
\newcommand{\visualizationlit}{\evallit{\faChartBar~Visualization literacy}\xspace}
\newcommand{\visuallit}{\evallit{\faEye~Visual literacy}\xspace}

\newcommand{\analysislitlow}{\goallit{\faLightbulb~analysis literacy*}\xspace}
\newcommand{\tasklitlow}{\goallit{\faTasks~task literacy*}\xspace}
\newcommand{\toollitlow}{\goallit{\faTools~tool literacy*}\xspace}
\newcommand{\orchlitlow}{\execlit{\faCogs~workflow literacy*}\xspace}
\newcommand{\intlitlow}{\execlit{\faHandPointer~interaction literacy}\xspace}
\newcommand{\devlitlow}{\execlit{\faMouse~device literacy}\xspace}
\newcommand{\evaluationlitlow}{\evallit{\faSearchPlus~insight literacy*}\xspace}
\newcommand{\visualizationlitlow}{\evallit{\faChartBar~visualization literacy}\xspace}
\newcommand{\visuallitlow}{\evallit{\faEye~visual literacy}\xspace}

\definecolor{boxpurple}{HTML}{97447a}
\definecolor{boxlight}{HTML}{F4ECF7}

\newcommand{\competencybox}[2]{%
  \begin{framed}
  \noindent
  \colorbox{boxpurple}{%
    \parbox{0.98\linewidth}{%
      \color{white}\bfseries #1
    }%
  }%
  \par
  \noindent
  \colorbox{boxlight}{%
    \parbox{0.98\linewidth}{%
      #2
    }%
  }%
  \end{framed}
}
\AtBeginDocument{%
  }

\copyrightyear{2026}
\acmYear{2026}
\setcopyright{cc}
\setcctype{by}
\acmConference[CHI '26]{Proceedings of the 2026 CHI Conference on Human Factors in Computing Systems}{April 13--17, 2026}{Barcelona, Spain}
\acmBooktitle{Proceedings of the 2026 CHI Conference on Human Factors in Computing Systems (CHI '26), April 13--17, 2026, Barcelona, Spain}
\acmPrice{}
\acmDOI{10.1145/3772318.3793423}
\acmISBN{979-8-4007-2278-3/2026/04}

\begin{document}

\newcommand{\definitionAbbrv}{the ability to explore data through interactive visualizations; to engage with data and visualization at different levels of abstraction during an often iterative process that includes formulating goals, performing interactions, and evaluating their results}

\newcommand{\definitiontwo}{Interactive visualization literacy is \definitionAbbrv.}

\title[A Multiliteracy Model for Interactive Visualization Literacy]{A Multiliteracy Model for Interactive Visualization Literacy: Definitions, Literacies, and Steps for Future Research}

\author{Gabriela Molina Le\'{o}n}
\orcid{0000-0002-9223-2022}
\affiliation{%
  \institution{Aarhus University}
  \city{Aarhus}
  \country{Denmark}}
\email{leon@cs.au.dk}

\author{Benjamin Bach}
\orcid{0000-0002-9201-7744}
\affiliation{%
  \institution{Inria Bordeaux}
  \city{Bordeaux}
  \country{France}}
\email{benjamin.bach@inria.fr}

\author{Matheus Valentim}
\orcid{0000-0003-0860-2084}
\affiliation{%
  \institution{Aarhus University}
  \city{Aarhus}
  \country{Denmark}}
\affiliation{%
  \institution{University of Illinois}
  \city{Urbana-Champaign}
  \state{IL}
  \country{USA}}
\email{mavalentim.b@gmail.com}

\author{Niklas Elmqvist}
\orcid{0000-0001-5805-5301}
\affiliation{%
  \institution{Aarhus University}
  \city{Aarhus}
  \country{Denmark}}
\email{elm@cs.au.dk}

\renewcommand{\shortauthors}{Le\'{o}n et al.}

\begin{abstract}
  This paper presents a theoretical model for interactive visualization literacy to describe how people use interactive data visualizations and systems. 
Literacies have become an important concept in describing modern life skills, with visualization literacy generally referring to the use and interpretation of data visualizations.
However, prior work on visualization literacy overlooks interaction and its associated challenges, despite it being an intrinsic aspect of using visualizations.
Based on existing theoretical frameworks, we derive a two-dimensional model that combines four well-known literacies with five novel ones.
We found evidence for our model through analyzing existing visualization systems as well as through observations from an exploratory study involving such systems.
We conclude by outlining steps towards measuring, evaluating, designing for, and teaching interactive visualization literacy.
\end{abstract}

\ccsdesc[500]{Human-centered computing~Visualization theory, concepts and paradigms}
\ccsdesc[500]{Human-centered computing~HCI theory, concepts and models}

\keywords{Interactive visualization literacy, visualization literacy, interactive visualization, action cycle}
\begin{teaserfigure}
  \includegraphics[width=\textwidth]{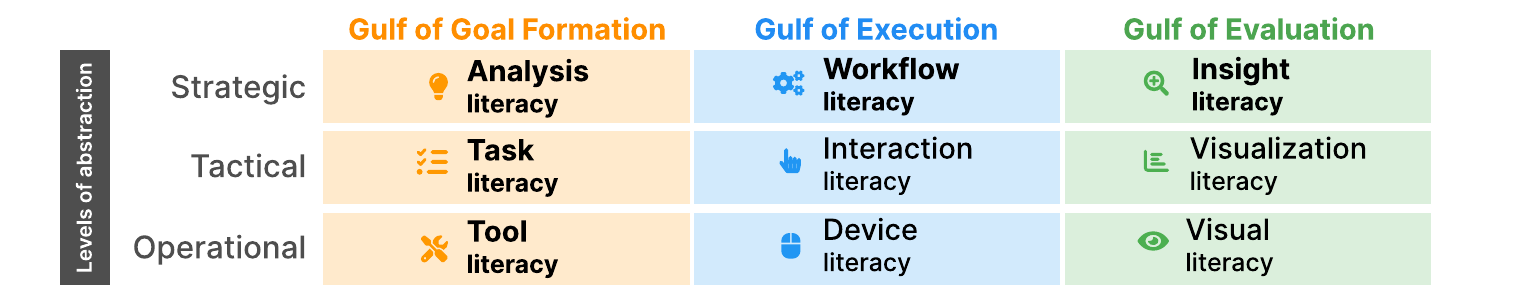}
  \caption{\textbf{Interactive Visualization Literacy model.}
        \normalfont
        Our multiliteracy model describes the visualization action cycle, through three levels of abstraction and three gulfs of visualization interaction.
        The model brings together nine \textit{multiliteracies} (novel ones in \textbf{bold}) that transcend the monolithic view of an all-encompassing visualization literacy for interactive visualization systems.}
  \Description{Matrix-based illustration of the model, representing the three levels as three rows, and the three gulfs as color-coded columns. Tags for each literacy are positioned according to the levels and gulfs, with the time in the x-axis indicating that people go through each gulf in a time sequence.}
  \label{fig:teaser}
\end{teaserfigure}


\maketitle

\section{Introduction}

Using visualizations is an inherently interactive process~\cite{dimara20}.
On a technical level, many computer-based visualizations, shared online or in immersive environments, provide extensive means for interacting with the data through parameter and view changes.
On a cognitive level, interaction allows a person to ask themselves questions while consulting a visualization, generate hypotheses, compare multiple visualizations, or seek additional background information~\cite{klein2006making1, klein2006making2, pirolli2005sensemaking}.
However, interaction can pose a challenge to people wishing to use visualizations to their full extent~\cite{boy16}, such as leading them to miss out on insights, uncritically accept the ``first impression'' of a visualization, or adhere to confirmation bias without questioning the data.
In short, deliberate and conscious engagement with visualization needs a broad approach and relates to diverse abilities, such as critical thinking~\cite{kahneman2011thinking, ge23}, numeracy~\cite{wb24}, and data literacy~\cite{dignazio16}. 

\textit{Visualization literacy} has been defined ~\cite{boy14} as a person's ability to read and create visualizations.
However, while many researchers acknowledge the interactive nature of working with visualizations (e.g.,~\cite{vanWijk2005Value, pike09, dimara20}) and despite explicit calls for a theory of \textit{interactive visualization literacy}~\cite{bach18, bach24}, there is no common definition of such an extended literacy concept and its implications for interactive visualization design and education.
Based on definitions for (static) visualization literacy, we can define interactive visualization literacy (IVL) as

\begin{quote}
\textit{\definitionAbbrv.}
\end{quote}

This definition is derived from a theoretical model we create in this paper to explain what happens during an interaction cycle and what stages, processes, and activities a person is engaged with while interacting with a visualization (Sect.~\ref{sec:action-cycle}).
The model is informed by existing theoretical work from the fields of visualization and human-computer interaction, modeling levels of abstraction as well as gaps (or gulfs~\cite{norman86}), and interaction costs~\cite{lam08}.
Combining these two dimensions, levels of \abstractions and \gulfs, creates a $3\times3$ space that describes nine distinct literacies a person calls upon when interacting with visualizations (Sect.~\ref{sec:literacies}).
We call these nine literacies \textit{multiliteracies} and, consequently, our model a \textit{multiliteracy model} (Fig.~\ref{fig:teaser}).
This extends Norman's initial focus on interface design to people's skills that are transferable beyond the design of a single system, and considers aspects specific to visualization, as suggested by Lam~\cite{lam08}.
Multiliteracies are a common concept in literacy pedagogy~\cite{newlondon1996multiliteracies} and align well with earlier work that describes visualization literacy as a ``multidimensional construct''~\cite{pandey23}.
Five of these multiliteracies are newly derived from our model and fill clear gaps in our understanding of visualization literacy.
Beyond these nine, we also acknowledge additional \textit{foundational literacies} that support sensemaking using visualization, such as data~\cite{dignazio16}, digital~\cite{rushkoff10}, and statistical literacy~\cite{gigerenzer02}.

We scrutinized our model and the resulting multiliteracies through a double approach: describing the literacies in existing example systems in data-driven storytelling, visual analytics, and immersive \& multimodal analytics; as well as through an observational study of \camerarevision{five}{nine} people using some of these systems. 
Our model and its findings provide \camerarevision{an exciting}{a} base to conjecture about the complexity of interactive visualizations and working with visualizations as part of\camerarevision{ interactive and}{} agile analyses workflows.
Consequently, we discuss a research agenda around 
\textit{i)} assessing IVL, 
\textit{ii)} evaluating interactive systems,
\textit{iii)} designing systems with IVL in mind, and 
\textit{iv)} education for IVL.

\section{Background}
\label{sec:rw}

We review prior work on visualization literacy, theories of interaction, and interaction challenges in data visualization.

\subsection{Visualization Literacy and Related Abilities}
\label{sec:rw-vl}

Boy \etal\cite{boy14} defined \emph{visualization literacy (VL)} as ``\textit{the ability to confidently use a data visualization to translate questions specified in the data domain into visual queries in the visual domain, as well as interpreting visual patterns in the visual domain as properties in the data domain.}''
Following this definition, researchers have variously described VL as the ability or skill to make meaning~\cite{boerner16}, extract information~\cite{lee17}, to read and interpret the data~\cite{lee17}, to interpret patterns, trends, and correlations~\cite{boerner16}, and to critically interpret and construct visualizations~\cite{solen22}.
Ge \etal\cite{ge23} also suggested that VL has a critical thinking aspect that involves reasoning about erroneous or misleading visualizations.

To assess someone's visualization literacy, several tests exist (\eg~\cite{lee17, pandey23, cui24}). These tend to involve only \textit{static} visualizations due to the risk of people having a limited ability to detect interactivity~\cite{lee17}. 
Yet, 
Card et al.'s~\cite{card99} seminal definition of visualization includes the use of \emph{interactive} visual representations to amplify cognition, making interaction intrinsic to visualization.
Other researchers likewise emphasize interaction in visualization as well~\cite{vanWijk2005Value, pike09, dimara20}, which suggests an interaction gap in visualization literacy research.
While work that explores \textit{interactive visualization literacy} exists~\cite{alper17, firat22}, \textbf{the community has yet to systematically study the abilities required for interaction.}
For example, recently, Bach et al.~\cite{bach24} called for the visualization community to reflect on advanced interactive visualization tools, highlighting the gap between current educational approaches and the skills needed for real-world visual analytics.

Generally, VL research increasingly indicates that using visualizations involves more abilities than those research has focused on so far~\cite{pandey23, ge25avec, creamer24}.
For example, visualization students should learn to deconstruct and critique visualizations, but these skills are not captured in existing assessments yet~\cite{hedayati25uni}.
Accordingly, Hedayati et al.~\cite{hedayati24reconcept} proposed to reconceptualize visualization literacy by considering multiple competencies and processes, but without considering interaction.
\textbf{With our model, we aim at understanding what interactive visualization literacy is and how it relates to existing notions of visualization literacy.}

Influential to our model is the \textbf{multiliteracies approach}, proposed by the New London Group~\cite{newlondon1996multiliteracies} in the field of general pedagogy.
It proposes a framework for going beyond monolithic views of literacies, arguing that we should no longer think of literacy as a single language-based skill, but rather as a set of multiple literacies connected to cultural diversity, context, and computer-based communication.
This approach led others to acknowledge the relationship between standard literacy and other literacies applicable to visualizations, such as \emph{visual literacy}~\cite{kress20, Dondis1974} and \emph{critical literacy}~\cite{luke12}, or numeracy~\cite{xexakis21}. 
For example, \emph{Data literacy} which has been defined by D'Ignazio and Bhargava~\cite{dignazio16} as \textit{``the ability to read, work with, analyze and argue with data as part of a larger inquiry process,''} forming another foundational layer for visualization.
Or, \emph{statistical literacy}~\cite{Rumsey2002} which is the ability to understand statistical concepts and reasoning and which becomes crucial when visualizations represent statistical transforms or uncertainty.
\emph{Visual literacy}~\mbox{\cite{Dondis1974, Barry1997}}, the ability to interpret visual information in general, underpins all visualization comprehension.

If visualizations are interactive, more literacies become relevant, such as computer literacy~\cite{kegel19} and device literacy~\cite{velghe14}.
Bach~\cite{bach18} coined the concept of \emph{interaction literacy} for visualization as \textit{``the ability to explore and interrogate data through visualization interfaces,''} proposing that it exists on a spectrum from basic interface understanding to comprehension of visualization models.
From these arguments, we derive a first criterion for a model for interactive visualization literacy:


\competencybox
{\faTasks~C1: Heterogeneous competencies}
{Effective visualization interaction requires diverse skill sets spanning
perceptual, cognitive, motor, and analytical domains that operate simultaneously.}

\subsection{Theories of Interaction}
\label{sec:rw-theories}

A core idea that led to the development of the field of human-computer interaction (HCI) is that interactive systems can be difficult to use~\cite{hornbaek25book}.
Don Norman, one of its pioneers, introduced the \emph{human action cycle}~\cite{norman88} as a model that describes how humans interact with computers.
The seven-stage model describes interaction as a goal-directed dialogue~\cite{hornbaek25book} where a person sets a goal, plans their actions, specifies them, and then executes them.
After the action, the person perceives the change, interprets the result, and evaluates whether the goal was achieved.
The model groups the seven stages around a gulf of execution and a gulf of evaluation, which help assess how user interfaces support task performance through a design that considers affordances, visibility, and feedback.
The cycle is built on work with Hutchins and Hollan~\cite{hutchins85} and on Shneiderman's concept of direct manipulation~\cite{shneiderman82dm}, where users interact with visual representations of objects rather than through command languages.

Despite decades of research, there are still open questions regarding what interaction means in the field of HCI~\cite{hornbaek19}.
Hornbæk and Oulasvirta~\cite{hornbaek17} identified seven different concepts of interaction in HCI literature, including interaction as dialogue, highlighting the multifaceted nature of this fundamental concept.
Jacob et al.~\cite{jacob08} proposed reality-based interaction as a unifying concept for post-WIMP interfaces, emphasizing how they leverage users' pre-existing knowledge of the everyday, non-digital world.
Overall, the distributed cognition framework~\cite{Hutchins_1995_CognitionInTheWild}, introduced to HCI by Hollan et al.~\cite{hollan00}, offers a theoretical foundation that considers the design of interaction technologies with a human-centered focus.
This perspective views cognitive processes as distributed across individuals, artifacts, and time rather than confined to individual minds. These theories point to the cyclical nature of interaction and its interdependencies with multiple elements, leading to a second criterion for our definition.


\competencybox
{\faTasks~C2: Iterative processes}
{Interaction happens over time, in a highly iterative and often agile way consisting of creating goals and evaluating these goals.
These processes involve different artifacts: interfaces, devices, visualizations, and users.
}

\subsection{Interaction for Visualization}

In visualization, interaction has also been considered an ambiguous term, leading Dimara and Perin~\mbox{\cite{dimara20}} to review existing notions and propose a new definition of interaction in visualization as \emph{``the interplay between a person and a data interface involving a data-related intent, at least one action from the person and an interface reaction that is perceived as such.''}
There are multiple taxonomies that help us understand what kind of interactions are relevant to visualization.
For example, Yi et al.~\cite{yi07} identified seven categories of interaction intents to classify what users want to achieve~\cite{DBLP:conf/infovis/AmarES05}, emphasizing the goal-directed nature of interactions. Heer and Shneiderman's taxonomy~\cite{heer12} proposes 12 task types assuming \textit{``a basic familiarity with visualization design,''} highlighting how interaction taxonomies often presume a level of visualization skills without explicitly addressing how these skills relate to interaction. 
Eventually, Liu and Stasko~\cite{liu10} investigated how users develop mental models during interaction and found that the ability to interact depends not only on motor skills but also on \textbf{cognitive representations} that develop through visualization experience.

Elmqvist et al.~\cite{elmqvist11} proposed \emph{fluid interaction}, arguing that effective visualizations should leverage direct manipulation~\cite{DBLP:journals/computer/Shneiderman83} and bridge Norman's gulfs~\cite{norman86} to support seamless transitions between different analytical states.
Building also upon Norman's action cycle~\cite{norman88}, Lam~\cite{lam08} proposed three types of \textbf{interaction costs for visualization}: 
physical-motion costs (gulf of execution), 
cognitive-interpretation costs (gulf of evaluation), and 
cognitive-decision costs (gulf of goal formation).
This framework adapts Norman's cycle to visualization contexts, highlighting the cognitive demands unique to data exploration.
It also connects well to Pirolli and Card's sensemaking loop~\cite{pirolli2005sensemaking}, as well as the cost structure of this loop Russell et al.~\cite{russell1993cost} contributed. The gulfs imply notions of ``bridging gaps'' that we address in our model.

Past research has documented various challenges when interacting with visualizations.
A fundamental issue is simply recognizing that a visualization supports interaction.
Boy et al.~\cite{boy16} found that the majority of participants in their experiment failed to discover that visualizations were interactive and proposed \emph{suggested interactivity} cues---graphical elements designed to signal interactive areas.
This aligns with Norman's gulf of execution~\cite{norman86}, where people must recognize potential actions before executing them.

Beyond discovery, users face challenges in knowing what interactions to perform and how to execute them.
Kwon et al.~\cite{kwon11} identified four types of roadblocks novices face with visual analytics systems: failure to choose the appropriate view for the task, failure to execute required interactions, failure to interpret the visualization, and failure to match expectations with system functionalities.
AlKadi et al.~\cite{alkadi23} identified barriers analysts face during interactive exploration, such as finding the right level of abstraction for their data, lacking specific analysis goals, and having preconceived ideas that limit exploration. 
Nobre et al.~\cite{nobre24} identified conceptual and operational barriers with static visualizations, such as misunderstanding visual encodings and reading incorrect values from axes.
These roadblocks and barriers highlight how successful interaction requires both conceptual understanding and operational skills.
So, while interaction supports insight development~\cite{pike09}, it requires more skills and may lead to cognitive overload on advanced tasks~\cite{mosca21bayesian}.
The way interaction techniques are implemented can also lead to user challenges.
For instance, interactive computations can lead to high latency, which reduces the rate at which people make observations and generate hypotheses~\cite{liu14latency}.

These findings all point to a need for more comprehensive interaction frameworks that bring these fragmented pieces together.
The challenges and barriers listed above seem to involve different levels of cognitive and technical skills.
We see this as our third and last criterion for our model:


\competencybox
{\faTasks~C3: Engagement on different levels}
{Interactive visualization involves actions spanning different temporal and cognitive scales~\cite{bertin10, kahneman2011thinking}, from millisecond motor responses to long-term analytical strategies.
}

\section{Method}

Our author team is composed of researchers with a combined 20+ years of expertise in interaction design, visualization literacy, multimodal interaction, and immersive analytics, as well as with diverse levels of experience on designing, implementing, evaluating, and teaching data visualization.
Inspired by the ongoing discussions on whether people use interactive visualizations to their full extent~\cite{aisch17, bach24, boy16}, what the \emph{visualization literacy} term covers~\cite{creamer24, hedayati24reconcept}, and the role of other personal characteristics~\cite{leon25, cabouat24}, we decided to investigate what abilities are involved in visualization interaction.

We developed the model in six phases.
In phase I, we \textit{exchanged personal experiences}, reflecting on projects and collaborations, the investigated interaction choices we made, how to foster interaction skills, and the challenges people faced while solving diverse visualization tasks.
This led to recognizing the personal knowledge we had developed and identifying diverse challenges and skills that traditional visualization literacy did not yet cover.

For phase II, we conducted a \textit{literature search} on interaction definitions~\cite{hornbaek17, dimara20}, affordances~\cite{norman88}, suggested interactivity~\cite{boy16}, interaction literacy~\cite{bach18}, and interactive visualization literacy~\cite{firat22}.
While prior work hints at different skills and interaction components, it does not explain how these skills are interconnected.
Thus, we proceeded to collect and curate a list of models (\eg~\cite{jansen13}), taxonomies (\eg \cite{yi07}), typologies (\eg~\cite{brehmer13}), frameworks (\eg~\cite{lam08}), and related approaches (\eg~\cite{nobre24}) that aim to explain how people use visualizations, related challenges, and literacy gaps.
We extended the list with work on literacies relevant to visualization, which we briefly summarized in Sect.~\ref{sec:rw-vl}. 
We grouped concepts and discussed their interdependencies in multiple iterations.
This phase led to recognizing the need for a multiliteracy model that describes the necessary skills for interaction and supports design according to those skills.

In phase III, we \textit{developed an initial coding} of interactive examples based on six visualization scenarios: visual analytics, data-driven storytelling, visualization authoring, visualization annotation, and immersive \& multimodal environments.
We gathered 15 real-world systems and research prototypes from these scenarios, focusing on the diversity of interactive features and\camerarevision{ target}{} audiences.
Then, three authors coded them qualitatively according to the abilities they may require.
\camerarevision{}{There were two coders per system, including at least one senior researcher. All coders discussed any disagreements in a follow-up meeting.}
In phase IV, we \textit{iteratively created the model}, based on the criteria identified through prior work (Sec.~\ref{sec:rw}) and the initial coding. 
In phase V, we focused on \textit{conducting a definitive coding of the examples}, following convergence on the final model. The goal was to scrutinize the model, focusing on three contrasting scenarios described in Sec.~\ref{sec:examples} and the appendix.
\camerarevision{}{Each coder was responsible for one example per scenario.}
The final and sixth phase was \textit{validating the model} through an observational study (Sec.~\ref{sec:study}).

Discussions took place over 10 months, supported by biweekly meetings of the author team (in person, virtual, and hybrid) and asynchronous work on shared documents and physical and digital whiteboards. 
We included sketches and notes we produced while developing the model in the supplemental material.

\section{A Multiliteracy Model for Interactive Visualization Literacy}
\label{sec:action-cycle}

Our review of prior work surfaced three important criteria (C1---3) to define and describe interaction visualization literacy (IVL).
To capture these criteria, we propose a \textit{multiliteracy model} that consists of two major components:
(1) a \textbf{visualization action cycle} based on Norman's human action cycle~\cite{norman86} that supports multiple levels of abstraction (C3) as well as continuous and cyclical interaction over time (C2); and
(2) a set of nine \textbf{multiliteracies} that capture the diverse and heterogeneous skills required to interact with visualizations (C1).

Our goal is to inform interventions to overcome eventual barriers and support skills and skill development.
The model consists of two dimensions, spanning a 3$\times$3 space of multiliteracies, as illustrated in Fig.~\ref{fig:teaser}.
\camerarevision{In the following sections, we first}{Next, we} describe the visualization action cycle and its dimensions, followed by the multiliteracies
involved.

\iftrack{}
\else
\subsection{The Visualization Action Cycle}
\fi

We capture the temporal and cyclical nature (C2) of our model for interactive visualization literacy by adopting and extending Norman's \textit{human action cycle}~\cite{norman86}.
Norman proposed the cycle, described in Sect.\ref{sec:rw-theories}, as a theoretical tool to guide designers in creating computer systems that support people in interacting efficiently, through careful design choices that consider multiple aspects, such as affordances and feedback.
While design remains a fundamental element that shapes interactive experiences, our model focuses on the system users and their abilities, which are not necessarily covered by the design and that extend beyond a single design. 
We advocate a proactive approach, similar to VL, thinking of long-term processes that go beyond the interaction with a single system.

While Norman implicitly acknowledges the existence of interaction at different levels of scale (C3), we explicitly introduce three such levels in our adaptation: \textit{strategic}, \textit{tactical}, and \textit{operational}.
\camerarevision{Furthermore, in}{In} Norman's model, a person interacting with a computer overcomes specific gaps---or gulfs---related to perception (\eg understanding a system message or reading a chart) or motor actions (\eg selecting an object or writing text).
Norman calls these the Gulfs of Execution and Evaluation.
To \camerarevision{this}{these}, we add a third gulf: the Gulf of Goal Formation, as described by Lam~\cite{lam08}.
Combining these three gulfs at three different levels of abstraction yields a multi-scale, cyclical, and temporal \textit{visualization action cycle} as shown in Fig.~\ref{fig:multicycles}.

Norman's model~\cite{norman86} emphasizes that people need to understand the interactive object to map its properties to their goals and actions.
The same holds for interactive visualizations: interacting requires reading and interpreting the visualization, which is the core idea of visualization literacy~\cite{pandey23}. 
Note that it does not matter where an interaction starts; \ie at the stage of evaluating a visualization (gulf of evaluation), at the moment of triggering a visualization change (gulf of execution), or way earlier while planning and considering an interactive visualization (gulf of goal formation).
While Lam~\mbox{\cite{lam08}} suggests starting with goal formation, evaluating the current state beforehand may be more suitable for visualizations, \camerarevision{while}{whereas} people may quickly jump to acting depending on the context.
Still, the goal formulation, the execution, and the evaluation require additional abilities beyond mere interpretation, as implied by the traditional view of VL.
Thus, Norman's model helps us think systematically about the abilities involved in visualization interaction.

We structure the \textsc{Gulfs} as one dimension and the levels of \textsc{Abstraction} as another.
The two dimensions can hence be imagined as a two-dimensional orthogonal space in which the cross-product generates the nine distinct multiliteracies that we discuss in Sect.~\ref{sec:literacies}.
The metaphor of gulfs also implies that there are two main ways of narrowing any of these gulfs; better education for improving the corresponding literacies, and better interfaces supporting people.

\begin{table}
    \caption{\textbf{The human action cycle.}
    \normalfont The original seven stages and two gulfs in Norman's model for interactive systems~\cite{norman86}.
    We have also added the Gulf of Goal Formation that was present as a stage in Norman's framework, but was elevated to a gulf by Lam~\cite{lam08}.
    }
    \label{tab:action-cycle}
    \begin{tabular}{lll}
    \toprule
    \textbf{Stage} & \textbf{Process} & \textbf{Task}\\
    \midrule
    \rowcolor{goallitcol!20}
    \multicolumn{3}{l}{\textbf{Gulf of Goal Formation}}\\ 
    1.\ Forming goal & Cognitive & What do I want to achieve?\\
    \rowcolor{execlitcol!20}
    \multicolumn{3}{l}{\textbf{Gulf of Execution}}\\ 
    2.\ Forming intention & Cognitive & What plan will achieve goal?\\
    3.\ Specifying action & Cognitive & What actions are needed?\\
    4.\ Executing action & Physical & Carrying out actions\\
    \rowcolor{evallitcol!20}
    \multicolumn{3}{l}{\textbf{Gulf of Evaluation}}\\ 
    5.\ Perceiving state & Perceptual	& Detecting response\\
    6.\ Interpreting state & Cognitive & Making sense of feedback\\
    7.\ Evaluating result & Cognitive & Judging goal completion\\
    \bottomrule
    \end{tabular}
\end{table}

\subsection{Dimension 1: Gulfs of Visualization Interaction}

In our model, gulfs represent gaps a user has to overcome to interact.
Norman's gulfs~\cite{norman88} provide an ideal theory for this which we have translated directly to IVL.
The gulf of execution represents the difference between a given state of a visualization (or an analysis) and the desired state.
The gulf of evaluation is the gap between that new state and its interpretation.
Traversing both of these gulfs, Norman describes seven \textit{action stages} to describe the process of interacting and leading a user through the gulfs.
In IVL, these stages represent the different steps of interacting with a system as we detail in the following. \autoref{tab:action-cycle} lists these stages for the three gulfs in the IVL model. 

The \textbf{Gulf of Execution} describes the gap between a person's intent and the necessary input actions supported by the system. 
Hence, a person would define intentions for their analysis and interactions, then specify their concrete actions and visualization states, and eventually execute their analysis steps and actions with the system. 
A narrow gulf of execution will make it easier for a person to perform interactions.
In contrast, a visualization with few interactive capabilities may block potential analyses.
So far, VL research has not considered user input, except for the assessment of encoding abilities~\cite{ge25avec} and the use of linked views to explain specific visualizations \cite{firat23treemap}.
By drawing attention to the gulf of execution, we extend the theoretical scope to include any input-related abilities.

The \textbf{Gulf of Evaluation} describes the gap between the system's (output) state and the person's understanding of that state.
For interactive visualization, this gulf is a measure of the expressiveness and effectiveness of the interface in supporting the understanding of its changes.
According to Norman's stages, a user perceives the visualization state, then interprets what the visualization is saying about the data, and finally assesses whether these findings help with their goal.
A narrow gulf of evaluation means the interface effectively conveys the data, visualization, and state changes to the person. 
A poorly designed chart may hinder a person from understanding the data, further preventing them from performing their analysis.
Most VL work to date is situated in this gulf, as the focus has been on interpretation~\cite{lee17, cui24}.
However, the gulf of evaluation goes beyond that: from perceiving visual properties to judging the value of the visualization for reaching a goal.
This gulf helps assess whether the visualization provides enough affordances for people to obtain and derive knowledge~\cite{burns20}.

\textbf{Gulf of Goal Formation}---%
However, the gulfs of execution and evaluation miss one crucial aspect of interactive visualization.
In their multi-level typology, Brehmer and Munzer~\cite{brehmer13} relate Norman's gulf of execution as to \textit{how} a person interacts with a visualization.
They refer to Lam's gulf of goal formation~\cite{lam08} as to \textit{why} a person uses a visualization and the tasks they wish to perform.
With that gulf, Lam recognizes the high decision costs inherent in data analytics and sensemaking by proposing a third gulf in relation to the process of identifying interaction goals.
Hence, we add the gulf of goal formation, which provides us with more expressivity to describe abilities related to motivating and defining goals in using interactive visualizations.
Narrowing this gulf means supporting people formulate meaningful questions when they lack a clear plan (\textit{1.\ Forming goal}).
Goal formation has not yet been studied in visualization literacy work because assessments usually involve predefined tasks (e.g., \cite{lee17, ge23}). 

The gulfs originally described by Norman, which we translated to interactive visualization, are strictly speaking \textbf{not} the same gulfs at each of the levels as shown in Fig.~\ref{fig:multicycles}.
Norman's original gulfs describe information flows between the person and the interface and back.
The gulfs we introduce describe ability and knowledge gaps.
We find the metaphor of gulfs highly appropriate to talk about challenges, as well as literacies, and technology to overcome them, so we stick to the terminology of gulfs.
However, we call Norman's gulfs \textbf{\intergulfs}, and our novel gulfs, at multiple levels of abstraction, we call \textbf{\intragulfs}.
While Norman's gulfs are meant to be bridged by designers, we look at abilities that allow anyone to use any interactive visualization system to their full extent.
Design flaws can cause usability issues, but interaction abilities allow people to identify those flaws.
Awareness of those abilities helps us distinguish between design shortcomings and insufficient abilities as the cause of the issue.
Moreover, if we need a complex visualization design for a complex problem, how can we help people make use of it?
\camerarevision{}{
Designing visualization systems with interaction abilities in mind is not too different from the ability-based design approach of Wobbrock et al.~\cite{wobbrock11}, which aims to focus on designing according to what people can do, or from participatory design interventions that prioritize the literacy practices of the community \cite{taylor18}. 
Norman \cite{norman88} acknowledges that people can commit errors in any stage of the cycle (\eg choosing an inappropriate goal), so we use the cycle structure to study the abilities related to such errors systematically. 
}

The action cycle and these three gulfs nicely explain the flow of information, \ie how interaction is \textit{technically} transmitting information from the person to a visualization and back.
However, the gulfs and stages do not explain higher-level cognitive mechanisms involved in interacting:
\textit{Is this visualization appropriate for my goals?}
\textit{What would that ideal visualization be?}
\textit{What shall I do next?} etc.
In other words, the traditional action cycle only concerns \textit{one} linear sequence where one action leads to another in a continuous flow of information.
Yet, the above questions hint towards multiple abstraction levels, where a user sets out the overall goal of the analysis and proceeds to implement it.

\subsection{Dimension 2: Levels of Abstraction}
\label{sec:levels}

That sensemaking processes exist at multiple levels has already been acknowledged in prior models of graph comprehension~\cite{bertin10, russell1993cost, pirolli2005sensemaking}.
Inspired by Brehmer and Munzner's multi-level typology of visualization tasks~\cite{brehmer13}, we thus integrate three levels of abstraction into our model.
These levels are necessary to explain the differences between the cognitive actions a person is engaged in when analyzing data and a visualization as a whole, as opposed to perceiving single elements and performing single actions to leverage specific interaction techniques.
However, rather than calling these levels \textit{low-level}, \textit{mid-level}, and \textit{high-level} as suggested in Brehmer and Munzner's typology, or \textit{elementary}, \textit{intermediate}, and \textit{overall} as suggested by Bertin~\cite{bertin10}, we adopt terms from decision support that better express the semantics of our levels (e.g., used in dashboard design~\cite{few2006information, sarikaya2018we} in relation to visualization, analysis, and decision making): \textbf{operational}, \textbf{tactical}, and \textbf{strategic}.
Each level deals with different \textit{scopes} of actions, including the time frame and area of influence.
This notion of scopes is consistent with the dual-process theory of cognition~\cite{kahneman2011thinking}, which distinguishes between fast (intuitive) and slow (deliberate) thinking processes at different levels.
Furthermore, we interpret interaction according to the definition of Dimara and Perin~\cite{dimara20}, assuming that the person interacting has a data-related intent, and the interface reaction will lead to either reaching that goal or planning a new attempt.
The gulfs repeat at each abstraction level with their respective scopes (Fig.~\ref{fig:multicycles}).

\subsubsection{Operational Level}

The lowest abstraction level concerns anything related to \textit{technically} executing an interaction, \ie a state change.
Operations are the immediate interactions with the interface and its input modalities.
The person needs to identify the interaction technique (\eg a lasso selection) and then execute the respective motor actions.
Operations that involve long physical interaction may require constant updating and adjusting to succeed.

The operational cycle (Fig.~\ref{fig:multicycles}-bottom) consists of a person controlling and manipulating a visualization system through a specific input device.
The gulf of execution describes performing interaction techniques to manage the motor actions required for manipulations.
The gulf of evaluation describes the understanding of a technique's effect on the visualization.
This level corresponds to the activity of \textit{encountering a visualization} in the NOVIS model~\cite{lee16} because it is where people perceive the first impression of the visualization as an image, before attempting to make sense of it.
In the model of Hedayati et al.~\cite{hedayati24reconcept}, the operational level corresponds to bottom-up factors, such as recognizing text objects.
This level extends beyond these models, as it also considers the low-level actions people perform to interact and their familiarity with the specific tool.

An example for the operational cycle is the application of an interactive semantic lens on a visualization.
The lens needs to be moved to a new position (execution), its updated content interpreted (evaluation), and potentially a new movement has to be started (goal formation).
Actions on this cycle can span from a few milliseconds to a few seconds.

\subsubsection{Tactical Level}

A tactic (Fig.~\ref{fig:multicycles}-center) is a focused sequence of steps designed to accomplish an immediate objective. 
In the context of IVL, the tactical level concerns a concrete set of actions to obtain a desired visualization state.
This level is at the boundary of cognition and technical implementation; it can include the set of intermediate states, e.g., a clustering or filtering in an exploratory scenario (execution).
Consequently, the person assesses whether the new state is fit for their task (evaluation).
Eventually, a new task is selected to be addressed next (goal formation).  

The tactical level is where data becomes part of the process, \ie where the person plans, acts, and interprets, translating between the visual domain and the data domain.
This level includes a wide range of tasks that associate user intent with interaction techniques, such as attribute-based filtering, sorting elements, and tracking changes over time, as defined in Yi et al.'s taxonomy~\cite{yi07}.
The scope of the tactical level can encompass more than one visualization and can span several seconds up to a minute.

\subsubsection{Strategic Level}

\begin{figure}[htb]
    \centering
    \includegraphics[width=\linewidth]{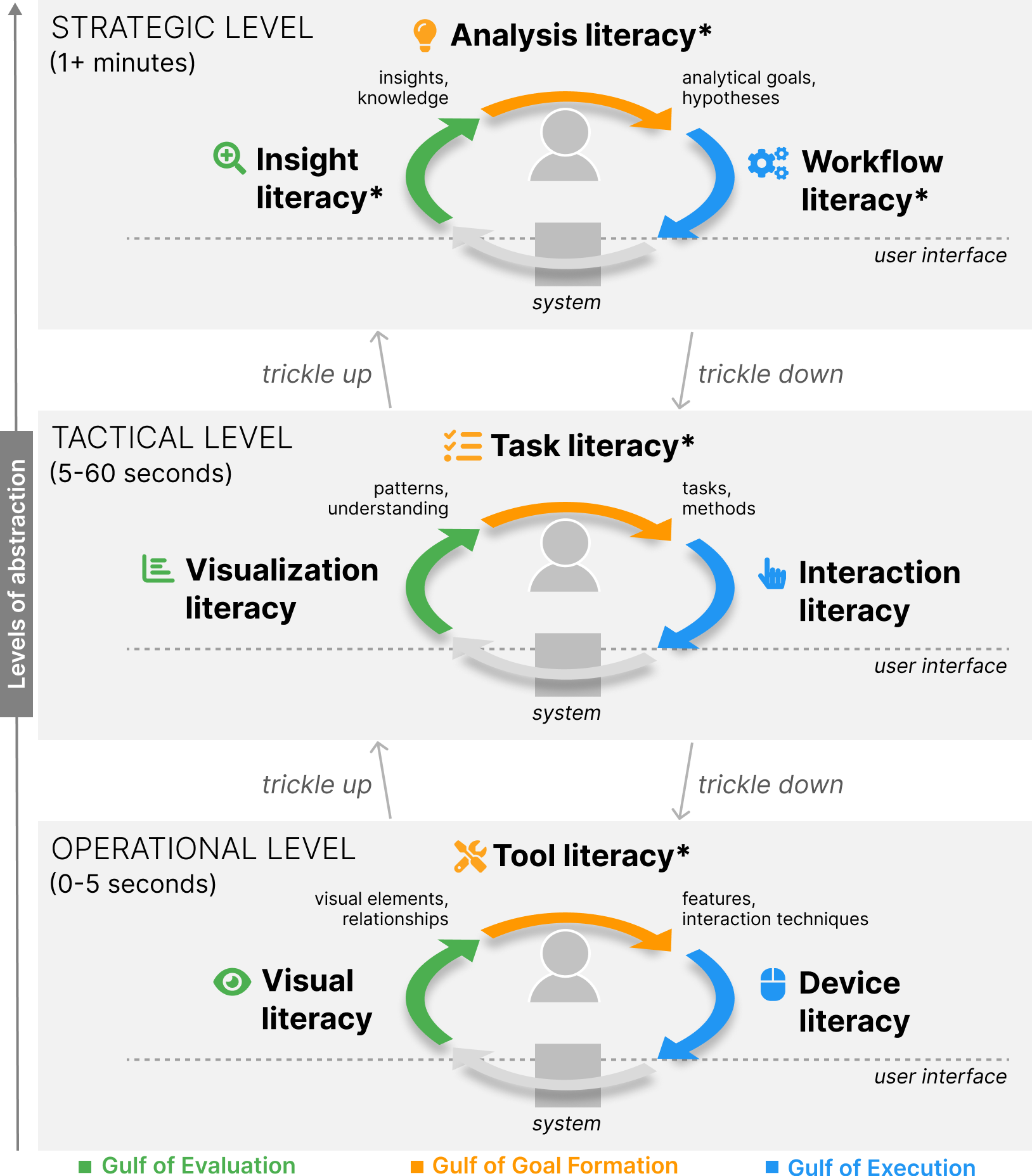}
    \caption{\textbf{A multi-level visualization action cycle.} 
      \normalfont
      The model expands the three gulfs of interaction (arrows) across three levels of abstraction (panels). 
      Interaction involves leveraging specific literacies to overcome the gulfs at each level.
      People plan, act through the interface, or understand the current visualization state with different outcomes and time frames at each level.
      They can start at any part of the cycle.
      The arrows between levels indicate that what happens at one level influences what happens in the others.
      }
    \Description{Illustration composed of three rectangles, ordered vertically. Each rectangle shows an arrow-based cycle with a user and a visualization system at its center, connecting three literacy concepts at the same level of abstraction. Arrows between the rectangles indicate the connections across levels.}
    \label{fig:multicycles}
\end{figure}

At the highest level of abstraction (Fig.~\ref{fig:multicycles}-top), a strategy is a structured approach to achieve long-term or overarching goals.
It involves setting clear objectives and priorities, identifying the actions required to reach those objectives, and allocating resources to carry out those actions effectively. 
In the context of IVL, this level represents the high-level steering of a visualization analysis task, which includes planning, organizing, and performing interactions to implement a specific approach; for example, to decide on an analysis goal and method (e.g., comparing clusters), a (set of) visualization techniques (e.g., a multidimensional projection), or a visualization tool (e.g., R).
People need to define the visualization state that can potentially achieve their goals and make a plan to obtain this state.

At this level, we evaluate the understanding of the data, as well as question and report the main insights. Knowledge is generated at the strategic level.
The scope\camerarevision{ of this level}{} encompasses the entire process of interactively analyzing data and can span from minutes up to\camerarevision{ several}{} days or months. 
It may involve different visualizations and datasets. 

In Fig.~\ref{fig:multicycles}, the \intragulfs\. are moving information \textit{vertically} between levels of abstractions as shown by the dashed arrows.
These gulfs exist between turning goals into tasks and interactions, as well as interpreting visualizations and matching goals.
The figure also shows that actions at all levels are composable and hierarchical.
When interacting, the user is constantly bridging the \intragulfs\. at each level: making strategic plans and goals; performing tactical tasks and making decisions; and executing motor actions and interpreting output. 
In the process, the user will also attempt to match the interpretation of a low-level result with higher-level goals (\intergulfs\.).
In other words, interacting with visualization systems involves continuous exchanges between each adjacent level in the action cycle and the continuous bridging of \intergulfs\. and \intragulfs\..

\section{Multiliteracies}
\label{sec:literacies}

The problems and associated costs that Lam~\cite{lam08} identified on each stage of the interaction cycle indicate that people leverage multiple abilities to interact with visualizations.
The visualization action cycle suggests assigning these abilities to the three gulfs (goal formation, execution, and evaluation) at three abstraction levels (strategic, tactical, and operational).
This approach transforms the monolithic view of a single ``visualization literacy'' into a total of nine \textit{visualization multiliteracies}~\cite{newlondon1996multiliteracies}.
Each literacy addresses specific cognitive and practical challenges users face when working with interactive visualizations.
Four literacies have established foundations in the literature, while five represent novel formulations.
We mark each literacy we believe novel to our model with an asterisk*.
In the following, we group the literacies by the gulfs they relate to.

\subsection{Goal Formation Literacies}

Goal formation literacies address the cognitive processes involved in formulating goals at different abstraction levels, from high-level analytical strategies to immediate operational objectives, enabling users to bridge the gulf between their analytical intent and the chart's capabilities. They help users decide what actions to perform.
    
\begin{itemize}

    \item\analysislit (strategic-goal formation) is the ability to formulate high-level analytical goals and frameworks for understanding complex data through visualization.
    This includes developing appropriate mental models of the data, generating testable hypotheses, and constructing meaningful narratives from insights.
    This connects directly to Pirolli and Card's sensemaking loop~\cite{pirolli2005sensemaking}, where analysts develop schemas and hypotheses about complex problems.
    Unlike general sensemaking, analysis literacy specifically focuses on planning---deciding what analytical approach and inquiries will reveal meaningful insights.

    \item\tasklit (tactical-goal formation) is the ability to decompose high-level analytical goals into concrete, achievable tasks and derive effective sequences of operations.
    This builds on Brehmer and Munzner's task typology~\cite{brehmer13} and Yi et al.'s interaction taxonomy~\cite{yi07}, focusing on selecting appropriate mid-level operations.
    Task literacy thus centers on analytical reasoning: understanding which visualization tasks (filtering, comparing, summarizing) will answer specific questions.
    Low task literacy can also lead to the conceptual barrier of misunderstanding a given task~\cite{nobre24}.

    \item\toollit (operational-goal formation) is the ability to determine how to achieve a specific analytical goal given a specific visualization tool.
    This includes understanding what specific tool functions accomplish, recognizing when to apply particular interactions, and formulating efficient pathways to access needed capabilities.
    It can be remedied by low-level visualization guidance~\cite{ceneda2016characterizing}, such as Chundury's contextual help~\cite{chundury2023contextual} and Boy et al.'s suggested interactivity cues~\cite{boy16}: recognizing what actions are possible and selecting the right one.
    This literacy operates at the level of individual interface elements rather than analytical strategies.

\end{itemize}

\subsection{Execution Literacies} 

Execution literacies encompass the skills needed to implement analytical goals into concrete actions across different levels of complexity, from coordinating complex analytical workflows to precisely manipulating interfaces, effectively bridging the gulf between intention and action.
They support the execution of interactive actions.
    
\begin{itemize}

    \item\orchlit (strategic-execution) - The ability to coordinate and execute complex, multi-stage analysis using interactive visualizations.
    This includes managing analysis workflows across multiple visualizations, integrating insights from different data perspectives, and adapting analytical approaches based on progressive discoveries.
    Unlike task literacy (which plans what to do), workflow literacy focuses on process management of executing long-term analytical campaigns: managing data provenance, coordinating insights across tools, and maintaining analytical coherence as investigations evolve.

    \item\intlit (tactical-execution) - The ability to effectively employ and combine appropriate visualization interaction techniques to achieve specific analytical goals~\cite{bach18}.
    Building on Bach's definition~\cite{bach18} and Yi et al.'s taxonomy~\cite{yi07}, this involves understanding how interaction techniques combine to support analytical reasoning.
    This includes mastery of standard interactive visualization operations like filtering, brushing, linking, zooming, and selecting, as well as understanding how different interactions complement each other. 
    Unlike device literacy (which handles how to manipulate interfaces), interaction literacy focuses on which interactions achieve analytical goals and how they work together.

    \item\devlit (operational-execution) - The ability to efficiently operate the input mechanisms that control visualization interfaces.
    This connects to Velghe's device literacy concept~\cite{velghe14}, and includes fluency with input devices (mouse, keyboard, touchscreen, \etc), understanding their affordances and constraints, and executing precise motor actions for effective interaction.
    In other words, unlike interaction literacy (which chooses analytical operations), device literacy handles the low-level sensorimotor skills needed to manipulate interface controls precisely and efficiently.

\end{itemize}

\subsection{Evaluation Literacies}

Evaluation literacies involve the skills needed to interpret, understand, and derive meaning from visualization outputs at varying levels of abstraction, from recognizing basic visual patterns to synthesizing complex insights across multiple views, helping users bridge the visual representation and the data.
They help the interpretation of the visualization system at its current state, to either support the next action or reach the final result.

\begin{itemize}

    \item\evaluationlit (strategic-evaluation) - The ability to critically assess sequences of visualizations against high-level analytical goals, connecting visualization discoveries to broader domain knowledge and theoretical frameworks.
    This includes validating visualization insights against existing understanding, identifying contradictions or confirmations, and integrating findings into broader theoretical or practical frameworks.
    It connects to the highest levels of Bloom's taxonomy~\cite{bloom1956taxonomy}---evaluation and synthesis---and specifically addresses the ``knowledge recall'' and ``questioning the frame'' activities in Lee et al.'s NOVIS model~\cite{lee16}.
        
    \item\visualizationlit (tactical-evaluation) - The ability to interpret and derive meaningful patterns, trends, and relationships from data visualizations~\cite{boy14, boerner16}.
    This includes understanding standard visualization techniques and encodings, recognizing significant data features (clusters, outliers, trends), and connecting visual patterns to their data implications.
    Although this literacy was initially generalized as \textit{retrieving information} from any visualization~\cite{lee16}, it can be assessed per technique~\cite{firat22parallel, zoss2018network}.

    \item\visuallit (operational-evaluation) - The ability to accurately perceive and interpret basic visual elements and encodings.
    While this concept generally covers any visual artifacts, we focus on the artifacts that correspond to visualization elements in the context of IVL.
    This includes perceiving color, distinguishing between shapes, and accurately perceiving relative positions, sizes, and other visual attributes.
    Rooted in Dondis's visual literacy theory~\cite{Dondis1974} and Barry's visual intelligence concept~\cite{Barry1997}, this handles the perceptual foundation that enables all higher-level interpretation.
    This literacy is important when \textit{encountering a visualization}~\cite{lee16} to help create a first impression of the visualization or to recognize new elements after interacting.
    
\end{itemize}

Across levels, literacies build upon each other; \ie recognizing a cluster requires perceiving the marks that form the group.
However, there is no strict hierarchy because, for example, a colorblind person can excel at the mini-VLAT test thanks to the use of colorblind-safe charts~\cite{pandey23}.
The scope of each literacy goes across multiple tasks, devices, and techniques, but each visualization system requires only being skillful with specific ones.
\textit{Interactive visualization literacy} is the ensemble of the nine multiliteracies embedded into the action of traversing the two dimensions of the model.

Beyond the literacies in our model, several \textit{foundational literacies }serve as prerequisites for successfully navigating the visualization action cycle.
Besides the traditional foundations, such as standard \foundlit{literacy} and \foundlit{numeracy}, there are technology-oriented foundations, such as \foundlit{computer literacy}~\cite{kegel19}, and data-oriented foundations, such as \foundlit{data literacy}~\cite{dignazio16} and \foundlit{statistical literacy}~\cite{Rumsey2002}. Next, we contrast our multiliteracy model with real-world systems.

\section{Scrutinizing the Model}
\label{sec:scenarios}

To scrutinize our model, this section reports on 
\textit{(a)} using the model to describe interaction with systems for data-driven storytelling, visual analytics, and immersive \& multimodal analytics, as well as 
\textit{(b)} an observational study with \camerarevision{five}{nine} participants, classifying their interactions with any of the systems in \textit{(a)}.

\subsection{Case Studies: Describing Multiliteraties in Existing Systems}
\label{sec:examples}

For each scenario, we selected three examples from researchers and practitioners where a demo, source code, or videos were openly available so that we could examine the system closely.
In the following, we present a representative case from the first two scenarios, describing the system interventions provided to shrink each gulf, and discussing the literacies people require to interact with the system.
We included the remaining scenario and cases in the appendix.

\subsubsection{Data-driven Storytelling}
\label{sec:storytelling}

Data-driven storytelling is often designed for the general public and combines text with custom visualizations and animation. While interaction is usually kept simple, allowing people to 
navigate the story linearly,
at times interaction is used to branch from that linear story and explore on demand~\cite{segel2010narrative, wang2021interactive}.

\paragraph{The Atlas of Sustainable Development Goals (2023).}

This award-winning project of the World Bank~\mbox{\cite{pirlea23}} tells an interactive visual story for each of the 17
Sustainable Development Goals (SDGs).
Reading and interacting with the visualizations helps readers understand the challenges of reaching the goal, e.g., the impact of the COVID-19 pandemic on education access (\analysislitlow).
The structured narrative requires \tasklitlow as readers must decompose the complex sustainability topics into manageable analytical tasks.
Each story uses scrollytelling to develop the story as the reader navigates, thus \devlitlow is essential for knowing how to scroll with the mouse, touchscreen, or keyboard.
Visualizations are presented one at a time, sometimes with animations (e.g., a choropleth map transforming into a scatterplot), and support standard interaction techniques, such as selecting a country of interest by clicking, requiring \intlitlow to effectively use these features.
There is often no indication that the visualization is interactive, demanding \toollitlow for the reader to recognize available functions.
The use of less-known visualizations, such as parallel coordinates, shows the need for advanced \visualizationlitlow. 
When comparing performance across countries or examining trends over time, readers apply \evaluationlitlow to connect insights to broader sustainability goals. 
The project also relies on readers having basic \foundlit{computer literacy}, \foundlit{data literacy}, and \foundlit{critical thinking} to properly contextualize the presented information.

\begin{figure}
    \centering
    \includegraphics[width=\columnwidth]{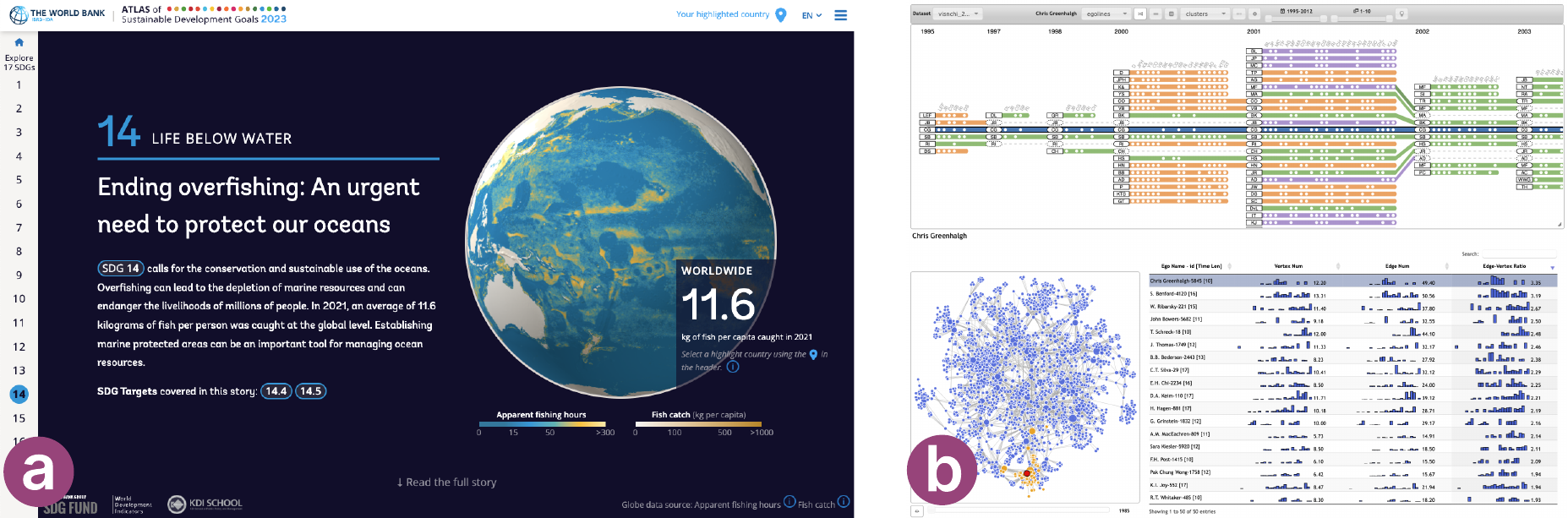}
    \caption{\textbf{Examples from the validation scenarios.}
    \normalfont
    (a) the Atlas of Sustainable Development Goals~\cite{pirlea23}, image \copyright The World Bank Group;
    (b) EgoLines~\cite{zhao16}, demo image \copyright Zhao et al.
    }
    \Description{Two screenshots ordered horizontally. The first one shows an interactive globe visualizing fishing data, accompanied by an explanatory text on the goal about ending overfishing. The second one shows a web-based system with three different views (the egolines visualization technique, a node-link diagram, and an interactive table). Each screenshot belongs to one of the example systems mentioned in Section 6.1.
    }
    \label{fig:scenarios}
\end{figure}

\subsubsection{Visual Analytics}

Visual analytics systems combine visualizations with automated analysis~\cite{keim08}.
They are usually designed for experts and consist of multiple visualization techniques and views with different levels of complexity.

\paragraph{EgoLines (2016).}

Zhao et al.'s~\cite{zhao16} EgoLines system (Fig.~\ref{fig:scenarios}b) is designed for egocentric analysis of dynamic networks.
To use the system, users need to understand network data (\foundlit{data literacy}), network analysis (\foundlit{data science literacy}), and node-link diagrams (\visualizationlitlow).
They will also need to learn what system functions are associated with interaction widgets (\toollitlow).
Given that some interaction techniques do not have any explicit interface element associated that reveal affordances (e.g., when hovering over a node, its neighbors get highlighted), the person needs \intlitlow to be familiar with the interaction techniques and the actions required to perform them.
Interpreting the highlighting when it happens (e.g., system highlights the reoccurrences of one node over time and its matrix-based connections) requires both \visuallitlow to perceive the visual changes and \evaluationlitlow to understand their analytical significance.
Similarly, the person needs to realize the table is interactive (\intlitlow), which could be better supported through suggested interactivity cues~\cite{boy16}.
For making use of the multiple views in combination, the person needs to be aware that the views are coordinated through interaction (e.g., clicking on a name on the table changes what network is shown on the top view), requiring \orchlitlow to manage this complex multi-view workflow.
Similarly, they need to grasp that the overview \& detail views are connected through coordinated highlighting, requiring \tasklitlow to decompose high-level goals into specific view-based actions.
For a given task, \analysislitlow is necessary to formulate appropriate network analysis strategies and determine what visualization helps solve it (e.g., find shortest path on the detail view) and identify what sequence of steps are required to reach the solution.
Finally, the EgoLines system provides automated analysis features.
For instance, the person needs to be aware that there is a custom sorting algorithm and understand it to interpret the result, requiring a combination of \foundlit{statistical literacy} and \visualizationlitlow to correctly interpret the algorithm's output and its visual representation.

\subsection{Observational Study: Reporting on People's Multiliteracies}
\label{sec:study}

To further scrutinize our model, we conducted an observational study with nine participants, where we asked them to interact with two systems we had examined: the SDG Atlas~\cite{pirlea23} and EgoLines~\cite{zhao16}. 
Similar to the study of Lee et al.~\cite{lee16}, we asked participants to think aloud to understand their impressions, issues, and reasoning, while interacting with the systems to solve a series of tasks.
The study started with a pre-questionnaire on prior experiences and a visualization literacy test.
Then, we asked participants to solve three tasks with each system.
Before presenting the tasks, we provided a one-paragraph explanation of each system, taken from its respective documentation, but we did not provide any interaction training. 
Afterwards, we conducted a semi-structured interview asking participants about their strategies, relevant abilities, and any challenges they experienced (see supplemental material). 

We recruited nine participants (six women and three men).
All but one had at least a bachelor's degree.
Eight were in the age range of 18--35, and one was in the range of 35--49.
They worked in diverse fields: computer science (four), psychology (one), forestry (one), health (one), language education (one), and information sciences (one).
Everyone used interactive systems at least weekly. 
Six participants used interactive visualizations weekly, two monthly, and one less than once per month. 
Participants scored between 6 and 10 (out of 12) in the mini-VLAT test~\cite{pandey23}.
The visualization systems were new to all participants.
In the following, we summarize our insights from the analyses of our observations, the think-aloud statements, the recorded interactions, and the short interviews.

\subsubsection{Observed Literacies}

Poor \intlitlow was one of the most common struggles in the study.
For example, on the atlas, only three participants realized, without help, that they were able to scroll through each SDG page.
They confused the first visualization with the end of the page, without realizing that the visualization would change and move up if they scrolled further (i.e., scrollytelling).
This was evident through many quotes: \textit{``Oh yeah! I thought it stopped here, because I was scrolling and nothing was going down''} (P4); \textit{``in the beginning, I didn't know I could scroll that much.''}(P3); \textit{``You can always scroll down [...] Oh.''} (P2).
Participants often missed specific interactive features.
For instance, P2 and P6 did not realize they could use a search bar to look for a specific data item in a visualization.

In the visual analytics system, P3 and P4 struggled to understand the ego network visualization to describe the temporal evolution of a cluster (\visualizationlitlow):
\textit{``I'm not sure even what the connections are in the cluster or if they also have some time relation or if it's only in the I can sense in the first part that there's a time relation, like, the destination at least''} (P3); \textit{``This question (cluster) is hard for me''} (P4).
Everyone showed good levels of \devlitlow by dominating the mouse and keyboard as input devices, and P1 especially relied on the keyboard, often using Ctrl+F.

While solving the first atlas task, both P4 and P5 initially thought the globe at the top of each atlas page (see Fig.~\ref{fig:scenarios}a) was the main (or only) element to interact with\camerarevision{ on the atlas}{}.
Interestingly, there were cues indicating the opposite, which they seemed to ignore, such as text labels and the absence of country names on the globe.
This suggests an issue with \toollitlow, as they not only misunderstood which mechanisms govern the pages, but also overlooked multiple interface elements for filtering.
This misconception affected how they approached the task.
Since filtering by country was an important step to solve it, P4 stated that she was ``bad at geography,'' and went back to the home page often, searching for the country in the home page's globe---which had a search bar.
P5 opted to click on multiple countries on the globe until finding the right one. 

On the same task, P2 struggled to find the data values associated with some of the countries relevant to the task because the corresponding marks appeared below the X-axis line on a scatterplot---and this compromised her understanding of what she could interact with, thinking that the search bar was not working.
This suggests a trickle effect of how \visuallitlow issues may feed into \intlitlow in the next iteration of the cycle.

In the EgoLines system, P3 showed how he understands what he needs from the visualization, showing \toollitlow, and he seemed to understand how graph visualizations usually filter or change their display (by clicking on edges or on vertices) as a basic level of \visualizationlitlow.
However, he did not understand how to convert the analytical requirements into observable findings, showing a lower \intlitlow: \textit{``I've been looking so much at these cluster graphs, so it's like it looks interesting, but it's also like I feel like when I click on it, it doesn't really give me any information about, like, who is this person or what are the connections.''}

P3 struggled with \visualizationlitlow regarding the EgoLines visualization technique---\textit{``Who had the longest [coauthor] relationship with Mac and it's a bit hard for me to understand actually.”}---but managed to use a different chart, which he thought was simpler and more direct, to solve the task correctly.
Despite those struggles, he displayed \tasklitlow in managing to extract the task requirements and \evaluationlitlow answering it correctly.

On the atlas, P1 had issues regarding goal formation.
After identifying the visualization she had to interact with early on, she clearly understood what the visualization was representing, its axis, and its meaning (\visuallitlow, \visualizationlitlow, \evaluationlitlow).
She interacted skillfully, understanding how her actions impacted the visualization state (\intlitlow), yet she could not understand how to connect that to the question requirements, showing low \analysislitlow.
On EgoLines, P1 displayed \toollitlow, being one of the only two participants to understand the system features:
\textit{``Just playing around with the system,''} she stated before trying out different interface elements (\textit{``Can I change these?''}).
Yet, she did not quite grasp the visualization on the top view without help, showcasing low \visualizationlitlow.

On the text-heavy pages of the atlas, all participants relied on a word-seeking approach: they all described how, to navigate, they searched for specific terms that matched the questions of the task.
Despite all of them using such a tactic, P5--P7 and P1 recognized that Ctrl+F could help them, while P2--P4, P8, and P9 never used it.
The unawareness towards a highly relevant action showed issues with \devlitlow, and potentially, \toollitlow.
Unlike others, P9 showed considerable \orchlitlow. 
While P2, P3, and P4 explored different parts of the page non-linearly without a clear approach, P9 executed her search according to the structure 
(\textit{``First part was understanding how it was organized [...] so I could figure how to learn the things I wanna learn''}) 
and asked for better organization: \textit{``it could have a table of contents''}.
P6, P7 and P8 believed that EgoLines did not have enough labeling and titles: \textit{``There are no legends [...] so, I do not have a framework that I can use to read''} (P6). 
These occurrences showcase that effectiveness was not only influenced by literacies, but also by the design. 
Finally, P7 
discussed how they changed their strategies:
\textit{``First I'm gonna think what I'm looking at. So I don't make the same mistake as before.''}

\subsubsection{Interviews}
All participants reported having issues with both systems, with EgoLines being more challenging than the atlas. Three participants mentioned the text descriptions of the atlas as a helpful characteristic to understand the data, but P1 explained that reading took a long time. P4 explained that she struggled to interpret the task questions, but after searching for keywords in the atlas, she was able to understand the topic better. 

\section{Discussion}
\label{sec:discussion}

Scrutinizing the model through the examination of real-world systems and the study brought some first evidence about the scope of each multiliteracies and help us exemplify them.
We also did not find any new literacies or situations we could not explain through our model.
We take this as first evidence of the expressiveness and descriptiveness of our model. 
The findings also surfaced some challenges with respect to some of the multiliteracies, namely those on the most abstract strategic level such as \orchlitlow. 
We believe this might have been due to us giving participants clearly defined tasks and can only further be tested by studies in realistic settings with real analysts and their analysis. 
Next, we acknowledge the limitations of our work, we use our findings to discuss implications for design and education, and we compare and connect our work with relevant theories.

\subsection{Limitations}

Our model provides an overview of IVL, but is not exhaustive and does not consider all individual differences that may influence interaction.
While we acknowledge non-visual modalities and discuss accessibility barriers, the model does not cover scenarios with blind and low-vision users.
Furthermore, we focus on individuals and do not yet consider cooperative scenarios, where performance is likely to be shaped by the literacies of everyone involved.

In the case studies and the observational study, we reported on our analyses and participants' literacies in relation to specific real-world systems.
While the qualitative findings provide concrete examples, \camerarevision{}{more work is needed to prove the causal attribution of problems to literacy levels, and }measuring specific literacy levels would require assessment tests for each literacy.
\camerarevision{}{A large-scale study with standard visualizations and open questions, similar to the one of Nobre et al.~\cite{nobre24}, could offer insights into the user struggles and the underlying causes, which can then contribute to the development of heuristics.}
Participants interacted through mouse and keyboard, so extending the scope of devices, modalities, and paradigms would provide a more comprehensive picture of the literacies in action.
\camerarevision{}{Using multiple devices could be helpful, for example, to compare how participants solve two identical tasks with different devices, isolating the effect of device literacy. Once assessment tests are available, testing quantitatively whether low literacy yields lower performance metrics will be possible.}
Additionally, extending the participant sample regarding educational background and age would help address issues about underrepresented communities in VL evaluations~\cite{creamer24}.

We used the term \textit{literacy} for the abilities relevant to interactive visualization despite recent concerns about the term~\cite{solen25skillset}.
We speak of \textit{multiliteracies} to acknowledge the multiliteracies approach~\cite{newlondon1996multiliteracies} and other well-known concepts (e.g., device literacy~\cite{wyche16}), as well as to emphasize that thinking of IVL as a set of multiple abilities is more fitting than the monolithic view.
Still, other terminology issues remain to be addressed.

\subsection{Implications for Interaction Design and Education}
\label{sec:implications}

Most visualization systems are designed assuming people know how to leverage the interactions available, although the target audience may not be able to do so without training.
This is especially critical when visualizations are meant for the general public or novices%
: \textit{``the best technology is of little value if people do not comprehend it''}~\mbox{\cite{gigerenzer02}}.
With our model, we wish to encourage both interaction designers to consider the abilities of the people they are designing for and educators to think about how to support skill development for IVL.
First, the model can help identify potential system flaws and obstacles, such as the level of interactivity, audience skills, interaction objectives, etc.
Second, the model can help suggest interventions to support specific literacies and to onboard people without sufficient knowledge.

\subsubsection{Implications for Design} 
\label{sec:design_implications}
We derive the following implications for visualization and interaction design.

\begin{enumerate}
    \item \textbf{Communicate interaction affordances} and suggested interactivity cues for discoverability through explicit explanations~\cite{boy16} or visual badges~\cite{edelsbrunner2026visualization}.
    In-situ help can assist through on-demand toolkits and messages~\cite{chundury2023contextual}.
    \item \textbf{Explain interactions}, e.g., through small visual representations of interaction~\cite{spence14}, cheat-sheets~\cite{wang2020cheat}, or comics~\cite{boucher2025interactive} explaining workflows, concepts, and interactions.
    \item \textbf{Use bespoke onboarding methods}, such as videos~\cite{stoiber2021design}, interactive tours~\cite{dhanoa25dtour}, or examples of interactive exploration workflows, potentially automated~\cite{li2023networknarratives}.
    \item \textbf{Leverage direct manipulation}, coupled with confirmatory feedback (\eg sounds or animations), to minimize Norman's gulfs~\cite{hollan00}, help people assess whether their action was processed, and support their \visuallitlow.
    \item \textbf{Explain interaction techniques} to reduce mental and motor efforts~\cite{lam08} (\devlitlow). For instance, through \textit{landmarking} for the current zoom factor.
    \item \textbf{Create or use design standards}. Interface libraries, such as Material Design\footnote{\url{https://m2.material.io/design/communication/data-visualization.html\#behavior}}, recommend specific implementations of interactive charts.
    Standards facilitate interoperability and a smooth skill transfer to new contexts.
    \item \textbf{Provide alternative interactions}, e.g., through multimodality, for when a person does not have enough \devlitlow to use a modality associated with a specific device.
    \item \textbf{Support strategic actions} for \analysislitlow,\\ \orchlitlow, and \evaluationlitlow.
    Examples could include personal notebooks~\cite{willett2015understanding} and history functions~\cite{heer2008graphical} to help people develop and evaluate strategies.
\end{enumerate}

\camerarevision{}{
    It is worth noting that designing with multiliteracies in mind should not lead to assuming that if a person struggles to interact, their potential low literacy is at fault.
    Norman \cite{norman88} has often emphasized that usability problems are due to faulty designs, and still, he also acknowledged that a good design may not be good enough in activities that require certain skills. This reveals a challenge around the diagnosis of interaction problems. Our model makes relevant skills explicit to help identify the problem's cause. 
    Human-centered design is about \emph{understanding the people} who will use the system \cite{hornbaek25book}, so we encourage designers to consider people's literacies, as they consider people's domain knowledge \cite{freedman02} or personality \cite{leon25}.
    }

\subsubsection{Implications for Education} 
In alignment with Bach et al.~\cite{bach24}, we consider education in the widest sense, including scenarios beyond the traditional classroom and university teaching, such as self-paced learning.

\begin{enumerate}
    \item \textbf{Emphasize strategic literacies} to foster an understanding of visualization and interaction that goes beyond the technical manipulation of interfaces (\orchlitlow).
    This is especially true for scenarios where no system is involved (yet), such as planning an analysis (\analysislitlow) and validating how specific discoveries connect with prior knowledge (\evaluationlitlow).
    \item \textbf{Foster the idea of interactive visualization} by making interaction a learning goal, include it into a curriculum and consider visualization as `interactive by nature'.
    Even if a visualization is static, people can ask questions (\tasklitlow), formulate hypotheses (\analysislitlow), and seek details (\intlitlow). 
    \item \textbf{Engage learners with interactive visualization tools}, such as Tableau, or environments to build interactive charts, such as Observable.
    Immersive environments could further the idea of agency and interactivity.
    \item \textbf{Assess for IVL} by testing learners' understanding across different levels and gulfs, supported by qualitative approaches, existing assessments \cite{wyche16, callow08}, and nascent proposals. 
\end{enumerate}

\subsection{Comparison to Other Models}

After presenting the IVL model, we here clarify how it connects with related models, typologies, and assessment studies. 
The abstraction levels of the IVL model are inspired by the structure of the typology of Brehmer and Munzner~\cite{brehmer13}, which characterizes how people use visualizations and was similarly influenced by Norman's work~\cite{norman88}, but does not address interaction techniques directly.
Bertin~\cite{bertin10} defined three levels of graph comprehension that go from perceiving visual elements to connecting domain knowledge with the visualized data, which overlap with the IVL levels, but he does not consider interaction or other high-level cognitive activities.
Similar to ours, the model of Jansen and Dragicevic~\cite{jansen13} describes interaction with visualization systems, but focuses on \textit{beyond the desktop} cases, including physicalization.
Our model covers all computer-based systems, but does not include physical ones.
Keim et al.~\cite{keim08} and El-Assady et al.~\cite{elassady19} describe a visual analytics pipeline and loops incorporating relevant human knowledge for interaction.
Our model takes visual analytics into account, but considers other scenarios as well (see Sect.~\ref{sec:scenarios}) and extends the knowledge scope beyond data literacy and domain expertise.

Regarding visualization literacy research, the NOVIS model of Lee et al.~\cite{lee16} was one of the first models proposed by the visualization community.
The model acknowledges bottom-up and top-down processes and considers the iterative nature of sensemaking.
Similarly, our model presents a visualization cycle inspired by the work of Norman \cite{norman88} and Lam \cite{lam08}, structured in levels that interact with one another.
Lee \etal focus on the activities of novices when they first encounter a visualization, but point out that with more time and motivation, people would engage in additional activities, such as generating hypotheses, which \analysislitlow supports.
Moreover, the visualizations of their study supported basic interactions, but the authors classified interaction-related activities only as \textit{miscellaneous}, without providing details.
They argue that their model could be combined with others, such as the sensemaking loop of Pirolli and Card~\cite{pirolli2005sensemaking}, to be more comprehensive, and mention the exploration of personal skills as a future research direction. Our model brings these concepts together, focusing on the abilities relevant to interaction.

The models of Lee et al.~\cite{lee16} and Hedayati et al.~\cite{hedayati24reconcept} are based on the idea that a person contructs and refines a conceptual frame over time.
Our model incorporates the temporal aspect but is interaction-driven and, thus, focuses on the goals that drive the person's actions. The frame construction and refinement are supported by evaluation literacies.
As interaction leads to visualization changes, constructing or refining frames is part of each iteration of the cycle.

Several studies on visualization literacy have hinted at the existence of multiple abilities connected to visualization literacy~\cite{pandey23, ge25avec,hedayati24reconcept}, mostly focusing on a single one in the context of static visualizations. 
Pandey and Ottley \cite{pandey23} argue that visualization literacy is multidimensional.
In a similar fashion, we propose to see interactive visualization literacy as a composition of interconnected literacies.
The connection between interaction abilities and visualization abilities has already been discussed in assessment work.
For instance, the authors of the AVEC assessment~\cite{ge25avec} acknowledge that the experience a person has with an authoring tool may influence what visualizations they can create.
The limitation that we address is that these findings did not connect to interaction theories.
We believe that the ability to visually encode data on a computer requires both execution and evaluation literacies because authoring requires interacting and interpreting the resulting visualization correctly.
Recently, Hedayati and Kay~\cite{hedayati25uni} concluded that there are cognitively complex abilities (e.g., visualization critique) that are not yet captured in existing assessments.
The IVL model positions them at the strategic level (e.g., \evaluationlitlow).
The authors also mention statistical literacy as part of the range of visualization skills, but they do not elaborate on how these skills may connect.  
We classify statistical literacy and critical thinking~\cite{ge23} as foundational literacies to describe the scope of literacies relevant to interactive visualizations.

Previous studies have identified problems related to interactive systems and literacy, but without contextualizing them in the interaction process.
Nobre et al.~\cite{nobre24} provided a classification of visualization literacy barriers.
The IVL model helps map these barriers to more specific literacies, e.g., ``misunderstanding the task'' with \tasklitlow.
The model also allows connecting the interaction roadblocks identified by Kwon et al.~\cite{kwon11} with single literacies, e.g., ``failure to execute appropriate interaction'' with \intlitlow.

Prior work on visualization and HCI has introduced three of the literacies we describe.
Beyond \visualizationlitlow, which we have discussed in depth, Bach~\cite{bach18} defined \intlitlow through visualization projects.
Velghe~\cite{velghe14} defined \devlitlow in the context of an ethnographic study involving mobile phones.
We extend her definition to consider any devices present in interactive visualization systems.
Overall, we build upon their work by situating these literacies in the interaction cycle and connecting them with other abilities required to use interactive visualizations.

\section{Research Agenda}
\label{sec:agenda}

After defining the scope of interactive visualization literacy, we propose a research agenda for future work.

\subsection{Assessing IVL}

Assessing IVL would help assess what issues people struggle with while interacting. 
Recent work has extended the scope of visualization abilities considered in measurement instruments~\mbox{\cite{ge24workshop, hedayati25uni}}, but has yet to consider interactivity.
We believe that the IVL of a person for a given system should be calculated based on the tasks, devices, and visualization techniques the system supports.
A potential assessment test for IVL should adapt to the system because the tool properties, input devices, and the visualizations can vary largely (e.g., network visualization literacy~\mbox{\cite{zoss2018network}}).
The multiliteracy model gives us a handy tool for assessing IVL by distinguishing different literacies whose measures we could aggregate.

Existing VL assessments cover only one out of nine literacies and should be tested against interactive scenarios.
New assessments could reduce or extend existing ones to differentiate between the well-known \mbox{\visualizationlitlow} and the related \mbox{\visuallitlow} and \mbox{\evaluationlitlow}.
We can rely on methods to assess other known literacies, such as \visuallitlow~\cite{callow08} and \devlitlow~\cite{wyche16}. 
Task typologies~\cite{brehmer13}, detected barriers~\cite{Yalcin2016cognitive, nobre24} and interaction roadblocks~\cite{kwon11} could inform methods for \tasklitlow and \intlitlow, 
while cognitive science methods~\cite{Yalcin2016cognitive} and extensive work on sensemaking~\cite{keim08, elassady19} can inform assessments of \analysislitlow, \orchlitlow, and \evaluationlitlow.
A follow-up of CALVI~\cite{ge23} could help determine how \evaluationlitlow is connected with misinformation.
While we can imagine specific tests about features for \toollitlow, assessing someone's ability to formulate goals and to divide them into tasks is more challenging because tasks have multiple interpretations in the literature (e.g.,~\cite{brehmer13, DBLP:conf/infovis/AmarES05, amar04}).
Likewise, given the extended time frame of an interactive analysis versus reading a single static visualization, assessments for IVL are likely to require longitudinal measures and tests~\mbox{\cite{wang23}}.
Interpreting these thoughts in light of the current literature on VL, 
we also suggest that assessing IVL will most likely be a highly \textit{qualitative} undertaking that has to take into account nuances, contexts, specific tasks, and the interactions between all the literacies involved.

\subsection{Evaluating a System's Capability of Scaffolding IVL}

Interaction gulfs can either be bridged by a person's IVL or by a system design explicitly shrinking them.
Thus, in system evaluations, the IVL of participants will have an impact on the outcome.
Yet, evaluation studies do not tend to consider how capable a person is of interacting before asking them to solve a task interactively~\cite{Liu2020}.
But individual differences \textit{do} play a significant role in visualization performance~\cite{green10, Liu2020}, and multiliteracies widen the scope of the differences to consider.
This leads us to ask \textit{how do \toollitlow, \devlitlow, \evaluationlitlow, \foundlit{statistical literacy}, influence the performance and experience of people with interactive visualizations?}
\textit{Where do we expect users to be in the literacy spectrum?}
\textit{How does the design support people in shrinking the gulfs?}
We should expand the range of variables to consider in current evaluation practices and be aware of the relevant literacies to differentiate between the effects of human abilities and the effects of the interaction design.

So then, \textit{how to assess the capabilities of an interactive visualization system to support IVL?}
We could visualize interaction logs and calculate indicators~(e.g., \mbox{\cite{wang23}}) to deliver interaction insights on the fly.
Think-aloud sessions and interviews could further help understand how literacies may influence visualization experiences.
Long-term case studies~\cite{shneiderman06milcs} could allow observing how people's IVL evolve over time, as they leverage the system in their own environment. 
Finally, extending heuristic evaluations~\cite{wall19heuristic} with heuristics associated with specific literacies could help assess how a design affects the gulfs (e.g., a system that supports deliberate search tasks shrinks the gulf of goal formation).
However, the question of assessing how well a system supports interaction skills in general remains largely open.
Answering it requires a target of what success is and what the current baseline for IVL in the population is.

\subsection{Designing Systems with IVL in Mind}

Literacies help people bridge the gulfs, but designers are responsible for the width of the gaps in the first place.
From the implications in Sect.~\ref{sec:design_implications}, open questions remain:
\textit{how should system design acknowledge and support interaction abilities?}
First, our model sets a framework that future work can apply to systematically investigate the interdependencies between design choices and human abilities.
Classifying tasks, tools, and devices according to the minimum literacy level they require would facilitate design choices for different audiences.
In case people struggle with one or more literacies, \textit{can we provide alternatives to complex interaction techniques that require less literacy?
How would that transformation work?}
Accessibility standards may provide examples of how to do that.
Such approaches could include interactive tutorials~\cite{chundury2023contextual}, data tours~\cite{li2023networknarratives, mehta2017datatours}, pattern explanations~\cite{shu2024does}, onboarding~\cite{stoiber2019visualization}, and guidance~\mbox{\cite{ceneda2016characterizing}}.

Second, the model can help address \textit{accessibility barriers}~\cite{marriott21lee} through design audits and interventions.
For instance, color vision deficiency primarily impacts \visuallitlow and
motor impairments affect \devlitlow.
Similarly, cognitive impairments can affect goal formation, requiring support for analytical planning or mixed-initiative interaction~\cite{horvitz99}.
By mapping barriers to gulfs, our model enables targeted interventions and accessible designs.

\subsection{Educating for IVL}

The line between \textit{applying} an interaction technique and \textit{learning} a technique is naturally fine; by using a technique, one learns, and by learning a technique, one can use it more efficiently and effectively.
Hence, some of the approaches cited \camerarevision{in the last subsection}{above} (\eg onboarding) can help increase a person's IVL simply by demonstrating and explaining.
Still, we need dedicated means to communicate the conceptual and cognitive means related to using interactive tools~\cite{alkadi23, boucher2025interactive}.

Echoing similar calls for VL~\cite{bach24}, we need learning goals for IVL.
Bloom's seminal pyramids~\cite{bloom1956taxonomy} could offer some clues about such goals.
For example, as part of the cognitive goals, Bloom lists ``analyze'' and ``evaluate'' as rather high in the hierarchy, based on lower-level goals such as ``comprehension'' (\visualizationlitlow{}?).
Likewise, he defines psychomotor goals which could be related to abilities required for \devlitlow (``set'', ``guided response'') and \visuallitlow (``perceive''), and could help describe higher-order processes for IVL such as ``adapt'' techniques to new contexts or ``originate'' novel ways of using an existing interaction technique.
Our mulitileracies will hopefully help inform dedicated education interventions either as part of systems or as part of visualization education resources~\cite{wang2020cheat}.

\section{Conclusion}

We introduced a definition and a theoretical model for \emph{interactive visualization literacy}, describing and identifying the abilities a person requires for interacting with data visualizations.
We characterized the visualization action cycle across two dimensions---\gulfs of visualization interaction and levels of \abstractions---that situate and describe nine distinct literacies required to interact with visualization systems.
Our model extends the understanding of visualization research around literacies, so far focused on single static visualizations, to consider interaction---an intrinsic element of visualizations. 
We scrutinize our model with nine case studies and an observational study, and propose a research agenda on how to assess, evaluate, design, and educate, taking IVL into account.
With this paper, we hope to start a conversation in the visualization community on how human abilities shape interactive experiences.

\begin{acks}
This work was partially supported by Villum Investigator grant VL-54492 by Villum Fonden.
Any opinions, findings, and conclusions expressed in this material are those of the authors and do not necessarily reflect the views of the funding agency.
\end{acks}

\bibliographystyle{ACM-Reference-Format}
\bibliography{ivl}


\begin{thebibliography}{121}


\ifx \showCODEN    \undefined \def \showCODEN     #1{\unskip}     \fi
\ifx \showISBNx    \undefined \def \showISBNx     #1{\unskip}     \fi
\ifx \showISBNxiii \undefined \def \showISBNxiii  #1{\unskip}     \fi
\ifx \showISSN     \undefined \def \showISSN      #1{\unskip}     \fi
\ifx \showLCCN     \undefined \def \showLCCN      #1{\unskip}     \fi
\ifx \shownote     \undefined \def \shownote      #1{#1}          \fi
\ifx \showarticletitle \undefined \def \showarticletitle #1{#1}   \fi
\ifx \showURL      \undefined \def \showURL       {\relax}        \fi
\providecommand\bibfield[2]{#2}
\providecommand\bibinfo[2]{#2}
\providecommand\natexlab[1]{#1}
\providecommand\showeprint[2][]{arXiv:#2}

\bibitem[Aisch(2017)]%
        {aisch17}
\bibfield{author}{\bibinfo{person}{Gregor Aisch}.} \bibinfo{year}{2017}\natexlab{}.
\newblock \bibinfo{booktitle}{\emph{In Defense of Interactive Graphics}}.
\newblock
\urldef\tempurl%
\url{https://www.vis4.net/blog/in-defense-of-interactive-graphics/}
\showURL{%
Retrieved November 19, 2025 from \tempurl}


\bibitem[AlKadi et~al\mbox{.}(2023)]%
        {alkadi23}
\bibfield{author}{\bibinfo{person}{Mashael AlKadi}, \bibinfo{person}{Vanessa Serrano}, \bibinfo{person}{James Scott-Brown}, \bibinfo{person}{Catherine Plaisant}, \bibinfo{person}{Jean-Daniel Fekete}, \bibinfo{person}{Uta Hinrichs}, {and} \bibinfo{person}{Benjamin Bach}.} \bibinfo{year}{2023}\natexlab{}.
\newblock \showarticletitle{Understanding Barriers to Network Exploration with Visualization: A Report from the Trenches}.
\newblock \bibinfo{journal}{\emph{{{IEEE} Transactions on Visualization and Computer Graphics}}} \bibinfo{volume}{29}, \bibinfo{number}{1} (\bibinfo{year}{2023}), \bibinfo{pages}{907--917}.
\newblock
\href{https://doi.org/10.1109/TVCG.2022.3209487}{doi:\nolinkurl{10.1109/TVCG.2022.3209487}}


\bibitem[Alper et~al\mbox{.}(2017)]%
        {alper17}
\bibfield{author}{\bibinfo{person}{Basak Alper}, \bibinfo{person}{Nathalie~Henry Riche}, \bibinfo{person}{Fanny Chevalier}, \bibinfo{person}{Jeremy Boy}, {and} \bibinfo{person}{Metin Sezgin}.} \bibinfo{year}{2017}\natexlab{}.
\newblock \showarticletitle{Visualization Literacy at Elementary School}. In \bibinfo{booktitle}{\emph{Proceedings of the {ACM} Conference on Human Factors in Computing Systems}}. \bibinfo{publisher}{{ACM}}, \bibinfo{address}{New York, NY, USA}, \bibinfo{pages}{5485--5497}.
\newblock
\showISBNx{9781450346559}
\href{https://doi.org/10.1145/3025453.3025877}{doi:\nolinkurl{10.1145/3025453.3025877}}


\bibitem[Amar and Stasko(2004)]%
        {amar04}
\bibfield{author}{\bibinfo{person}{Robert Amar} {and} \bibinfo{person}{John Stasko}.} \bibinfo{year}{2004}\natexlab{}.
\newblock \showarticletitle{A Knowledge Task-Based Framework for Design and Evaluation of Information Visualizations}. In \bibinfo{booktitle}{\emph{Proceedings of the {IEEE} Conference on Information Visualization}}. \bibinfo{publisher}{{IEEE Computer Society}}, \bibinfo{address}{Los Alamitos, CA, USA}, \bibinfo{pages}{143--150}.
\newblock
\href{https://doi.org/10.1109/INFVIS.2004.10}{doi:\nolinkurl{10.1109/INFVIS.2004.10}}


\bibitem[Amar et~al\mbox{.}(2005)]%
        {DBLP:conf/infovis/AmarES05}
\bibfield{author}{\bibinfo{person}{Robert~A. Amar}, \bibinfo{person}{James Eagan}, {and} \bibinfo{person}{John~T. Stasko}.} \bibinfo{year}{2005}\natexlab{}.
\newblock \showarticletitle{Low-Level Components of Analytic Activity in Information Visualization}. In \bibinfo{booktitle}{\emph{Proceedings of the {IEEE} Conference on Information Visualization}}. \bibinfo{publisher}{{IEEE Computer Society}}, \bibinfo{address}{Los Alamitos, CA, USA}, \bibinfo{pages}{111--117}.
\newblock
\href{https://doi.org/10.1109/INFVIS.2005.1532136}{doi:\nolinkurl{10.1109/INFVIS.2005.1532136}}


\bibitem[Bach(2018)]%
        {bach18}
\bibfield{author}{\bibinfo{person}{Benjamin Bach}.} \bibinfo{year}{2018}\natexlab{}.
\newblock \showarticletitle{Ceci n'est pas la data: Towards a Notion of Interaction Literacy for Data Visualization.}. In \bibinfo{booktitle}{\emph{Proc.\ AVI Workshop on Visual Interfaces for Big Data Environments in Industrial Applications}}. \bibinfo{publisher}{{ACM}}, \bibinfo{address}{New York, NY, USA}, \bibinfo{pages}{1--3}.
\newblock
\urldef\tempurl%
\url{https://ceur-ws.org/Vol-2108/invited1.pdf}
\showURL{%
\tempurl}


\bibitem[Bach et~al\mbox{.}(2024)]%
        {bach24}
\bibfield{author}{\bibinfo{person}{Benjamin Bach}, \bibinfo{person}{Mandy Keck}, \bibinfo{person}{Fateme Rajabiyazdi}, \bibinfo{person}{Tatiana Losev}, \bibinfo{person}{Isabel Meirelles}, \bibinfo{person}{Jason Dykes}, \bibinfo{person}{Robert~S. Laramee}, \bibinfo{person}{Mashael AlKadi}, \bibinfo{person}{Christina Stoiber}, \bibinfo{person}{Samuel Huron}, \bibinfo{person}{Charles Perin}, \bibinfo{person}{Luiz Morais}, \bibinfo{person}{Wolfgang Aigner}, \bibinfo{person}{Doris Kosminsky}, \bibinfo{person}{Magdalena Boucher}, \bibinfo{person}{Søren Knudsen}, \bibinfo{person}{Areti Manataki}, \bibinfo{person}{Jan Aerts}, \bibinfo{person}{Uta Hinrichs}, \bibinfo{person}{Jonathan~C. Roberts}, {and} \bibinfo{person}{Sheelagh Carpendale}.} \bibinfo{year}{2024}\natexlab{}.
\newblock \showarticletitle{Challenges and Opportunities in Data Visualization Education: A Call to Action}.
\newblock \bibinfo{journal}{\emph{{{IEEE} Transactions on Visualization and Computer Graphics}}} \bibinfo{volume}{30}, \bibinfo{number}{1} (\bibinfo{year}{2024}), \bibinfo{pages}{649--660}.
\newblock
\href{https://doi.org/10.1109/TVCG.2023.3327378}{doi:\nolinkurl{10.1109/TVCG.2023.3327378}}


\bibitem[Barry(1997)]%
        {Barry1997}
\bibfield{author}{\bibinfo{person}{Ann~Marie Barry}.} \bibinfo{year}{1997}\natexlab{}.
\newblock \bibinfo{booktitle}{\emph{Visual Intelligence: Perception, Image, and Manipulation in Visual Communication}}.
\newblock \bibinfo{publisher}{State University of New York Press}, \bibinfo{address}{New York}.
\newblock
\showISBNx{9780791434352}


\bibitem[Bertin(2010)]%
        {bertin10}
\bibfield{author}{\bibinfo{person}{Jacques Bertin}.} \bibinfo{year}{2010}\natexlab{}.
\newblock \bibinfo{booktitle}{\emph{Semiology of Graphics: Diagrams, Networks, Maps} (\bibinfo{edition}{1} ed.)}.
\newblock \bibinfo{publisher}{ESRI Press}, \bibinfo{address}{Redlands, CA, USA}.
\newblock


\bibitem[Bloom et~al\mbox{.}(1956)]%
        {bloom1956taxonomy}
\bibfield{author}{\bibinfo{person}{Benjamin Bloom}, \bibinfo{person}{Max Engelhart}, \bibinfo{person}{Edward Furst}, \bibinfo{person}{Walker Hill}, {and} \bibinfo{person}{David Krathwohl}.} \bibinfo{year}{1956}\natexlab{}.
\newblock \bibinfo{booktitle}{\emph{Taxonomy of Educational Objectives: The Classification of Educational Goals}}.
\newblock \bibinfo{publisher}{Longmans, Green and Co.}, \bibinfo{address}{London, UK}.
\newblock


\bibitem[Boucher et~al\mbox{.}(2025)]%
        {boucher2025interactive}
\bibfield{author}{\bibinfo{person}{Magdalena Boucher}, \bibinfo{person}{Mashael AlKadi}, \bibinfo{person}{Benjamin Bach}, {and} \bibinfo{person}{Wolfgang Aigner}.} \bibinfo{year}{2025}\natexlab{}.
\newblock \showarticletitle{Instructional Comics for Self-Paced Learning of Data Visualization Tools and Concepts}.
\newblock \bibinfo{journal}{\emph{{Computer Graphics Forum}}} \bibinfo{volume}{44}, \bibinfo{number}{3}, Article \bibinfo{articleno}{e70130} (\bibinfo{year}{2025}), \bibinfo{numpages}{12}~pages.
\newblock
\href{https://doi.org/10.1111/cgf.70130}{doi:\nolinkurl{10.1111/cgf.70130}}


\bibitem[Boy et~al\mbox{.}(2016)]%
        {boy16}
\bibfield{author}{\bibinfo{person}{Jeremy Boy}, \bibinfo{person}{Louis Eveillard}, \bibinfo{person}{Françoise Detienne}, {and} \bibinfo{person}{Jean-Daniel Fekete}.} \bibinfo{year}{2016}\natexlab{}.
\newblock \showarticletitle{Suggested Interactivity: Seeking Perceived Affordances for Information Visualization}.
\newblock \bibinfo{journal}{\emph{{{IEEE} Transactions on Visualization and Computer Graphics}}} \bibinfo{volume}{22}, \bibinfo{number}{1} (\bibinfo{year}{2016}), \bibinfo{pages}{639--648}.
\newblock
\href{https://doi.org/10.1109/TVCG.2015.2467201}{doi:\nolinkurl{10.1109/TVCG.2015.2467201}}


\bibitem[Boy et~al\mbox{.}(2014)]%
        {boy14}
\bibfield{author}{\bibinfo{person}{Jeremy Boy}, \bibinfo{person}{Ronald~A. Rensink}, \bibinfo{person}{Enrico Bertini}, {and} \bibinfo{person}{Jean-Daniel Fekete}.} \bibinfo{year}{2014}\natexlab{}.
\newblock \showarticletitle{A Principled Way of Assessing Visualization Literacy}.
\newblock \bibinfo{journal}{\emph{{{IEEE} Transactions on Visualization and Computer Graphics}}} \bibinfo{volume}{20}, \bibinfo{number}{12} (\bibinfo{year}{2014}), \bibinfo{pages}{1963--1972}.
\newblock
\href{https://doi.org/10.1109/TVCG.2014.2346984}{doi:\nolinkurl{10.1109/TVCG.2014.2346984}}


\bibitem[Brehmer and Munzner(2013)]%
        {brehmer13}
\bibfield{author}{\bibinfo{person}{Matthew Brehmer} {and} \bibinfo{person}{Tamara Munzner}.} \bibinfo{year}{2013}\natexlab{}.
\newblock \showarticletitle{A Multi-Level Typology of Abstract Visualization Tasks}.
\newblock \bibinfo{journal}{\emph{{{IEEE} Transactions on Visualization and Computer Graphics}}} \bibinfo{volume}{19}, \bibinfo{number}{12} (\bibinfo{year}{2013}), \bibinfo{pages}{2376--2385}.
\newblock
\href{https://doi.org/10.1109/TVCG.2013.124}{doi:\nolinkurl{10.1109/TVCG.2013.124}}


\bibitem[Burns et~al\mbox{.}(2020)]%
        {burns20}
\bibfield{author}{\bibinfo{person}{Alyxander Burns}, \bibinfo{person}{Cindy Xiong}, \bibinfo{person}{Steven Franconeri}, \bibinfo{person}{Alberto Cairo}, {and} \bibinfo{person}{Narges Mahyar}.} \bibinfo{year}{2020}\natexlab{}.
\newblock \showarticletitle{How to evaluate data visualizations across different levels of understanding}. In \bibinfo{booktitle}{\emph{Proceedings of the Workshop on Beyond Time and Errors in Visualization}}. \bibinfo{publisher}{{IEEE Computer Society}}, \bibinfo{address}{Los Alamitos, CA, USA}, \bibinfo{pages}{19--28}.
\newblock
\href{https://doi.org/10.1109/BELIV51497.2020.00010}{doi:\nolinkurl{10.1109/BELIV51497.2020.00010}}


\bibitem[Börner et~al\mbox{.}(2016)]%
        {boerner16}
\bibfield{author}{\bibinfo{person}{Katy Börner}, \bibinfo{person}{Adam Maltese}, \bibinfo{person}{Russell~Nelson Balliet}, {and} \bibinfo{person}{Joe Heimlich}.} \bibinfo{year}{2016}\natexlab{}.
\newblock \showarticletitle{Investigating aspects of data visualization literacy using 20 information visualizations and 273 science museum visitors}.
\newblock \bibinfo{journal}{\emph{Information Visualization}} \bibinfo{volume}{15}, \bibinfo{number}{3} (\bibinfo{year}{2016}), \bibinfo{pages}{198--213}.
\newblock
\href{https://doi.org/10.1177/1473871615594652}{doi:\nolinkurl{10.1177/1473871615594652}}


\bibitem[Cabouat et~al\mbox{.}(2024)]%
        {cabouat24}
\bibfield{author}{\bibinfo{person}{Anne-Flore Cabouat}, \bibinfo{person}{Tingying He}, \bibinfo{person}{Florent Cabric}, \bibinfo{person}{Tobias Isenberg}, {and} \bibinfo{person}{Petra Isenberg}.} \bibinfo{year}{2024}\natexlab{}.
\newblock \showarticletitle{{Position paper: A case to study the relationship between data visualization readability and visualization literacy}}. In \bibinfo{booktitle}{\emph{Proceedings of the ACM CHI Workshop Toward a More Comprehensive Understanding of Visualization Literacy}}. \bibinfo{publisher}{{ACM}}, \bibinfo{address}{New York, NY, USA}, \bibinfo{numpages}{11}~pages.
\newblock
\urldef\tempurl%
\url{https://hal.science/hal-04523790}
\showURL{%
\tempurl}


\bibitem[Callow(2008)]%
        {callow08}
\bibfield{author}{\bibinfo{person}{Jon Callow}.} \bibinfo{year}{2008}\natexlab{}.
\newblock \showarticletitle{Show Me: Principles for Assessing Students' Visual Literacy}.
\newblock \bibinfo{journal}{\emph{The Reading Teacher}} \bibinfo{volume}{61}, \bibinfo{number}{8} (\bibinfo{year}{2008}), \bibinfo{pages}{616--626}.
\newblock
\href{https://doi.org/10.1598/RT.61.8.3}{doi:\nolinkurl{10.1598/RT.61.8.3}}


\bibitem[Card et~al\mbox{.}(1999)]%
        {card99}
\bibfield{author}{\bibinfo{person}{Stuart~K. Card}, \bibinfo{person}{Jock Mackinlay}, {and} \bibinfo{person}{Ben Shneiderman}.} \bibinfo{year}{1999}\natexlab{}.
\newblock \bibinfo{booktitle}{\emph{Readings in Information Visualization: Using Vision to Think}}.
\newblock \bibinfo{publisher}{Morgan Kaufmann}, \bibinfo{address}{San Francisco, CA, USA}.
\newblock


\bibitem[Cassidy and Eachus(2002)]%
        {cassidy02}
\bibfield{author}{\bibinfo{person}{Simon Cassidy} {and} \bibinfo{person}{Peter Eachus}.} \bibinfo{year}{2002}\natexlab{}.
\newblock \showarticletitle{Developing the Computer User Self-Efficacy (Cuse) Scale: Investigating the Relationship between Computer Self-Efficacy, Gender and Experience with Computers}.
\newblock \bibinfo{journal}{\emph{Journal of Educational Computing Research}} \bibinfo{volume}{26}, \bibinfo{number}{2} (\bibinfo{year}{2002}), \bibinfo{pages}{133--153}.
\newblock
\showeprint{https://doi.org/10.2190/JGJR-0KVL-HRF7-GCNV}
\href{https://doi.org/10.2190/JGJR-0KVL-HRF7-GCNV}{doi:\nolinkurl{10.2190/JGJR-0KVL-HRF7-GCNV}}


\bibitem[Ceneda et~al\mbox{.}(2016)]%
        {ceneda2016characterizing}
\bibfield{author}{\bibinfo{person}{Davide Ceneda}, \bibinfo{person}{Theresia Gschwandtner}, \bibinfo{person}{Thorsten May}, \bibinfo{person}{Silvia Miksch}, \bibinfo{person}{Hans-J{\"o}rg Schulz}, \bibinfo{person}{Marc Streit}, {and} \bibinfo{person}{Christian Tominski}.} \bibinfo{year}{2016}\natexlab{}.
\newblock \showarticletitle{Characterizing guidance in visual analytics}.
\newblock \bibinfo{journal}{\emph{{{IEEE} Transactions on Visualization and Computer Graphics}}} \bibinfo{volume}{23}, \bibinfo{number}{1} (\bibinfo{year}{2016}), \bibinfo{pages}{111--120}.
\newblock
\href{https://doi.org/10.1109/TVCG.2016.2598468}{doi:\nolinkurl{10.1109/TVCG.2016.2598468}}


\bibitem[Chatzimparmpas et~al\mbox{.}(2020)]%
        {chatzimparmpas20}
\bibfield{author}{\bibinfo{person}{Angelos Chatzimparmpas}, \bibinfo{person}{Rafael~M. Martins}, {and} \bibinfo{person}{Andreas Kerren}.} \bibinfo{year}{2020}\natexlab{}.
\newblock \showarticletitle{{t-viSNE}: Interactive Assessment and Interpretation of {t-SNE} Projections}.
\newblock \bibinfo{journal}{\emph{{{IEEE} Transactions on Visualization and Computer Graphics}}} \bibinfo{volume}{26}, \bibinfo{number}{8} (\bibinfo{year}{2020}), \bibinfo{pages}{2696--2714}.
\newblock
\href{https://doi.org/10.1109/TVCG.2020.2986996}{doi:\nolinkurl{10.1109/TVCG.2020.2986996}}


\bibitem[Chundury et~al\mbox{.}(2023)]%
        {chundury2023contextual}
\bibfield{author}{\bibinfo{person}{Pramod Chundury}, \bibinfo{person}{Mehmet~Adil Yal{\c{c}}in}, \bibinfo{person}{Jonathan Crabtree}, \bibinfo{person}{Anup Mahurkar}, \bibinfo{person}{Lisa~M. Shulman}, {and} \bibinfo{person}{Niklas Elmqvist}.} \bibinfo{year}{2023}\natexlab{}.
\newblock \showarticletitle{Contextual in situ help for visual data interfaces}.
\newblock \bibinfo{journal}{\emph{Information Visualization}} \bibinfo{volume}{22}, \bibinfo{number}{1} (\bibinfo{year}{2023}), \bibinfo{pages}{69--84}.
\newblock
\href{https://doi.org/10.1177/14738716221120064}{doi:\nolinkurl{10.1177/14738716221120064}}


\bibitem[Creamer et~al\mbox{.}(2024)]%
        {creamer24}
\bibfield{author}{\bibinfo{person}{Mackenzie Creamer}, \bibinfo{person}{Lace Padilla}, {and} \bibinfo{person}{Michelle Borkin}.} \bibinfo{year}{2024}\natexlab{}.
\newblock \bibinfo{title}{Finding Gaps in Modern Visualization Literacy}.
\newblock
\href{https://doi.org/10.31219/osf.io/jy9v2}{doi:\nolinkurl{10.31219/osf.io/jy9v2}}


\bibitem[Cui et~al\mbox{.}(2024)]%
        {cui24}
\bibfield{author}{\bibinfo{person}{Yuan Cui}, \bibinfo{person}{Lily~W. Ge}, \bibinfo{person}{Yiren Ding}, \bibinfo{person}{Fumeng Yang}, \bibinfo{person}{Lane Harrison}, {and} \bibinfo{person}{Matthew Kay}.} \bibinfo{year}{2024}\natexlab{}.
\newblock \showarticletitle{Adaptive Assessment of Visualization Literacy}.
\newblock \bibinfo{journal}{\emph{{{IEEE} Transactions on Visualization and Computer Graphics}}} \bibinfo{volume}{30}, \bibinfo{number}{1} (\bibinfo{year}{2024}), \bibinfo{pages}{628--637}.
\newblock
\href{https://doi.org/10.1109/TVCG.2023.3327165}{doi:\nolinkurl{10.1109/TVCG.2023.3327165}}


\bibitem[Dhanoa et~al\mbox{.}(2025)]%
        {dhanoa25dtour}
\bibfield{author}{\bibinfo{person}{Vaishali Dhanoa}, \bibinfo{person}{Andreas Hinterreiter}, \bibinfo{person}{Vanessa Fediuk}, \bibinfo{person}{Niklas Elmqvist}, \bibinfo{person}{Eduard Gröller}, {and} \bibinfo{person}{Marc Streit}.} \bibinfo{year}{2025}\natexlab{}.
\newblock \showarticletitle{D-Tour: Semi-Automatic Generation of Interactive Guided Tours for Visualization Dashboard Onboarding}.
\newblock \bibinfo{journal}{\emph{{{IEEE} Transactions on Visualization and Computer Graphics}}} \bibinfo{volume}{31}, \bibinfo{number}{1} (\bibinfo{year}{2025}), \bibinfo{pages}{721--731}.
\newblock
\href{https://doi.org/10.1109/TVCG.2024.3456347}{doi:\nolinkurl{10.1109/TVCG.2024.3456347}}


\bibitem[D'Ignazio and Bhargava(2016)]%
        {dignazio16}
\bibfield{author}{\bibinfo{person}{Catherine D'Ignazio} {and} \bibinfo{person}{Rahul Bhargava}.} \bibinfo{year}{2016}\natexlab{}.
\newblock \showarticletitle{{DataBasic}: Design principles, tools and activities for data literacy learners}.
\newblock \bibinfo{journal}{\emph{J. Community Inform.}} \bibinfo{volume}{12}, \bibinfo{number}{3} (\bibinfo{year}{2016}), \bibinfo{pages}{83--107}.
\newblock
\urldef\tempurl%
\url{https://hdl.handle.net/1721.1/123450}
\showURL{%
\tempurl}


\bibitem[Dimara and Perin(2020)]%
        {dimara20}
\bibfield{author}{\bibinfo{person}{Evanthia Dimara} {and} \bibinfo{person}{Charles Perin}.} \bibinfo{year}{2020}\natexlab{}.
\newblock \showarticletitle{What is Interaction for Data Visualization?}
\newblock \bibinfo{journal}{\emph{{{IEEE} Transactions on Visualization and Computer Graphics}}} \bibinfo{volume}{26}, \bibinfo{number}{1} (\bibinfo{year}{2020}), \bibinfo{pages}{119--129}.
\newblock
\href{https://doi.org/10.1109/TVCG.2019.2934283}{doi:\nolinkurl{10.1109/TVCG.2019.2934283}}


\bibitem[Dondis(1974)]%
        {Dondis1974}
\bibfield{author}{\bibinfo{person}{Donis~A. Dondis}.} \bibinfo{year}{1974}\natexlab{}.
\newblock \bibinfo{booktitle}{\emph{A Primer of Visual Literacy}}.
\newblock \bibinfo{publisher}{MIT Press}, \bibinfo{address}{Boston, MA, USA}.
\newblock


\bibitem[Edelsbrunner et~al\mbox{.}(2026)]%
        {edelsbrunner2026visualization}
\bibfield{author}{\bibinfo{person}{Valentin Edelsbrunner}, \bibinfo{person}{Jinrui Wang}, \bibinfo{person}{Alexis Pister}, \bibinfo{person}{Tomas Vancisin}, \bibinfo{person}{Sian Phillips}, \bibinfo{person}{Min Chen}, {and} \bibinfo{person}{Benjamin Bach}.} \bibinfo{year}{2026}\natexlab{}.
\newblock \showarticletitle{Visualization Badges: Communicating Design and Provenance through Graphical Labels Alongside Visualizations}.
\newblock \bibinfo{journal}{\emph{{{IEEE} Transactions on Visualization and Computer Graphics}}} \bibinfo{volume}{32}, \bibinfo{number}{1} (\bibinfo{year}{2026}), \bibinfo{numpages}{11}~pages.
\newblock
\href{https://doi.org/10.1109/TVCG.2025.3634782}{doi:\nolinkurl{10.1109/TVCG.2025.3634782}}
\newblock
\shownote{to appear}.


\bibitem[El-Assady et~al\mbox{.}(2019)]%
        {elassady19}
\bibfield{author}{\bibinfo{person}{Mennatallah El-Assady}, \bibinfo{person}{Fabian Sperrle}, \bibinfo{person}{Oliver Deussen}, \bibinfo{person}{Daniel Keim}, {and} \bibinfo{person}{Christopher Collins}.} \bibinfo{year}{2019}\natexlab{}.
\newblock \showarticletitle{Visual Analytics for Topic Model Optimization based on User-Steerable Speculative Execution}.
\newblock \bibinfo{journal}{\emph{{{IEEE} Transactions on Visualization and Computer Graphics}}} \bibinfo{volume}{25}, \bibinfo{number}{1} (\bibinfo{year}{2019}), \bibinfo{pages}{374--384}.
\newblock
\href{https://doi.org/10.1109/TVCG.2018.2864769}{doi:\nolinkurl{10.1109/TVCG.2018.2864769}}


\bibitem[Elmqvist et~al\mbox{.}(2011)]%
        {elmqvist11}
\bibfield{author}{\bibinfo{person}{Niklas Elmqvist}, \bibinfo{person}{Andrew~Vande Moere}, \bibinfo{person}{Hans-Christian Jetter}, \bibinfo{person}{Daniel Cernea}, \bibinfo{person}{Harald Reiterer}, {and} \bibinfo{person}{TJ Jankun-Kelly}.} \bibinfo{year}{2011}\natexlab{}.
\newblock \showarticletitle{Fluid interaction for information visualization}.
\newblock \bibinfo{journal}{\emph{Information Visualization}} \bibinfo{volume}{10}, \bibinfo{number}{4} (\bibinfo{year}{2011}), \bibinfo{pages}{327--340}.
\newblock
\href{https://doi.org/10.1177/1473871611413180}{doi:\nolinkurl{10.1177/1473871611413180}}


\bibitem[Few(2006)]%
        {few2006information}
\bibfield{author}{\bibinfo{person}{Stephen Few}.} \bibinfo{year}{2006}\natexlab{}.
\newblock \bibinfo{booktitle}{\emph{Information Dashboard Design}}.
\newblock \bibinfo{publisher}{O'Reilly Media, Inc.}, \bibinfo{address}{Sebastopol, CA, USA}.
\newblock


\bibitem[Firat et~al\mbox{.}(2022a)]%
        {firat22parallel}
\bibfield{author}{\bibinfo{person}{Elif~E. Firat}, \bibinfo{person}{Alena Denisova}, \bibinfo{person}{Max~L. Wilson}, {and} \bibinfo{person}{Robert~S. Laramee}.} \bibinfo{year}{2022}\natexlab{a}.
\newblock \showarticletitle{P-Lite: A study of parallel coordinate plot literacy}.
\newblock \bibinfo{journal}{\emph{Visual Informatics}} \bibinfo{volume}{6}, \bibinfo{number}{3} (\bibinfo{year}{2022}), \bibinfo{pages}{81--99}.
\newblock
\href{https://doi.org/10.1016/j.visinf.2022.05.002}{doi:\nolinkurl{10.1016/j.visinf.2022.05.002}}


\bibitem[Firat et~al\mbox{.}(2022b)]%
        {firat22}
\bibfield{author}{\bibinfo{person}{Elif~E. Firat}, \bibinfo{person}{Alark Joshi}, {and} \bibinfo{person}{Robert~S. Laramee}.} \bibinfo{year}{2022}\natexlab{b}.
\newblock \showarticletitle{Interactive visualization literacy: The state-of-the-art}.
\newblock \bibinfo{journal}{\emph{Information Visualization}} \bibinfo{volume}{21}, \bibinfo{number}{3} (\bibinfo{year}{2022}), \bibinfo{pages}{285--310}.
\newblock
\href{https://doi.org/10.1177/14738716221081831}{doi:\nolinkurl{10.1177/14738716221081831}}


\bibitem[Firat et~al\mbox{.}(2023)]%
        {firat23treemap}
\bibfield{author}{\bibinfo{person}{Elif~E. Firat}, \bibinfo{person}{Colm Lang}, \bibinfo{person}{Bhumika Srinivas}, \bibinfo{person}{Ilena Peng}, \bibinfo{person}{Robert~S. Laramee}, {and} \bibinfo{person}{Alark Joshi}.} \bibinfo{year}{2023}\natexlab{}.
\newblock \showarticletitle{{A Constructivism-based Approach to Treemap Literacy in the Classroom}}. In \bibinfo{booktitle}{\emph{Proceedings of Eurographics Education Papers}}. \bibinfo{publisher}{The Eurographics Association}, \bibinfo{address}{Eindhoven, The Netherlands}, \bibinfo{pages}{9--16}.
\newblock
\href{https://doi.org/10.2312/eged.20231016}{doi:\nolinkurl{10.2312/eged.20231016}}


\bibitem[Freedman and Shah(2002)]%
        {freedman02}
\bibfield{author}{\bibinfo{person}{Eric~G. Freedman} {and} \bibinfo{person}{Priti Shah}.} \bibinfo{year}{2002}\natexlab{}.
\newblock \showarticletitle{Toward a Model of Knowledge-Based Graph Comprehension}. In \bibinfo{booktitle}{\emph{Diagrammatic Representation and Inference}}, \bibfield{editor}{\bibinfo{person}{Mary Hegarty}, \bibinfo{person}{Bernd Meyer}, {and} \bibinfo{person}{N.~Hari Narayanan}} (Eds.). \bibinfo{publisher}{Springer Berlin Heidelberg}, \bibinfo{address}{Berlin, Heidelberg}, \bibinfo{pages}{18--30}.
\newblock
\showISBNx{978-3-540-46037-4}


\bibitem[Ge et~al\mbox{.}(2023)]%
        {ge23}
\bibfield{author}{\bibinfo{person}{Lily~W. Ge}, \bibinfo{person}{Yuan Cui}, {and} \bibinfo{person}{Matthew Kay}.} \bibinfo{year}{2023}\natexlab{}.
\newblock \showarticletitle{{CALVI}: Critical Thinking Assessment for Literacy in Visualizations}. In \bibinfo{booktitle}{\emph{Proceedings of the {ACM} Conference on Human Factors in Computing Systems}}. \bibinfo{publisher}{{ACM}}, \bibinfo{address}{New York, NY, USA}, \bibinfo{pages}{815:1--815:18}.
\newblock
\href{https://doi.org/10.1145/3544548.3581406}{doi:\nolinkurl{10.1145/3544548.3581406}}


\bibitem[Ge et~al\mbox{.}(2025)]%
        {ge25avec}
\bibfield{author}{\bibinfo{person}{Lily~W. Ge}, \bibinfo{person}{Yuan Cui}, {and} \bibinfo{person}{Matthew Kay}.} \bibinfo{year}{2025}\natexlab{}.
\newblock \showarticletitle{AVEC: An Assessment of Visual Encoding Ability in Visualization Construction}. In \bibinfo{booktitle}{\emph{Proceedings of the {ACM} Conference on Human Factors in Computing Systems}}. \bibinfo{publisher}{{ACM}}, \bibinfo{address}{New York, NY, USA}, Article \bibinfo{articleno}{1166}, \bibinfo{numpages}{16}~pages.
\newblock
\showISBNx{9798400713941}
\href{https://doi.org/10.1145/3706598.3713364}{doi:\nolinkurl{10.1145/3706598.3713364}}


\bibitem[Ge et~al\mbox{.}(2024)]%
        {ge24workshop}
\bibfield{author}{\bibinfo{person}{Lily~W. Ge}, \bibinfo{person}{Maryam Hedayati}, \bibinfo{person}{Yuan Cui}, \bibinfo{person}{Yiren Ding}, \bibinfo{person}{Karen Bonilla}, \bibinfo{person}{Alark Joshi}, \bibinfo{person}{Alvitta Ottley}, \bibinfo{person}{Benjamin Bach}, \bibinfo{person}{Bum~Chul Kwon}, \bibinfo{person}{David~N. Rapp}, \bibinfo{person}{Evan Peck}, \bibinfo{person}{Lace~M. Padilla}, \bibinfo{person}{Michael Correll}, \bibinfo{person}{Michelle~A. Borkin}, \bibinfo{person}{Lane Harrison}, {and} \bibinfo{person}{Matthew Kay}.} \bibinfo{year}{2024}\natexlab{}.
\newblock \showarticletitle{Toward a More Comprehensive Understanding of Visualization Literacy}. In \bibinfo{booktitle}{\emph{Extended Abstracts of the ACM Conference on Human Factors in Computing Systems}}. \bibinfo{publisher}{{ACM}}, \bibinfo{address}{New York, NY, USA}, Article \bibinfo{articleno}{494}, \bibinfo{numpages}{7}~pages.
\newblock
\href{https://doi.org/10.1145/3613905.3636289}{doi:\nolinkurl{10.1145/3613905.3636289}}


\bibitem[Gigerenzer(2002)]%
        {gigerenzer02}
\bibfield{author}{\bibinfo{person}{Gerd Gigerenzer}.} \bibinfo{year}{2002}\natexlab{}.
\newblock \bibinfo{booktitle}{\emph{Calculated Risks: How to Know When Numbers Deceive You}}.
\newblock \bibinfo{publisher}{Simon \& Schuster}, \bibinfo{address}{New York}.
\newblock
\showISBNx{0743205561}
\urldef\tempurl%
\url{https://hdl.handle.net/11858/00-001M-0000-0025-919F-0}
\showURL{%
\tempurl}


\bibitem[Green and Fisher(2010)]%
        {green10}
\bibfield{author}{\bibinfo{person}{Tera~Marie Green} {and} \bibinfo{person}{Brian Fisher}.} \bibinfo{year}{2010}\natexlab{}.
\newblock \showarticletitle{Towards the Personal Equation of Interaction: The Impact of Personality Factors on Visual Analytics Interface Interaction}. In \bibinfo{booktitle}{\emph{Proceedings of the {IEEE} Conference on Visual Analytics Science and Technology}}. \bibinfo{publisher}{{IEEE Computer Society}}, \bibinfo{address}{Los Alamitos, CA, USA}, \bibinfo{pages}{203--210}.
\newblock
\href{https://doi.org/10.1109/VAST.2010.5653587}{doi:\nolinkurl{10.1109/VAST.2010.5653587}}


\bibitem[Group(2024)]%
        {wb24}
\bibfield{author}{\bibinfo{person}{World~Bank Group}.} \bibinfo{year}{2024}\natexlab{}.
\newblock \bibinfo{booktitle}{\emph{Foundational Learning}}.
\newblock
\newblock
\shownote{\url{https://www.worldbank.org/en/topic/education/brief/foundational-learning}}.


\bibitem[Hedayati et~al\mbox{.}(2024)]%
        {hedayati24reconcept}
\bibfield{author}{\bibinfo{person}{Maryam Hedayati}, \bibinfo{person}{Ayse Hunt}, {and} \bibinfo{person}{Matthew Kay}.} \bibinfo{year}{2024}\natexlab{}.
\newblock \bibinfo{title}{From pixels to practices: Reconceptualizing visualization literacy}.
\newblock
\href{https://doi.org/10.31219/osf.io/6mq42}{doi:\nolinkurl{10.31219/osf.io/6mq42}}


\bibitem[Hedayati and Kay(2025)]%
        {hedayati25uni}
\bibfield{author}{\bibinfo{person}{Maryam Hedayati} {and} \bibinfo{person}{Matthew Kay}.} \bibinfo{year}{2025}\natexlab{}.
\newblock \showarticletitle{What University Students Learn In Visualization Classes}.
\newblock \bibinfo{journal}{\emph{{{IEEE} Transactions on Visualization and Computer Graphics}}} \bibinfo{volume}{31}, \bibinfo{number}{1} (\bibinfo{year}{2025}), \bibinfo{pages}{1072--1082}.
\newblock
\href{https://doi.org/10.1109/TVCG.2024.3456291}{doi:\nolinkurl{10.1109/TVCG.2024.3456291}}


\bibitem[Heer et~al\mbox{.}(2008)]%
        {heer2008graphical}
\bibfield{author}{\bibinfo{person}{Jeffrey Heer}, \bibinfo{person}{Jock Mackinlay}, \bibinfo{person}{Chris Stolte}, {and} \bibinfo{person}{Maneesh Agrawala}.} \bibinfo{year}{2008}\natexlab{}.
\newblock \showarticletitle{Graphical histories for visualization: Supporting analysis, communication, and evaluation}.
\newblock \bibinfo{journal}{\emph{{{IEEE} Transactions on Visualization and Computer Graphics}}} \bibinfo{volume}{14}, \bibinfo{number}{6} (\bibinfo{year}{2008}), \bibinfo{pages}{1189--1196}.
\newblock
\href{https://doi.org/10.1109/TVCG.2008.137}{doi:\nolinkurl{10.1109/TVCG.2008.137}}


\bibitem[Heer and Shneiderman(2012)]%
        {heer12}
\bibfield{author}{\bibinfo{person}{Jeffrey Heer} {and} \bibinfo{person}{Ben Shneiderman}.} \bibinfo{year}{2012}\natexlab{}.
\newblock \showarticletitle{Interactive Dynamics for Visual Analysis: A taxonomy of tools that support the fluent and flexible use of visualizations}.
\newblock \bibinfo{journal}{\emph{Queue}} \bibinfo{volume}{10}, \bibinfo{number}{2} (\bibinfo{year}{2012}), \bibinfo{pages}{30--55}.
\newblock
\href{https://doi.org/10.1145/2133416.2146416}{doi:\nolinkurl{10.1145/2133416.2146416}}


\bibitem[Hollan et~al\mbox{.}(2000)]%
        {hollan00}
\bibfield{author}{\bibinfo{person}{James Hollan}, \bibinfo{person}{Edwin Hutchins}, {and} \bibinfo{person}{David Kirsh}.} \bibinfo{year}{2000}\natexlab{}.
\newblock \showarticletitle{Distributed cognition: toward a new foundation for human-computer interaction research}.
\newblock \bibinfo{journal}{\emph{{ACM} Transactions on Computer-Human Interaction}} \bibinfo{volume}{7}, \bibinfo{number}{2} (\bibinfo{date}{June} \bibinfo{year}{2000}), \bibinfo{pages}{174--196}.
\newblock
\href{https://doi.org/10.1145/353485.353487}{doi:\nolinkurl{10.1145/353485.353487}}


\bibitem[Horak et~al\mbox{.}(2018)]%
        {horak18}
\bibfield{author}{\bibinfo{person}{Tom Horak}, \bibinfo{person}{Sriram~Karthik Badam}, \bibinfo{person}{Niklas Elmqvist}, {and} \bibinfo{person}{Raimund Dachselt}.} \bibinfo{year}{2018}\natexlab{}.
\newblock \showarticletitle{When {David} Meets {Goliath}: Combining Smartwatches with a Large Vertical Display for Visual Data Exploration}. In \bibinfo{booktitle}{\emph{Proceedings of the {ACM} Conference on Human Factors in Computing Systems}}. \bibinfo{publisher}{{ACM}}, \bibinfo{address}{New York, NY, USA}, \bibinfo{pages}{19:1--19:13}.
\newblock
\href{https://doi.org/10.1145/3173574.3173593}{doi:\nolinkurl{10.1145/3173574.3173593}}


\bibitem[Horak et~al\mbox{.}(2019)]%
        {DBLP:conf/chi/HorakMKDE19}
\bibfield{author}{\bibinfo{person}{Tom Horak}, \bibinfo{person}{Andreas Mathisen}, \bibinfo{person}{Clemens~Nylandsted Klokmose}, \bibinfo{person}{Raimund Dachselt}, {and} \bibinfo{person}{Niklas Elmqvist}.} \bibinfo{year}{2019}\natexlab{}.
\newblock \showarticletitle{Vistribute: Distributing Interactive Visualizations in Dynamic Multi-Device Setups}. In \bibinfo{booktitle}{\emph{Proceedings of the {ACM} Conference on Human Factors in Computing Systems}}. \bibinfo{publisher}{{ACM}}, \bibinfo{address}{New York, NY, USA}, \bibinfo{pages}{616}.
\newblock
\href{https://doi.org/10.1145/3290605.3300846}{doi:\nolinkurl{10.1145/3290605.3300846}}


\bibitem[Hornb\ae{}k et~al\mbox{.}(2019)]%
        {hornbaek19}
\bibfield{author}{\bibinfo{person}{Kasper Hornb\ae{}k}, \bibinfo{person}{Aske Mottelson}, \bibinfo{person}{Jarrod Knibbe}, {and} \bibinfo{person}{Daniel Vogel}.} \bibinfo{year}{2019}\natexlab{}.
\newblock \showarticletitle{What Do We Mean by “Interaction”? {A}n Analysis of 35 Years of {CHI}}.
\newblock \bibinfo{journal}{\emph{{ACM} Transactions on Computer-Human Interaction}} \bibinfo{volume}{26}, \bibinfo{number}{4}, Article \bibinfo{articleno}{27} (\bibinfo{date}{July} \bibinfo{year}{2019}), \bibinfo{numpages}{30}~pages.
\newblock
\href{https://doi.org/10.1145/3325285}{doi:\nolinkurl{10.1145/3325285}}


\bibitem[Hornb\ae{}k and Oulasvirta(2017)]%
        {hornbaek17}
\bibfield{author}{\bibinfo{person}{Kasper Hornb\ae{}k} {and} \bibinfo{person}{Antti Oulasvirta}.} \bibinfo{year}{2017}\natexlab{}.
\newblock \showarticletitle{What Is Interaction?}. In \bibinfo{booktitle}{\emph{Proceedings of the {ACM} Conference on Human Factors in Computing Systems}}. \bibinfo{publisher}{{ACM}}, \bibinfo{address}{New York, NY, USA}, \bibinfo{pages}{5040--5052}.
\newblock
\href{https://doi.org/10.1145/3025453.3025765}{doi:\nolinkurl{10.1145/3025453.3025765}}


\bibitem[Hornbæk et~al\mbox{.}(2025)]%
        {hornbaek25book}
\bibfield{author}{\bibinfo{person}{Kasper Hornbæk}, \bibinfo{person}{Per~Ola Kristensson}, {and} \bibinfo{person}{Antti Oulasvirta}.} \bibinfo{year}{2025}\natexlab{}.
\newblock \bibinfo{booktitle}{\emph{Introduction to Human-Computer Interaction}}.
\newblock \bibinfo{publisher}{Oxford University Press}, \bibinfo{address}{Oxford, UK}.
\newblock
\href{https://doi.org/10.1093/oso/9780192864543.001.0001}{doi:\nolinkurl{10.1093/oso/9780192864543.001.0001}}


\bibitem[Horvitz(1999)]%
        {horvitz99}
\bibfield{author}{\bibinfo{person}{Eric Horvitz}.} \bibinfo{year}{1999}\natexlab{}.
\newblock \showarticletitle{Principles of Mixed-Initiative User Interfaces}. In \bibinfo{booktitle}{\emph{Proceedings of the {ACM} Conference on Human Factors in Computing Systems}}. \bibinfo{publisher}{{ACM}}, \bibinfo{address}{New York, NY, USA}, \bibinfo{pages}{159--166}.
\newblock
\href{https://doi.org/10.1145/302979.303030}{doi:\nolinkurl{10.1145/302979.303030}}


\bibitem[Hubenschmid et~al\mbox{.}(2021)]%
        {hubenschmid21}
\bibfield{author}{\bibinfo{person}{Sebastian Hubenschmid}, \bibinfo{person}{Johannes Zagermann}, \bibinfo{person}{Simon Butscher}, {and} \bibinfo{person}{Harald Reiterer}.} \bibinfo{year}{2021}\natexlab{}.
\newblock \showarticletitle{{STREAM}: Exploring the Combination of Spatially-Aware Tablets with Augmented Reality Head-Mounted Displays for Immersive Analytics}. In \bibinfo{booktitle}{\emph{Proceedings of the {ACM} Conference on Human Factors in Computing Systems}}. \bibinfo{publisher}{{ACM}}, \bibinfo{address}{New York, NY, USA}, \bibinfo{pages}{469:1--469:14}.
\newblock
\href{https://doi.org/10.1145/3411764.3445298}{doi:\nolinkurl{10.1145/3411764.3445298}}


\bibitem[Hutchins(1995)]%
        {Hutchins_1995_CognitionInTheWild}
\bibfield{author}{\bibinfo{person}{Edwin Hutchins}.} \bibinfo{year}{1995}\natexlab{}.
\newblock \bibinfo{booktitle}{\emph{Cognition in the Wild}}.
\newblock \bibinfo{publisher}{MIT Press}, \bibinfo{address}{Cambridge, MA, USA}.
\newblock


\bibitem[Hutchins et~al\mbox{.}(1985)]%
        {hutchins85}
\bibfield{author}{\bibinfo{person}{Edwin~L. Hutchins}, \bibinfo{person}{James~D. Hollan}, {and} \bibinfo{person}{Donald~A. Norman}.} \bibinfo{year}{1985}\natexlab{}.
\newblock \showarticletitle{Direct Manipulation Interfaces}.
\newblock \bibinfo{journal}{\emph{Human–Computer Interaction}} \bibinfo{volume}{1}, \bibinfo{number}{4} (\bibinfo{year}{1985}), \bibinfo{pages}{311--338}.
\newblock
\href{https://doi.org/10.1207/s15327051hci0104_2}{doi:\nolinkurl{10.1207/s15327051hci0104_2}}


\bibitem[Jacob et~al\mbox{.}(2008)]%
        {jacob08}
\bibfield{author}{\bibinfo{person}{Robert~J.K. Jacob}, \bibinfo{person}{Audrey Girouard}, \bibinfo{person}{Leanne~M. Hirshfield}, \bibinfo{person}{Michael~S. Horn}, \bibinfo{person}{Orit Shaer}, \bibinfo{person}{Erin~Treacy Solovey}, {and} \bibinfo{person}{Jamie Zigelbaum}.} \bibinfo{year}{2008}\natexlab{}.
\newblock \showarticletitle{Reality-based interaction: a framework for post-WIMP interfaces}. In \bibinfo{booktitle}{\emph{Proceedings of the {ACM} Conference on Human Factors in Computing Systems}}. \bibinfo{publisher}{{ACM}}, \bibinfo{address}{New York, NY, USA}, \bibinfo{pages}{201--210}.
\newblock
\showISBNx{9781605580111}
\href{https://doi.org/10.1145/1357054.1357089}{doi:\nolinkurl{10.1145/1357054.1357089}}


\bibitem[Jansen and Dragicevic(2013)]%
        {jansen13}
\bibfield{author}{\bibinfo{person}{Yvonne Jansen} {and} \bibinfo{person}{Pierre Dragicevic}.} \bibinfo{year}{2013}\natexlab{}.
\newblock \showarticletitle{An Interaction Model for Visualizations Beyond The Desktop}.
\newblock \bibinfo{journal}{\emph{{{IEEE} Transactions on Visualization and Computer Graphics}}} \bibinfo{volume}{19}, \bibinfo{number}{12} (\bibinfo{year}{2013}), \bibinfo{pages}{2396--2405}.
\newblock
\href{https://doi.org/10.1109/TVCG.2013.134}{doi:\nolinkurl{10.1109/TVCG.2013.134}}


\bibitem[Kahneman(2011)]%
        {kahneman2011thinking}
\bibfield{author}{\bibinfo{person}{Daniel Kahneman}.} \bibinfo{year}{2011}\natexlab{}.
\newblock \bibinfo{booktitle}{\emph{Thinking, Fast and Slow}}.
\newblock \bibinfo{publisher}{Farrar, Straus and Giroux}, \bibinfo{address}{New York}.
\newblock


\bibitem[Kegel et~al\mbox{.}(2019)]%
        {kegel19}
\bibfield{author}{\bibinfo{person}{R.H.P. Kegel}, \bibinfo{person}{M. {van Sinderen}}, {and} \bibinfo{person}{R.J. Wieringa}.} \bibinfo{year}{2019}\natexlab{}.
\newblock \showarticletitle{Towards More Individualized Interfaces: Automating the assessment of computer literacy}. In \bibinfo{booktitle}{\emph{Behavior Change Support Systems}}. \bibinfo{publisher}{CEUR}, \bibinfo{address}{Germany}, \bibinfo{numpages}{12}~pages.
\newblock


\bibitem[Keim et~al\mbox{.}(2008)]%
        {keim08}
\bibfield{author}{\bibinfo{person}{Daniel Keim}, \bibinfo{person}{Gennady Andrienko}, \bibinfo{person}{Jean-Daniel Fekete}, \bibinfo{person}{Carsten G{\"o}rg}, \bibinfo{person}{J{\"o}rn Kohlhammer}, {and} \bibinfo{person}{Guy Melan{\c{c}}on}.} \bibinfo{year}{2008}\natexlab{}.
\newblock \bibinfo{booktitle}{\emph{Visual Analytics: Definition, Process, and Challenges}}.
\newblock \bibinfo{publisher}{Springer Publishing}, \bibinfo{address}{New York, NY, USA}, \bibinfo{pages}{154--175}.
\newblock
\href{https://doi.org/10.1007/978-3-540-70956-5_7}{doi:\nolinkurl{10.1007/978-3-540-70956-5_7}}


\bibitem[Klein et~al\mbox{.}(2006a)]%
        {klein2006making1}
\bibfield{author}{\bibinfo{person}{Gary Klein}, \bibinfo{person}{Brian Moon}, {and} \bibinfo{person}{Robert~R. Hoffman}.} \bibinfo{year}{2006}\natexlab{a}.
\newblock \showarticletitle{Making Sense of Sensemaking 1: Alternative Perspectives}.
\newblock \bibinfo{journal}{\emph{IEEE Intelligent Systems}} \bibinfo{volume}{21}, \bibinfo{number}{4} (\bibinfo{year}{2006}), \bibinfo{pages}{70--73}.
\newblock
\href{https://doi.org/10.1109/MIS.2006.75}{doi:\nolinkurl{10.1109/MIS.2006.75}}


\bibitem[Klein et~al\mbox{.}(2006b)]%
        {klein2006making2}
\bibfield{author}{\bibinfo{person}{Gary Klein}, \bibinfo{person}{David~T. Woods}, \bibinfo{person}{Jeff~M. Bradshaw}, \bibinfo{person}{Robert~R. Hoffman}, {and} \bibinfo{person}{Paul~J. Feltovich}.} \bibinfo{year}{2006}\natexlab{b}.
\newblock \showarticletitle{Making Sense of Sensemaking 2: A Macrocognitive Model}.
\newblock \bibinfo{journal}{\emph{IEEE Intelligent Systems}} \bibinfo{volume}{21}, \bibinfo{number}{5} (\bibinfo{year}{2006}), \bibinfo{pages}{88--92}.
\newblock
\href{https://doi.org/10.1109/MIS.2006.100}{doi:\nolinkurl{10.1109/MIS.2006.100}}


\bibitem[Kress and Van~Leeuwen(2020)]%
        {kress20}
\bibfield{author}{\bibinfo{person}{Gunther Kress} {and} \bibinfo{person}{Theo Van~Leeuwen}.} \bibinfo{year}{2020}\natexlab{}.
\newblock \bibinfo{booktitle}{\emph{Reading Images: The Grammar of Visual Design} (\bibinfo{edition}{3rd} ed.)}.
\newblock \bibinfo{publisher}{Routledge}, \bibinfo{address}{London, UK}.
\newblock
\href{https://doi.org/10.4324/9781003099857}{doi:\nolinkurl{10.4324/9781003099857}}


\bibitem[Kwon et~al\mbox{.}(2011)]%
        {kwon11}
\bibfield{author}{\bibinfo{person}{Bum~chul Kwon}, \bibinfo{person}{Brian Fisher}, {and} \bibinfo{person}{Ji~Soo Yi}.} \bibinfo{year}{2011}\natexlab{}.
\newblock \showarticletitle{Visual analytic roadblocks for novice investigators}. In \bibinfo{booktitle}{\emph{Proceedings of the {IEEE} Conference on Visual Analytics Science and Technology}}. \bibinfo{publisher}{{IEEE Computer Society}}, \bibinfo{address}{Los Alamitos, CA, USA}, \bibinfo{pages}{3--11}.
\newblock
\href{https://doi.org/10.1109/VAST.2011.6102435}{doi:\nolinkurl{10.1109/VAST.2011.6102435}}


\bibitem[Lam(2008)]%
        {lam08}
\bibfield{author}{\bibinfo{person}{Heidi Lam}.} \bibinfo{year}{2008}\natexlab{}.
\newblock \showarticletitle{A Framework of Interaction Costs in Information Visualization}.
\newblock \bibinfo{journal}{\emph{{{IEEE} Transactions on Visualization and Computer Graphics}}} \bibinfo{volume}{14}, \bibinfo{number}{6} (\bibinfo{year}{2008}), \bibinfo{pages}{1149--1156}.
\newblock
\href{https://doi.org/10.1109/TVCG.2008.109}{doi:\nolinkurl{10.1109/TVCG.2008.109}}


\bibitem[Lee et~al\mbox{.}(2013)]%
        {lee13}
\bibfield{author}{\bibinfo{person}{Bongshin Lee}, \bibinfo{person}{Rubaiat~Habib Kazi}, {and} \bibinfo{person}{Greg Smith}.} \bibinfo{year}{2013}\natexlab{}.
\newblock \showarticletitle{{SketchStory}: Telling More Engaging Stories with Data through Freeform Sketching}.
\newblock \bibinfo{journal}{\emph{{{IEEE} Transactions on Visualization and Computer Graphics}}} \bibinfo{volume}{19}, \bibinfo{number}{12} (\bibinfo{year}{2013}), \bibinfo{pages}{2416--2425}.
\newblock
\href{https://doi.org/10.1109/TVCG.2013.191}{doi:\nolinkurl{10.1109/TVCG.2013.191}}


\bibitem[Lee et~al\mbox{.}(2016)]%
        {lee16}
\bibfield{author}{\bibinfo{person}{Sukwon Lee}, \bibinfo{person}{Sung-Hee Kim}, \bibinfo{person}{Ya-Hsin Hung}, \bibinfo{person}{Heidi Lam}, \bibinfo{person}{Youn-Ah Kang}, {and} \bibinfo{person}{Ji~Soo Yi}.} \bibinfo{year}{2016}\natexlab{}.
\newblock \showarticletitle{How do People Make Sense of Unfamiliar Visualizations?: A Grounded Model of Novice's Information Visualization Sensemaking}.
\newblock \bibinfo{journal}{\emph{{{IEEE} Transactions on Visualization and Computer Graphics}}} \bibinfo{volume}{22}, \bibinfo{number}{1} (\bibinfo{year}{2016}), \bibinfo{pages}{499--508}.
\newblock
\href{https://doi.org/10.1109/TVCG.2015.2467195}{doi:\nolinkurl{10.1109/TVCG.2015.2467195}}


\bibitem[Lee et~al\mbox{.}(2017)]%
        {lee17}
\bibfield{author}{\bibinfo{person}{Sukwon Lee}, \bibinfo{person}{Sung-Hee Kim}, {and} \bibinfo{person}{Bum~Chul Kwon}.} \bibinfo{year}{2017}\natexlab{}.
\newblock \showarticletitle{{VLAT}: Development of a Visualization Literacy Assessment Test}.
\newblock \bibinfo{journal}{\emph{{{IEEE} Transactions on Visualization and Computer Graphics}}} \bibinfo{volume}{23}, \bibinfo{number}{1} (\bibinfo{year}{2017}), \bibinfo{pages}{551--560}.
\newblock
\href{https://doi.org/10.1109/TVCG.2016.2598920}{doi:\nolinkurl{10.1109/TVCG.2016.2598920}}


\bibitem[Li et~al\mbox{.}(2023)]%
        {li2023networknarratives}
\bibfield{author}{\bibinfo{person}{Wenchao Li}, \bibinfo{person}{Sarah Sch{\"o}ttler}, \bibinfo{person}{James Scott-Brown}, \bibinfo{person}{Yun Wang}, \bibinfo{person}{Siming Chen}, \bibinfo{person}{Huamin Qu}, {and} \bibinfo{person}{Benjamin Bach}.} \bibinfo{year}{2023}\natexlab{}.
\newblock \showarticletitle{{NetworkNarratives}: Data tours for visual network exploration and analysis}. In \bibinfo{booktitle}{\emph{Proceedings of the {ACM} Conference on Human Factors in Computing Systems}}. \bibinfo{publisher}{{ACM}}, \bibinfo{address}{New York, NY, USA}, Article \bibinfo{articleno}{172}, \bibinfo{numpages}{15}~pages.
\newblock


\bibitem[Lintunen et~al\mbox{.}(2024)]%
        {lintunen24}
\bibfield{author}{\bibinfo{person}{Erik Lintunen}, \bibinfo{person}{Viljami Salmela}, \bibinfo{person}{Petri Jarre}, \bibinfo{person}{Tuukka Heikkinen}, \bibinfo{person}{Markku Kilpeläinen}, \bibinfo{person}{Markus Jokela}, {and} \bibinfo{person}{Antti Oulasvirta}.} \bibinfo{year}{2024}\natexlab{}.
\newblock \showarticletitle{Cognitive abilities predict performance in everyday computer tasks}.
\newblock \bibinfo{journal}{\emph{International Journal of Human-Computer Studies}}  \bibinfo{volume}{192} (\bibinfo{year}{2024}), \bibinfo{pages}{103354}.
\newblock
\showISSN{1071-5819}
\href{https://doi.org/10.1016/j.ijhcs.2024.103354}{doi:\nolinkurl{10.1016/j.ijhcs.2024.103354}}


\bibitem[Liu et~al\mbox{.}(2020)]%
        {Liu2020}
\bibfield{author}{\bibinfo{person}{Zhengliang Liu}, \bibinfo{person}{R.~Jordan Crouser}, {and} \bibinfo{person}{Alvitta Ottley}.} \bibinfo{year}{2020}\natexlab{}.
\newblock \showarticletitle{Survey on Individual Differences in Visualization}.
\newblock \bibinfo{journal}{\emph{{Computer Graphics Forum}}}  \bibinfo{volume}{39} (\bibinfo{date}{7} \bibinfo{year}{2020}), \bibinfo{pages}{693--712}.
\newblock
\href{https://doi.org/10.1111/cgf.14033}{doi:\nolinkurl{10.1111/cgf.14033}}


\bibitem[Liu and Heer(2014)]%
        {liu14latency}
\bibfield{author}{\bibinfo{person}{Zhicheng Liu} {and} \bibinfo{person}{Jeffrey Heer}.} \bibinfo{year}{2014}\natexlab{}.
\newblock \showarticletitle{The Effects of Interactive Latency on Exploratory Visual Analysis}.
\newblock \bibinfo{journal}{\emph{{{IEEE} Transactions on Visualization and Computer Graphics}}} \bibinfo{volume}{20}, \bibinfo{number}{12} (\bibinfo{year}{2014}), \bibinfo{pages}{2122--2131}.
\newblock
\href{https://doi.org/10.1109/TVCG.2014.2346452}{doi:\nolinkurl{10.1109/TVCG.2014.2346452}}


\bibitem[Liu and Stasko(2010)]%
        {liu10}
\bibfield{author}{\bibinfo{person}{Zhicheng Liu} {and} \bibinfo{person}{John Stasko}.} \bibinfo{year}{2010}\natexlab{}.
\newblock \showarticletitle{Mental Models, Visual Reasoning and Interaction in Information Visualization: A Top-down Perspective}.
\newblock \bibinfo{journal}{\emph{{{IEEE} Transactions on Visualization and Computer Graphics}}} \bibinfo{volume}{16}, \bibinfo{number}{6} (\bibinfo{year}{2010}), \bibinfo{pages}{999--1008}.
\newblock
\href{https://doi.org/10.1109/TVCG.2010.177}{doi:\nolinkurl{10.1109/TVCG.2010.177}}


\bibitem[Luke(2012)]%
        {luke12}
\bibfield{author}{\bibinfo{person}{Allan Luke}.} \bibinfo{year}{2012}\natexlab{}.
\newblock \showarticletitle{Critical Literacy: Foundational Notes}.
\newblock \bibinfo{journal}{\emph{Theory Into Practice}} \bibinfo{volume}{51}, \bibinfo{number}{1} (\bibinfo{year}{2012}), \bibinfo{pages}{4--11}.
\newblock
\showeprint{https://doi.org/10.1080/00405841.2012.636324}
\href{https://doi.org/10.1080/00405841.2012.636324}{doi:\nolinkurl{10.1080/00405841.2012.636324}}


\bibitem[Marriott et~al\mbox{.}(2021)]%
        {marriott21lee}
\bibfield{author}{\bibinfo{person}{Kim Marriott}, \bibinfo{person}{Bongshin Lee}, \bibinfo{person}{Matthew Butler}, \bibinfo{person}{Ed Cutrell}, \bibinfo{person}{Kirsten Ellis}, \bibinfo{person}{Cagatay Goncu}, \bibinfo{person}{Marti Hearst}, \bibinfo{person}{Kathleen McCoy}, {and} \bibinfo{person}{Danielle~Albers Szafir}.} \bibinfo{year}{2021}\natexlab{}.
\newblock \showarticletitle{Inclusive data visualization for people with disabilities: a call to action}.
\newblock \bibinfo{journal}{\emph{Interactions}} \bibinfo{volume}{28}, \bibinfo{number}{3} (\bibinfo{year}{2021}), \bibinfo{pages}{47--51}.
\newblock
\href{https://doi.org/10.1145/3457875}{doi:\nolinkurl{10.1145/3457875}}


\bibitem[Mehta et~al\mbox{.}(2017)]%
        {mehta2017datatours}
\bibfield{author}{\bibinfo{person}{Hrim Mehta}, \bibinfo{person}{Amira Chalbi}, \bibinfo{person}{Fanny Chevalier}, {and} \bibinfo{person}{Christopher Collins}.} \bibinfo{year}{2017}\natexlab{}.
\newblock \showarticletitle{{DataTours}: A data narratives framework}. In \bibinfo{booktitle}{\emph{Poster Proceedings of the IEEE Conference on Information Visualization}}. \bibinfo{publisher}{{IEEE Computer Society}}, \bibinfo{address}{Los Alamitos, CA, USA}, \bibinfo{numpages}{2}~pages.
\newblock
\urldef\tempurl%
\url{https://hal.science/hal-01655433}
\showURL{%
\tempurl}


\bibitem[Miller et~al\mbox{.}(2022)]%
        {miller22}
\bibfield{author}{\bibinfo{person}{Matthias Miller}, \bibinfo{person}{Daniel Fürst}, \bibinfo{person}{Hanna Hauptmann}, \bibinfo{person}{Daniel~A. Keim}, {and} \bibinfo{person}{Mennatallah El-Assady}.} \bibinfo{year}{2022}\natexlab{}.
\newblock \showarticletitle{Augmenting Digital Sheet Music through Visual Analytics}.
\newblock \bibinfo{journal}{\emph{{Computer Graphics Forum}}} \bibinfo{volume}{41}, \bibinfo{number}{1} (\bibinfo{year}{2022}), \bibinfo{pages}{301--316}.
\newblock
\href{https://doi.org/10.1111/cgf.14436}{doi:\nolinkurl{10.1111/cgf.14436}}


\bibitem[Molina~León et~al\mbox{.}(2025)]%
        {leon25}
\bibfield{author}{\bibinfo{person}{Gabriela Molina~León}, \bibinfo{person}{Anastasia Bezerianos}, \bibinfo{person}{Olivier Gladin}, {and} \bibinfo{person}{Petra Isenberg}.} \bibinfo{year}{2025}\natexlab{}.
\newblock \showarticletitle{Talk to the Wall: The Role of Speech Interaction in Collaborative Visual Analytics}.
\newblock \bibinfo{journal}{\emph{{{IEEE} Transactions on Visualization and Computer Graphics}}} \bibinfo{volume}{31}, \bibinfo{number}{1} (\bibinfo{year}{2025}), \bibinfo{pages}{941--951}.
\newblock
\href{https://doi.org/10.1109/TVCG.2024.3456335}{doi:\nolinkurl{10.1109/TVCG.2024.3456335}}


\bibitem[Mora and Jia(2024)]%
        {islovedying24}
\bibfield{author}{\bibinfo{person}{David Mora} {and} \bibinfo{person}{Michelle Jia}.} \bibinfo{year}{2024}\natexlab{}.
\newblock \bibinfo{title}{Is the love song dying?}
\newblock
\newblock
\shownote{The Pudding. \href{https://pudding.cool/2024/11/love-songs/}{https://pudding.cool/2024/11/love-songs/}}.


\bibitem[Mosca et~al\mbox{.}(2021)]%
        {mosca21bayesian}
\bibfield{author}{\bibinfo{person}{Ab Mosca}, \bibinfo{person}{Alvitta Ottley}, {and} \bibinfo{person}{Remco Chang}.} \bibinfo{year}{2021}\natexlab{}.
\newblock \showarticletitle{Does Interaction Improve Bayesian Reasoning with Visualization?}. In \bibinfo{booktitle}{\emph{Proceedings of the {ACM} Conference on Human Factors in Computing Systems}}. \bibinfo{publisher}{{ACM}}, \bibinfo{address}{New York, NY, USA}, \bibinfo{pages}{609:1--609:14}.
\newblock
\href{https://doi.org/10.1145/3411764.3445176}{doi:\nolinkurl{10.1145/3411764.3445176}}


\bibitem[Nobre et~al\mbox{.}(2024)]%
        {nobre24}
\bibfield{author}{\bibinfo{person}{Carolina Nobre}, \bibinfo{person}{Kehang Zhu}, \bibinfo{person}{Eric M\"{o}rth}, \bibinfo{person}{Hanspeter Pfister}, {and} \bibinfo{person}{Johanna Beyer}.} \bibinfo{year}{2024}\natexlab{}.
\newblock \showarticletitle{Reading Between the Pixels: Investigating the Barriers to Visualization Literacy}. In \bibinfo{booktitle}{\emph{Proceedings of the {ACM} Conference on Human Factors in Computing Systems}}. \bibinfo{publisher}{{ACM}}, \bibinfo{address}{New York, NY, USA}, \bibinfo{pages}{197:1--197:17}.
\newblock
\href{https://doi.org/10.1145/3613904.3642760}{doi:\nolinkurl{10.1145/3613904.3642760}}


\bibitem[Norman(1988)]%
        {norman88}
\bibfield{author}{\bibinfo{person}{Donald~A. Norman}.} \bibinfo{year}{1988}\natexlab{}.
\newblock \bibinfo{booktitle}{\emph{Design of Everyday Things}}.
\newblock \bibinfo{publisher}{Basic Books}, \bibinfo{address}{New York, NY, USA}.
\newblock
\showISBNx{978-0-465-06710-7}


\bibitem[Norman and Draper(1986)]%
        {norman86}
\bibfield{author}{\bibinfo{person}{Donald~A. Norman} {and} \bibinfo{person}{Stephen~W. Draper}.} \bibinfo{year}{1986}\natexlab{}.
\newblock \bibinfo{booktitle}{\emph{User Centered System Design: New Perspectives on Human-Computer Interaction}}.
\newblock \bibinfo{publisher}{Lawrence Erlbaum Associates Inc.}, \bibinfo{address}{Mahwah, NJ, USA}.
\newblock


\bibitem[{Organisation for Economic Co-operation and Development}(2014)]%
        {oecdindex14}
\bibfield{author}{\bibinfo{person}{{Organisation for Economic Co-operation and Development}}.} \bibinfo{year}{2014}\natexlab{}.
\newblock \bibinfo{title}{Better Life Index}.
\newblock
\newblock
\shownote{OECD website. \href{https://www.oecdbetterlifeindex.org/}{https://www.oecdbetterlifeindex.org/}}.


\bibitem[Pandey and Ottley(2023)]%
        {pandey23}
\bibfield{author}{\bibinfo{person}{Saugat Pandey} {and} \bibinfo{person}{Alvitta Ottley}.} \bibinfo{year}{2023}\natexlab{}.
\newblock \showarticletitle{Mini-{VLAT}: A Short and Effective Measure of Visualization Literacy}.
\newblock \bibinfo{journal}{\emph{{Computer Graphics Forum}}} \bibinfo{volume}{42}, \bibinfo{number}{3} (\bibinfo{year}{2023}), \bibinfo{pages}{1--11}.
\newblock
\href{https://doi.org/10.1111/cgf.14809}{doi:\nolinkurl{10.1111/cgf.14809}}


\bibitem[Pike et~al\mbox{.}(2009)]%
        {pike09}
\bibfield{author}{\bibinfo{person}{William~A. Pike}, \bibinfo{person}{John~T. Stasko}, \bibinfo{person}{Remco Chang}, {and} \bibinfo{person}{Theresa O'Connell}.} \bibinfo{year}{2009}\natexlab{}.
\newblock \showarticletitle{The Science of Interaction}.
\newblock \bibinfo{journal}{\emph{Information Visualization}} \bibinfo{volume}{8}, \bibinfo{number}{4} (\bibinfo{year}{2009}), \bibinfo{pages}{263--274}.
\newblock
\href{https://doi.org/10.1057/ivs.2009.22}{doi:\nolinkurl{10.1057/ivs.2009.22}}


\bibitem[Pirlea et~al\mbox{.}(2023)]%
        {pirlea23}
\bibfield{author}{\bibinfo{person}{A.~F. Pirlea}, \bibinfo{person}{U. Serajuddin}, \bibinfo{person}{D. Wadhwa}, {and} \bibinfo{person}{M. Welch}.} \bibinfo{year}{2023}\natexlab{}.
\newblock \bibinfo{title}{Atlas of {SDGs}}.
\newblock
\newblock
\shownote{\href{https://datatopics.worldbank.org/sdgatlas/}{https://datatopics.worldbank.org/sdgatlas/}}.


\bibitem[Pirolli and Card(2005)]%
        {pirolli2005sensemaking}
\bibfield{author}{\bibinfo{person}{Peter Pirolli} {and} \bibinfo{person}{Stuart Card}.} \bibinfo{year}{2005}\natexlab{}.
\newblock \showarticletitle{The Sensemaking Process and Leverage Points for Analyst Technology as Identified Through Cognitive Task Analysis}. In \bibinfo{booktitle}{\emph{Proceedings of the International Conference on Intelligence Analysis}}. \bibinfo{publisher}{MITRE Corporation}, \bibinfo{address}{McLean, VA, USA}, \bibinfo{pages}{2--4}.
\newblock
\urldef\tempurl%
\url{https://analysis.mitre.org/proceedings/Final_Papers_Files/206_Camera_Ready_Paper.pdf}
\showURL{%
\tempurl}


\bibitem[Rumsey(2002)]%
        {Rumsey2002}
\bibfield{author}{\bibinfo{person}{Deborah~J. Rumsey}.} \bibinfo{year}{2002}\natexlab{}.
\newblock \showarticletitle{Statistical Literacy as a Goal for Introductory Statistics Courses}.
\newblock \bibinfo{journal}{\emph{Journal of Statistics Education}} \bibinfo{volume}{10}, \bibinfo{number}{3} (\bibinfo{year}{2002}), \bibinfo{numpages}{12}~pages.
\newblock
\href{https://doi.org/10.1080/10691898.2002.11910678}{doi:\nolinkurl{10.1080/10691898.2002.11910678}}


\bibitem[Rushkoff(2010)]%
        {rushkoff10}
\bibfield{author}{\bibinfo{person}{Douglas Rushkoff}.} \bibinfo{year}{2010}\natexlab{}.
\newblock \bibinfo{booktitle}{\emph{Program Or Be Programmed: Ten Commands for a Digital Age}}.
\newblock \bibinfo{publisher}{OR Books, LLC}, \bibinfo{address}{New York}.
\newblock
\showISBNx{9781935928164}
\showLCCN{2012554862}
\urldef\tempurl%
\url{https://books.google.de/books?id=SB474JCHewcC}
\showURL{%
\tempurl}


\bibitem[Russell et~al\mbox{.}(1993)]%
        {russell1993cost}
\bibfield{author}{\bibinfo{person}{Daniel~M. Russell}, \bibinfo{person}{Mark~J. Stefik}, \bibinfo{person}{Peter Pirolli}, {and} \bibinfo{person}{Stuart~K. Card}.} \bibinfo{year}{1993}\natexlab{}.
\newblock \showarticletitle{The Cost Structure of Sensemaking}. In \bibinfo{booktitle}{\emph{Proceedings of the {ACM} Conference on Human Factors in Computing Systems}}. \bibinfo{publisher}{{ACM}}, \bibinfo{address}{New York, NY, USA}, \bibinfo{pages}{269--276}.
\newblock
\href{https://doi.org/10.1145/169059.169209}{doi:\nolinkurl{10.1145/169059.169209}}


\bibitem[Sarikaya et~al\mbox{.}(2018)]%
        {sarikaya2018we}
\bibfield{author}{\bibinfo{person}{Alper Sarikaya}, \bibinfo{person}{Michael Correll}, \bibinfo{person}{Lyn Bartram}, \bibinfo{person}{Melanie Tory}, {and} \bibinfo{person}{Danyel Fisher}.} \bibinfo{year}{2018}\natexlab{}.
\newblock \showarticletitle{What do we talk about when we talk about dashboards?}
\newblock \bibinfo{journal}{\emph{{{IEEE} Transactions on Visualization and Computer Graphics}}} \bibinfo{volume}{25}, \bibinfo{number}{1} (\bibinfo{year}{2018}), \bibinfo{pages}{682--692}.
\newblock


\bibitem[Segel and Heer(2010)]%
        {segel2010narrative}
\bibfield{author}{\bibinfo{person}{Edward Segel} {and} \bibinfo{person}{Jeffrey Heer}.} \bibinfo{year}{2010}\natexlab{}.
\newblock \showarticletitle{Narrative visualization: Telling stories with data}.
\newblock \bibinfo{journal}{\emph{{{IEEE} Transactions on Visualization and Computer Graphics}}} \bibinfo{volume}{16}, \bibinfo{number}{6} (\bibinfo{year}{2010}), \bibinfo{pages}{1139--1148}.
\newblock
\href{https://doi.org/10.1109/TVCG.2010.179}{doi:\nolinkurl{10.1109/TVCG.2010.179}}


\bibitem[Shneiderman(1982)]%
        {shneiderman82dm}
\bibfield{author}{\bibinfo{person}{Ben Shneiderman}.} \bibinfo{year}{1982}\natexlab{}.
\newblock \showarticletitle{The future of interactive systems and the emergence of direct manipulation}.
\newblock \bibinfo{journal}{\emph{Behaviour \& Information Technology}} \bibinfo{volume}{1}, \bibinfo{number}{3} (\bibinfo{year}{1982}), \bibinfo{pages}{237--256}.
\newblock
\href{https://doi.org/10.1080/01449298208914450}{doi:\nolinkurl{10.1080/01449298208914450}}


\bibitem[Shneiderman(1983)]%
        {DBLP:journals/computer/Shneiderman83}
\bibfield{author}{\bibinfo{person}{Ben Shneiderman}.} \bibinfo{year}{1983}\natexlab{}.
\newblock \showarticletitle{Direct Manipulation: {A} Step Beyond Programming Languages}.
\newblock \bibinfo{journal}{\emph{Computer}} \bibinfo{volume}{16}, \bibinfo{number}{8} (\bibinfo{year}{1983}), \bibinfo{pages}{57--69}.
\newblock
\href{https://doi.org/10.1109/MC.1983.1654471}{doi:\nolinkurl{10.1109/MC.1983.1654471}}


\bibitem[Shneiderman and Plaisant(2006)]%
        {shneiderman06milcs}
\bibfield{author}{\bibinfo{person}{Ben Shneiderman} {and} \bibinfo{person}{Catherine Plaisant}.} \bibinfo{year}{2006}\natexlab{}.
\newblock \showarticletitle{Strategies for evaluating information visualization tools: multi-dimensional in-depth long-term case studies}. In \bibinfo{booktitle}{\emph{Proceedings of the Workshop on Beyond Time and Errors in Visualization}}. \bibinfo{publisher}{{ACM}}, \bibinfo{address}{New York, NY, USA}, \bibinfo{numpages}{7}~pages.
\newblock
\href{https://doi.org/10.1145/1168149.1168158}{doi:\nolinkurl{10.1145/1168149.1168158}}


\bibitem[Shu et~al\mbox{.}(2024)]%
        {shu2024does}
\bibfield{author}{\bibinfo{person}{Xinhuan Shu}, \bibinfo{person}{Alexis Pister}, \bibinfo{person}{Junxiu Tang}, \bibinfo{person}{Fanny Chevalier}, {and} \bibinfo{person}{Benjamin Bach}.} \bibinfo{year}{2024}\natexlab{}.
\newblock \showarticletitle{Does This Have a Particular Meaning? Interactive Pattern Explanation for Network Visualizations}.
\newblock \bibinfo{journal}{\emph{{{IEEE} Transactions on Visualization and Computer Graphics}}} \bibinfo{volume}{25}, \bibinfo{number}{1} (\bibinfo{year}{2024}), \bibinfo{pages}{677--687}.
\newblock
\href{https://doi.org/10.1109/TVCG.2024.3456192}{doi:\nolinkurl{10.1109/TVCG.2024.3456192}}


\bibitem[Solen(2022)]%
        {solen22}
\bibfield{author}{\bibinfo{person}{Mara Solen}.} \bibinfo{year}{2022}\natexlab{}.
\newblock \bibinfo{title}{Scoping the Future of Visualization Literacy: A Review}.
\newblock \bibinfo{howpublished}{VisComm Workshop}.
\newblock
\href{https://doi.org/10.31219/osf.io/eypgm}{doi:\nolinkurl{10.31219/osf.io/eypgm}}


\bibitem[Solen et~al\mbox{.}(2025)]%
        {solen25skillset}
\bibfield{author}{\bibinfo{person}{Mara Solen}, \bibinfo{person}{Saugat Pandey}, \bibinfo{person}{Firas Moosvi}, \bibinfo{person}{Alvitta Ottley}, {and} \bibinfo{person}{Tamara Munzner}.} \bibinfo{year}{2025}\natexlab{}.
\newblock \bibinfo{booktitle}{\emph{Visualization Literacy or Skillset? Beyond the Analogy to Textual Literacy}}.
\newblock
\href{https://doi.org/10.31219/osf.io/9rc45_v1}{doi:\nolinkurl{10.31219/osf.io/9rc45_v1}}


\bibitem[Spence(2014)]%
        {spence14}
\bibfield{author}{\bibinfo{person}{Robert Spence}.} \bibinfo{year}{2014}\natexlab{}.
\newblock \bibinfo{booktitle}{\emph{Information Visualization: An Introduction} (\bibinfo{edition}{3rd} ed.)}.
\newblock \bibinfo{publisher}{Springer International Publishing}, \bibinfo{address}{Cham, Switzerland}.
\newblock
\showISBNx{978-3-319-07340-8}
\href{https://doi.org/10.1007/978-3-319-07341-5}{doi:\nolinkurl{10.1007/978-3-319-07341-5}}


\bibitem[Spiel et~al\mbox{.}(2019)]%
        {spiel19}
\bibfield{author}{\bibinfo{person}{Katta Spiel}, \bibinfo{person}{Oliver~L. Haimson}, {and} \bibinfo{person}{Danielle Lottridge}.} \bibinfo{year}{2019}\natexlab{}.
\newblock \showarticletitle{How to do better with gender on surveys: a guide for HCI researchers}.
\newblock \bibinfo{journal}{\emph{Interactions}} \bibinfo{volume}{26}, \bibinfo{number}{4} (\bibinfo{date}{June} \bibinfo{year}{2019}), \bibinfo{pages}{62---65}.
\newblock
\showISSN{1072-5520}
\href{https://doi.org/10.1145/3338283}{doi:\nolinkurl{10.1145/3338283}}


\bibitem[Stoiber et~al\mbox{.}(2019)]%
        {stoiber2019visualization}
\bibfield{author}{\bibinfo{person}{Christina Stoiber}, \bibinfo{person}{Florian Grassinger}, \bibinfo{person}{Margit Pohl}, \bibinfo{person}{Holger Stitz}, \bibinfo{person}{Marc Streit}, {and} \bibinfo{person}{Wolfgang Aigner}.} \bibinfo{year}{2019}\natexlab{}.
\newblock \bibinfo{booktitle}{\emph{Visualization onboarding: Learning how to read and use visualizations}}.
\newblock
\href{https://doi.org/10.31219/osf.io/c38ab}{doi:\nolinkurl{10.31219/osf.io/c38ab}}


\bibitem[Stoiber et~al\mbox{.}(2021)]%
        {stoiber2021design}
\bibfield{author}{\bibinfo{person}{Christina Stoiber}, \bibinfo{person}{Conny Walchshofer}, \bibinfo{person}{Florian Grassinger}, \bibinfo{person}{Holger Stitz}, \bibinfo{person}{Marc Streit}, {and} \bibinfo{person}{Wolfgang Aigner}.} \bibinfo{year}{2021}\natexlab{}.
\newblock \showarticletitle{Design and comparative evaluation of visualization onboarding methods}. In \bibinfo{booktitle}{\emph{Proceedings of the International Symposium on Visual Information Communication and Interaction}}. \bibinfo{publisher}{{ACM}}, \bibinfo{address}{New York, NY, USA}, Article \bibinfo{articleno}{11}, \bibinfo{numpages}{5}~pages.
\newblock
\href{https://doi.org/10.1145/3481549.3481558}{doi:\nolinkurl{10.1145/3481549.3481558}}


\bibitem[Taylor et~al\mbox{.}(2018)]%
        {taylor18}
\bibfield{author}{\bibinfo{person}{Jennyfer~Lawrence Taylor}, \bibinfo{person}{Alessandro Soro}, {and} \bibinfo{person}{Margot Brereton}.} \bibinfo{year}{2018}\natexlab{}.
\newblock \showarticletitle{New literacy theories for participatory design: lessons from three design cases with Australian Aboriginal communities}. In \bibinfo{booktitle}{\emph{Participatory Design Conference: Full Papers - Volume 1}} (Hasselt and Genk, Belgium). \bibinfo{publisher}{{ACM}}, \bibinfo{address}{New York, NY, USA}, Article \bibinfo{articleno}{11}, \bibinfo{numpages}{13}~pages.
\newblock
\showISBNx{9781450363716}
\href{https://doi.org/10.1145/3210586.3210588}{doi:\nolinkurl{10.1145/3210586.3210588}}


\bibitem[{The New London Group}(1996)]%
        {newlondon1996multiliteracies}
\bibfield{author}{\bibinfo{person}{{The New London Group}}.} \bibinfo{year}{1996}\natexlab{}.
\newblock \showarticletitle{A Pedagogy of Multiliteracies: Designing Social Futures}.
\newblock \bibinfo{journal}{\emph{Harvard Educational Review}} \bibinfo{volume}{66}, \bibinfo{number}{1} (\bibinfo{year}{1996}), \bibinfo{pages}{60--92}.
\newblock
\href{https://doi.org/10.17763/haer.66.1.17370n67v22j160u}{doi:\nolinkurl{10.17763/haer.66.1.17370n67v22j160u}}


\bibitem[van Wijk(2005)]%
        {vanWijk2005Value}
\bibfield{author}{\bibinfo{person}{Jarke~J. van Wijk}.} \bibinfo{year}{2005}\natexlab{}.
\newblock \showarticletitle{The Value of Visualization}. In \bibinfo{booktitle}{\emph{Proceedings of the {IEEE} Conference on Visualization}}. \bibinfo{publisher}{{IEEE Computer Society}}, \bibinfo{address}{Los Alamitos, CA, USA}, \bibinfo{pages}{79--86}.
\newblock
\href{https://doi.org/10.1109/VISUAL.2005.1532781}{doi:\nolinkurl{10.1109/VISUAL.2005.1532781}}


\bibitem[Velghe(2014)]%
        {velghe14}
\bibfield{author}{\bibinfo{person}{Fie Velghe}.} \bibinfo{year}{2014}\natexlab{}.
\newblock \showarticletitle{‘{I} wanna go in the phone’: literacy acquisition, informal learning processes, ‘voice’ and mobile phone appropriation in a {South African} township}.
\newblock \bibinfo{journal}{\emph{Ethnography and Education}} \bibinfo{volume}{9}, \bibinfo{number}{1} (\bibinfo{year}{2014}), \bibinfo{pages}{111--126}.
\newblock
\href{https://doi.org/10.1080/17457823.2013.836456}{doi:\nolinkurl{10.1080/17457823.2013.836456}}


\bibitem[Wall et~al\mbox{.}(2019)]%
        {wall19heuristic}
\bibfield{author}{\bibinfo{person}{Emily Wall}, \bibinfo{person}{Meeshu Agnihotri}, \bibinfo{person}{Laura Matzen}, \bibinfo{person}{Kristin Divis}, \bibinfo{person}{Michael Haass}, \bibinfo{person}{Alex Endert}, {and} \bibinfo{person}{John Stasko}.} \bibinfo{year}{2019}\natexlab{}.
\newblock \showarticletitle{A Heuristic Approach to Value-Driven Evaluation of Visualizations}.
\newblock \bibinfo{journal}{\emph{{{IEEE} Transactions on Visualization and Computer Graphics}}} \bibinfo{volume}{25}, \bibinfo{number}{1} (\bibinfo{year}{2019}), \bibinfo{pages}{491--500}.
\newblock
\href{https://doi.org/10.1109/TVCG.2018.2865146}{doi:\nolinkurl{10.1109/TVCG.2018.2865146}}


\bibitem[Wang et~al\mbox{.}(2023)]%
        {wang23}
\bibfield{author}{\bibinfo{person}{Jinrui Wang}, \bibinfo{person}{Mashael AlKadi}, {and} \bibinfo{person}{Benjamin Bach}.} \bibinfo{year}{2023}\natexlab{}.
\newblock \showarticletitle{Show Me My Users: A Dashboard Visualizing User Interaction Logs}. In \bibinfo{booktitle}{\emph{Proceedings of the {IEEE} Conference on Visualization}}. \bibinfo{publisher}{{IEEE Computer Society}}, \bibinfo{address}{Los Alamitos, CA, USA}, \bibinfo{pages}{156--160}.
\newblock
\href{https://doi.org/10.1109/VIS54172.2023.00040}{doi:\nolinkurl{10.1109/VIS54172.2023.00040}}


\bibitem[Wang et~al\mbox{.}(2021)]%
        {wang2021interactive}
\bibfield{author}{\bibinfo{person}{Zezhong Wang}, \bibinfo{person}{Hugo Romat}, \bibinfo{person}{Fanny Chevalier}, \bibinfo{person}{Nathalie~Henry Riche}, \bibinfo{person}{Dave Murray-Rust}, {and} \bibinfo{person}{Benjamin Bach}.} \bibinfo{year}{2021}\natexlab{}.
\newblock \showarticletitle{Interactive data comics}.
\newblock \bibinfo{journal}{\emph{{{IEEE} Transactions on Visualization and Computer Graphics}}} \bibinfo{volume}{28}, \bibinfo{number}{1} (\bibinfo{year}{2021}), \bibinfo{pages}{944--954}.
\newblock


\bibitem[Wang et~al\mbox{.}(2020)]%
        {wang2020cheat}
\bibfield{author}{\bibinfo{person}{Zezhong Wang}, \bibinfo{person}{Lovisa Sundin}, \bibinfo{person}{Dave Murray-Rust}, {and} \bibinfo{person}{Benjamin Bach}.} \bibinfo{year}{2020}\natexlab{}.
\newblock \showarticletitle{Cheat sheets for data visualization techniques}. In \bibinfo{booktitle}{\emph{Proceedings of the {ACM} Conference on Human Factors in Computing Systems}}. \bibinfo{publisher}{{ACM}}, \bibinfo{address}{New York, NY, USA}, \bibinfo{pages}{1--13}.
\newblock
\href{https://doi.org/10.1145/3313831.3376271}{doi:\nolinkurl{10.1145/3313831.3376271}}


\bibitem[Willett et~al\mbox{.}(2015)]%
        {willett2015understanding}
\bibfield{author}{\bibinfo{person}{Wesley Willett}, \bibinfo{person}{Pascal Goffin}, {and} \bibinfo{person}{Petra Isenberg}.} \bibinfo{year}{2015}\natexlab{}.
\newblock \showarticletitle{Understanding digital note-taking practice for visualization}.
\newblock \bibinfo{journal}{\emph{{IEEE} Computer Graphics and Applications}} \bibinfo{volume}{35}, \bibinfo{number}{4} (\bibinfo{year}{2015}), \bibinfo{pages}{38--51}.
\newblock
\href{https://doi.org/10.1109/MCG.2015.52}{doi:\nolinkurl{10.1109/MCG.2015.52}}


\bibitem[Wobbrock et~al\mbox{.}(2011)]%
        {wobbrock11}
\bibfield{author}{\bibinfo{person}{Jacob~O. Wobbrock}, \bibinfo{person}{Shaun~K. Kane}, \bibinfo{person}{Krzysztof~Z. Gajos}, \bibinfo{person}{Susumu Harada}, {and} \bibinfo{person}{Jon Froehlich}.} \bibinfo{year}{2011}\natexlab{}.
\newblock \showarticletitle{Ability-Based Design: Concept, Principles and Examples}.
\newblock \bibinfo{journal}{\emph{ACM Trans. Access. Comput.}} \bibinfo{volume}{3}, \bibinfo{number}{3}, Article \bibinfo{articleno}{9} (\bibinfo{date}{April} \bibinfo{year}{2011}), \bibinfo{numpages}{27}~pages.
\newblock
\showISSN{1936-7228}
\href{https://doi.org/10.1145/1952383.1952384}{doi:\nolinkurl{10.1145/1952383.1952384}}


\bibitem[Wyche et~al\mbox{.}(2016)]%
        {wyche16}
\bibfield{author}{\bibinfo{person}{Susan Wyche}, \bibinfo{person}{Charles Steinfield}, \bibinfo{person}{Tian Cai}, \bibinfo{person}{Nightingale Simiyu}, {and} \bibinfo{person}{Martha~E. Othieno}.} \bibinfo{year}{2016}\natexlab{}.
\newblock \showarticletitle{Reflecting on Video: Exploring the Efficacy of Video for Teaching Device Literacy in Rural Kenya}. In \bibinfo{booktitle}{\emph{Proceedings of the ACM Conference on Information and Communication Technologies and Development}}. \bibinfo{publisher}{{ACM}}, \bibinfo{address}{New York, NY, USA}, \bibinfo{pages}{8:1--8:10}.
\newblock
\href{https://doi.org/10.1145/2909609.2909667}{doi:\nolinkurl{10.1145/2909609.2909667}}


\bibitem[Xexakis and Trutnevyte(2021)]%
        {xexakis21}
\bibfield{author}{\bibinfo{person}{Georgios Xexakis} {and} \bibinfo{person}{Evelina Trutnevyte}.} \bibinfo{year}{2021}\natexlab{}.
\newblock \showarticletitle{Empirical testing of the visualizations of climate change mitigation scenarios with citizens: A comparison among {Germany}, {Poland}, and {France}}.
\newblock \bibinfo{journal}{\emph{Global Environmental Change}}  \bibinfo{volume}{70} (\bibinfo{year}{2021}), \bibinfo{pages}{102324}.
\newblock
\href{https://doi.org/10.1016/j.gloenvcha.2021.102324}{doi:\nolinkurl{10.1016/j.gloenvcha.2021.102324}}


\bibitem[Yal\c{c}in et~al\mbox{.}(2016)]%
        {Yalcin2016cognitive}
\bibfield{author}{\bibinfo{person}{M.~Adil Yal\c{c}in}, \bibinfo{person}{Niklas Elmqvist}, {and} \bibinfo{person}{Benjamin~B. Bederson}.} \bibinfo{year}{2016}\natexlab{}.
\newblock \showarticletitle{Cognitive Stages in Visual Data Exploration}. In \bibinfo{booktitle}{\emph{Proceedings of the Workshop on Beyond Time and Errors in Visualization}}. \bibinfo{publisher}{{ACM}}, \bibinfo{address}{New York, NY, USA}, \bibinfo{pages}{86--95}.
\newblock
\href{https://doi.org/10.1145/2993901.2993902}{doi:\nolinkurl{10.1145/2993901.2993902}}


\bibitem[Yi et~al\mbox{.}(2007)]%
        {yi07}
\bibfield{author}{\bibinfo{person}{Ji~Soo Yi}, \bibinfo{person}{Youn~ah Kang}, \bibinfo{person}{John Stasko}, {and} \bibinfo{person}{J.A. Jacko}.} \bibinfo{year}{2007}\natexlab{}.
\newblock \showarticletitle{Toward a Deeper Understanding of the Role of Interaction in Information Visualization}.
\newblock \bibinfo{journal}{\emph{{{IEEE} Transactions on Visualization and Computer Graphics}}} \bibinfo{volume}{13}, \bibinfo{number}{6} (\bibinfo{year}{2007}), \bibinfo{pages}{1224--1231}.
\newblock
\href{https://doi.org/10.1109/TVCG.2007.70515}{doi:\nolinkurl{10.1109/TVCG.2007.70515}}


\bibitem[Zhao et~al\mbox{.}(2016)]%
        {zhao16}
\bibfield{author}{\bibinfo{person}{Jian Zhao}, \bibinfo{person}{Michael Glueck}, \bibinfo{person}{Fanny Chevalier}, \bibinfo{person}{Yanhong Wu}, {and} \bibinfo{person}{Azam Khan}.} \bibinfo{year}{2016}\natexlab{}.
\newblock \showarticletitle{Egocentric Analysis of Dynamic Networks with {EgoLines}}. In \bibinfo{booktitle}{\emph{Proceedings of the {ACM} Conference on Human Factors in Computing Systems}}. \bibinfo{publisher}{{ACM}}, \bibinfo{address}{New York, NY, USA}, \bibinfo{pages}{5003--5014}.
\newblock
\showISBNx{9781450333627}
\href{https://doi.org/10.1145/2858036.2858488}{doi:\nolinkurl{10.1145/2858036.2858488}}


\bibitem[Zoss et~al\mbox{.}(2018)]%
        {zoss2018network}
\bibfield{author}{\bibinfo{person}{Angela Zoss}, \bibinfo{person}{Adam Maltese}, \bibinfo{person}{Stephen~Miles Uzzo}, {and} \bibinfo{person}{Katy B{\"o}rner}.} \bibinfo{year}{2018}\natexlab{}.
\newblock \showarticletitle{Network visualization literacy: Novel approaches to measurement and instruction}.
\newblock In \bibinfo{booktitle}{\emph{Network Science In Education}}. \bibinfo{publisher}{Springer International Publishing}, \bibinfo{address}{Cham, Switzerland}, \bibinfo{pages}{169--187}.
\newblock
\href{https://doi.org/10.1007/978-3-319-77237-0}{doi:\nolinkurl{10.1007/978-3-319-77237-0}}


\end{thebibliography}

\clearpage\newpage

\appendix

\section{Additional Case Studies}
\label{appendix:scenarios}
In this appendix, we present two additional case studies for each of the two scenarios presented in Sect.~\ref{sec:scenarios}, plus the three case studies for the immersive \& multimodal analytics scenario.

\begin{figure*}
    \centering
    \includegraphics[width=1\linewidth]{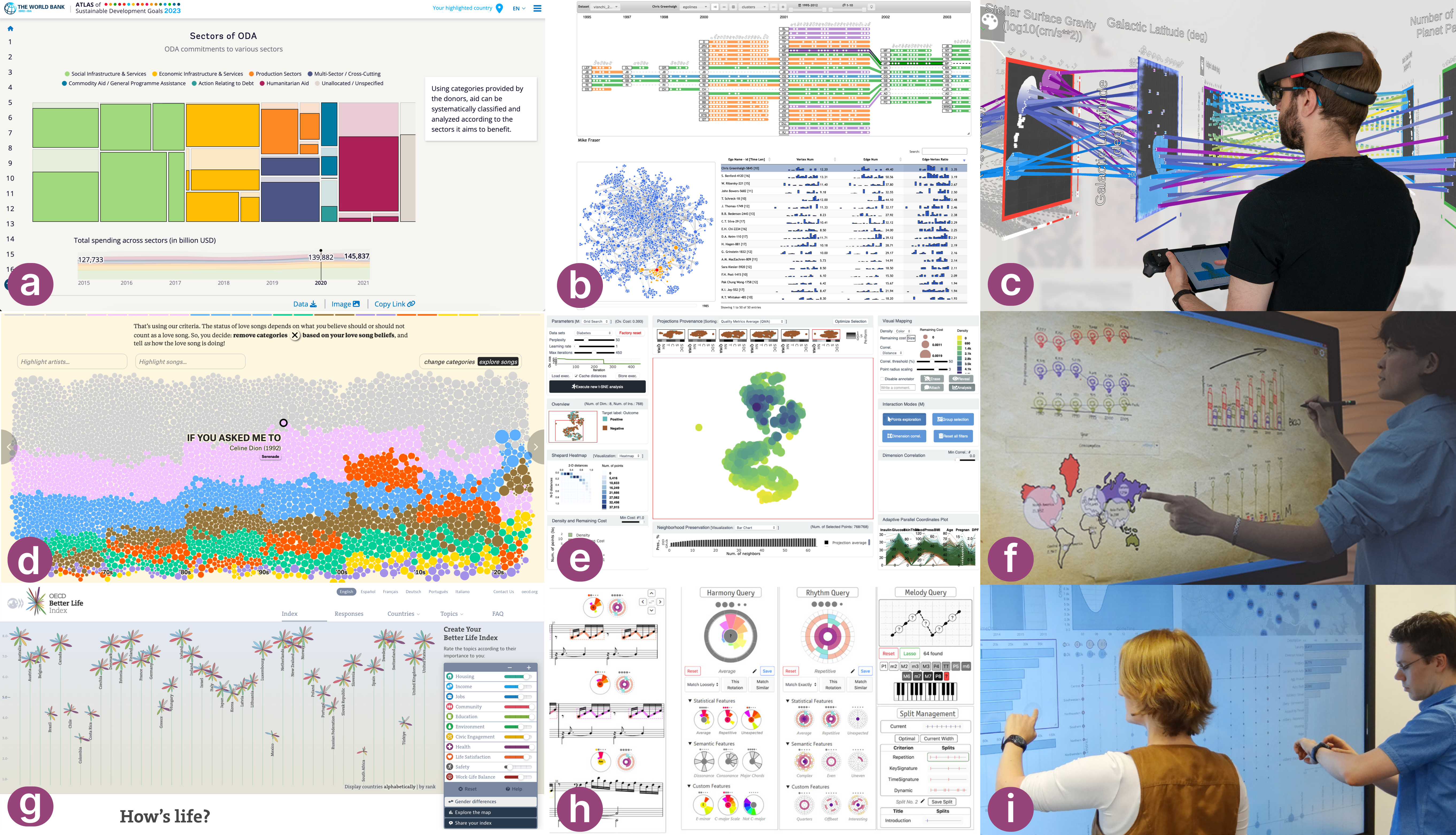}
    \caption{\textbf{Visualization systems as case studies.}
    \normalfont
    Data-driven storytelling: 
    (a) the Atlas of Sustainable Development Goals~\cite{pirlea23}, image \copyright The World Bank Group,
    (d) visual essay \textit{Is the Love Song Dying?}~\cite{islovedying24},
    (g) OECD Better Life Index~\cite{oecdindex14}, image \copyright OECD;
    visual analytics: 
    (b) EgoLines~\cite{zhao16},
    (e) t-viSNE~\cite{chatzimparmpas20}, system image \copyright ISOVIS group,
    (h) MusicVis~\cite{miller22}, image \copyright Eurographics available under \href{https://creativecommons.org/licenses/by-nc/4.0/}{CC BY-NC 4.0} license;
    immersive \& multimodal analytics: 
    (c) STREAM~\cite{hubenschmid21}, available under \href{https://creativecommons.org/licenses/by/4.0/}{CC BY 4.0} license,
    (f) SketchStory~\cite{lee13}, image \copyright IEEE used with permission,
    (i) David \& Goliath cross-device system~\cite{horak18}.
    Unless otherwise indicated, images are copyrighted by their respective authors.
    }
    \Description{Nine images ordered in a grid of three columns and three rows.
    The first one (a) shows an interactive treemap on top of a line chart, accompanied by an explanatory text on spending categories. 
    The second one (b) shows a web-based system with three different views (the egolines visualization technique, a node-link diagram, and an interactive table). 
    The third image (c) shows a person interacting with a 3D parallel coordinates visualization in an immersive environment.
    The fourth one (d) shows a unit chart representing the evolution of love song categories over time, with the categories encoded as colors. 
    The fifth image (e) is a screenshot of a web system with seven visualization views, accompanied by multiple sliders and buttons that enable users to adjust the visualizations.
    The sixth image (f) is a photo of someone writing on a large vertical display with a pen, while pointing with their finger to one of the four visualizations on the display.
    The seventh one (g) is the homepage of the OECD Better Life Index website, which includes a slider menu and a sequence of flower visualizations, representing the OECD countries. 
    The eighth one (h) shows another web system featuring three custom visualizations of music data, accompanied by multiple slides and drop-down menus that support exploration through harmony, rhythm, and melody queries.
    The ninth image (i) shows two people standing in front of a similar display. One person interacts with a bar chart by tapping their finger on it, while the other person touches the smartwatch on their wrist to also interact with the visualizations on the display.
    }
    \label{fig:suppl-scenarios}
\end{figure*}

\subsection{Data-driven Storytelling}

\subsubsection{\textit{Is the Love Song Dying?} (2024).}
This data-driven and award-winning article from The Pudding~\cite{islovedying24} (see Fig.~\ref{fig:suppl-scenarios}d) discusses the prevalence of love songs in popular music in a sequence of 26 story segments with a bottom-up approach.
The interactive visualizations show the temporal evolution and classification of 5,100 Billboard Top 10 hits, from 1958 to 2023.
In the first segment, an animation accompanied by text indicates that the reader can navigate back and forth through the story segments using the left and right arrow keys, clicking or tapping on the forward and back buttons (\devlitlow).
Jumping between segments is possible via a breadcrumb menu, which shows the progress of the reader through the story, and allows the reader to revisit previous segments to make decisions later (\tasklitlow).
To understand the story, readers need to apply \analysislitlow to connect individual data points to the broader cultural question about love songs' prevalence.
Interpreting the unit and area charts that appear in most segments requires \visualizationlitlow, while making overall assessments of the story's argument across several charts demands \evaluationlitlow.
Readers can identify songs by hovering over the circles that represent them in the unit chart, requiring both \intlitlow to execute these interactions and \visuallitlow to perceive the encoded attributes.
After completing the 24 segments of the main story, the text encourages readers to interact by modifying the classification of song types into the higher classification of love or non-love songs, according to their opinion.
Playing song samples via floating buttons can help understand the groups and reclassify, which leads to changes in the area chart.
This complex manipulation of multiple interactive elements to achieve analytical goals requires \orchlitlow.
Comparing the area heights at the beginning and the end of the temporal axis (\visuallitlow) helps answer the main question of the story, i.e., are there fewer love songs over time? (\evaluationlitlow).
Once all the song types are presented, readers can explore the complete dataset by hovering or looking for specific singers and songs via search fields.
The labels on a toggle menu indicate that specific tasks fit best with a visualization, i.e., \textit{change categories} with the unit chart and \textit{explore songs} with the area chart, which helps develop readers' \toollitlow by teaching them which visualizations are appropriate for the different analytical goals.
The essay also demands foundational \foundlit{media literacy} to understand how the piece's narrative framing influences data interpretation.

\subsubsection{OECD Better Life Index (2014).}

In this data-driven collection of stories~\cite{oecdindex14}, visitors can create, compare, and share a personal index reflecting their view on what makes a good life.
They can assign an importance rate to 11 topics associated with well-being, such as education, safety, and health.
This project provides data-driven stories per country and topic with interactive visualizations to explore via standard WIMP elements (requiring \foundlit{computer literacy}).
People can navigate through the text and visualizations with the mouse or keyboard (\devlitlow).
To create the index, the person needs to reflect on their personal view of welfare (\analysislitlow) and how they can describe it through weight assignment to each topic (\tasklitlow).
The tool presents standard visualizations such as symbol maps, dot plots, and bar charts---often as sparkline charts.
As Fig.~\ref{fig:suppl-scenarios}g shows, the overview visualization represents each country as a flower with the topics encoded as petals, combining a lollipop chart with glyphs, requiring sophisticated \visuallitlow and \visualizationlitlow to perceive and interpret these multi-attribute encodings.
The flowers are interactive via hover and click, even though no cues indicate that, except for an interactive text label naming sorting orders, demanding developed \toollitlow to be aware of the functions.
Clicking on the help button opens a 22-second video that explains the visual encoding of the flowers and how to use the sliders.
When the person changes the topic weights, the visualization updates by changing the size and saturation of the petals, requiring \intlitlow to manipulate these controls effectively.
Managing the complex workflow of setting multiple weights while evaluating visual outcomes requires \orchlitlow to coordinate these actions toward meaningful analysis.
If a person notices flowers are clickable (understanding interaction affordances), they can access a country page by clicking on the corresponding flower.
Still, such actions are supported by redundancy through the toolbar, which includes menus with topics and a country list. 
Some button labels suggest interaction, such as ``Explore the map.'' The ``Frequently Asked Questions'' page describes multiple ways of interacting, helping develop readers' \evaluationlitlow by encouraging them to link insights to their personal values and priorities.

\subsection{Visual Analytics}

\subsubsection{t-viSNE (2020).}

t-viSNE~\cite{chatzimparmpas20} is a web-based system for assessing and interpreting dimensionality reduction projections, shown in Fig.~\ref{fig:suppl-scenarios}e.
Its multiple coordinated views (\evaluationlitlow) expose quality metrics, parameter effects, and dimension relationships to help data analysts (\foundlit{statistical literacy}) understand and trust complex high-dimensional visualizations designed by the authors, such as a Shepard heatmap and adaptive parallel coordinates (\visualizationlitlow).
The system was designed for people with previous knowledge about dimensionality reduction methods (\foundlit{data science literacy}) who can explore high-dimensional patterns through interaction (\orchlitlow), e.g., drawing a polyline with the mouse (\devlitlow) in the scatterplot showing a projection of interest.
The system supports multiple interaction modes (\intlitlow), requiring users to develop sophisticated \toollitlow to understand which system functions are appropriate for different analytical goals.
Users must apply \analysislitlow to formulate effective strategies for exploring the high-dimensional space and \evaluationlitlow to critically assess whether the projection accurately represents the underlying data relationships.
The use of coordinated views demands \tasklitlow for users to decompose their complex analytical goals into sequences of specific visualization operations.
Since the system visualizes abstract mathematical concepts, users also need advanced \visuallitlow to accurately perceive subtle patterns in the novel visual representations that might indicate meaningful structures in the high-dimensional data (\visualizationlitlow).

\subsubsection{MusicVis (2022).}

As visible in Fig.~\ref{fig:suppl-scenarios}h, MusicVis~\cite{miller22} is a web application meant for desktop computers (\devlitlow).
The authors designed novel glyph visualizations for music analysts, and thus, \foundlit{domain knowledge} is highly relevant for interpreting musical patterns and structures.
The system uses multiple views, where interactive highlighting and brushing \& linking are possible, requiring well-developed \intlitlow to effectively employ these techniques.
Moreover, some interaction techniques (e.g., lasso selection) are only available after switching to a specific interaction mode, demanding also \toollitlow to understand the available system functions and their contextual availability.
The authors acknowledge that the system has a steep learning curve because it supports multiple analysis tasks (\orchlitlow) and leverages abstract glyph visualizations, which were hard to grasp for some study participants (\visualizationlitlow).
Visual querying through the glyphs is the main way of interacting, which requires discovering the interaction technique despite minimal affordances (\intlitlow), challenging the users' \tasklitlow to determine how to execute specific analytical objectives.
After interacting, the person needs to perceive the changes across views (\visuallitlow) and interpret the effect in relation to their musical analysis goals (\evaluationlitlow).
Users must also formulate appropriate strategies for musical exploration and pattern discovery (\analysislitlow) to effectively use the system's advanced capabilities for tasks like identifying recurring motifs or analyzing harmonic structures.

\subsection{Immersive \& Multimodal Analytics}
This scenario consists of post-WIMP visualizations, often adapted to novel technologies, involving devices and modalities different than the standard mouse and keyboard.

\subsubsection{STREAM (2021).}

Interacting with this immersive analytics system~\mbox{\cite{hubenschmid21}} requires sophisticated \devlitlow regarding the use of a tablet and a head-mounted display (HMD).
The person can interact via speech commands, touch gestures on the tablet, head-gaze, and proxemic interaction via the HMD. 
Thus, being able to handle each of these input modalities individually and in combination is necessary (\devlitlow), as well as being familiar with the interface elements on the tablet and the actions that trigger a reaction (e.g., holding two fingers on the tablet to activate speech recognition).
The system demands \orchlitlow to coordinate multiple modalities effectively toward analytical goals, managing the complex workflow across devices.
Additionally, the prototype offers several interaction modes (\intlitlow), which the person needs to understand to avoid confusion (e.g., in the focus mode, tablet and HMD interactions are disabled).
Besides the input devices and modalities, the person needs to know how to interpret the 3D environment in augmented reality (\foundlit{spatial literacy}).
The prototype enables people to explore a dataset via 3D parallel coordinates that consist of linked 2D scatterplots (shown in Fig.~\ref{fig:suppl-scenarios}c), requiring \analysislitlow to formulate appropriate exploration strategies in this complex visualization environment.
Users need to be familiar with these visualization techniques (\visualizationlitlow) and the concept of connected views, while also applying \evaluationlitlow to critically assess findings across different spatial perspectives.
The system's novel interaction paradigm requires users to develop \toollitlow to recognize which system functions are appropriate for different analytical tasks, and \tasklitlow to decompose high-level goals into specific sequences of multimodal operations.

\subsubsection{SketchStory (2013).}

Next, we look at SketchStory~\cite{lee13}, a system that leverages multimodal interaction to create visual stories on a digital whiteboard (see Fig.~\ref{fig:suppl-scenarios}f).
People can sketch and annotate visualizations with a pen and manipulate the visualizations with touch (\devlitlow).
While the touch gestures are standard (e.g., dragging to move a chart), the pen gestures are less known.
For instance, drawing an \textit{L} leads to generating a Cartesian chart (e.g., line chart), and drawing a \textit{rotated L} generates a world map.
Thus, the person needs to be familiar with the gesture vocabulary the system recognizes (\intlitlow) and must develop \toollitlow to understand which drawing patterns trigger specific system functions.
Additionally, the system can create glyph visualizations based on freeform sketches in real time.
While the automatic completion helps people to finish the visualizations, they still need \visualizationlitlow to use it properly and \visuallitlow to perceive how their sketches transform into formal visualizations.
When the person creates multiple visualizations, they can be used as multiple coordinated views and support dynamic filtering through visual keywords, requiring \orchlitlow to coordinate these views toward coherent storytelling.
Through the creation and refinement of visualizations, the person creates a visual story, demanding \analysislitlow to formulate an effective narrative strategy and \evaluationlitlow to assess whether the created visualizations effectively communicate the intended story.
The system also requires \tasklitlow to decompose the storytelling goal into specific visualization creation and annotation steps.

\subsubsection{David \& Goliath cross-device visualization (2018).}

Finally, we examine the David \& Goliath cross-device visualization system of Horak et al.~\cite{horak18, DBLP:conf/chi/HorakMKDE19}, combining large interactive displays with smartwatches (see Fig.~\ref{fig:suppl-scenarios}i).
The system is designed to support fluid data exploration by allowing analysts to extract, manipulate, and preview data items and visualization properties across devices.
The goal is to support users in developing cross-device workflows. The devices have complementary roles: the smartwatch is for personal interactions, while the large display is meant for public or group work.
Users can create sets of data points of interest and then add or remove them from the smartwatch.
People need to be able to apply set operations (\foundlit{data literacy}), so they can make sense of the data through set manipulation.
As insights emerge, they should determine which sets to combine, requiring \analysislitlow to formulate effective exploration strategies and \evaluationlitlow to assess the significance of emerging patterns.
People can perform touch gestures on the large display (e.g., tapping on marks to select) and on the smartwatch (e.g., swiping along the arm axis to add a set to the large display), demanding sophisticated \intlitlow to employ these multi-device techniques effectively.
Rotating the smartwatch bezel supports navigation, and haptic feedback confirms successful interactions through vibration (\devlitlow). The novel interaction context requires \toollitlow to understand which functions are available across different devices.
The multiple coordinated views on the large display contain bar charts, time-series filled area charts, and a bubble matrix (\visualizationlitlow), requiring \visuallitlow to accurately perceive the visual patterns and their changes across the different views. 
Managing the complex workflow between personal and shared spaces demands \orchlitlow to coordinate actions toward analytical goals. \tasklitlow is necessary to decompose high-level exploration objectives into specific cross-device operations.

\section{Study Materials}
\label{appendix:study}

In the following, we provide the questions and tasks of the observational study presented in Sect.~6.2. For the questions based on related work, we cite the source, although it was not indicated to the participants in the study.

\subsection{Pre-questionnaire}
\begin{enumerate}
    \item Please rate your experience with computers (Cassidy and Eachus \cite{cassidy02})
    \begin{itemize}
        \item None
        \item Very limited
        \item Some experience
        \item Quite a lot
        \item Extensive
    \end{itemize}
    \item What is the average number of hours per week you use a computer? (Lintunen et al. \cite{lintunen24})
    \begin{itemize}
        \item \textit{(insert number)} hours per week [0, 168]
    \end{itemize}
    \item How often do you use interactive systems?
    \begin{itemize}
        \item Daily
        \item Weekly
        \item Monthly
        \item Less than once per month
        \item Never
    \end{itemize}
    \item Please select your familiarity with data visualizations (Pandey and Ottley \cite{pandey23})
    \begin{itemize}
        \item I have never created a visualization
        \item I am somewhat familiar
        \item I have created visualization systems before
    \end{itemize}
    \item How often do you use interactive visualizations?
    \begin{itemize}
        \item Daily
        \item Weekly
        \item Monthly
        \item Less than once per month
        \item Never
    \end{itemize}
    \item Please select your highest level of education.
    \begin{itemize}
        \item High school diploma
        \item Bachelor’s degree
        \item Master’s degree
        \item Doctoral degree
    \end{itemize}
    \item What is your age group? (adapted from Lintunen et al. \cite{lintunen24})
    \begin{itemize}
        \item 18---34
        \item 35---49
        \item 50 or more
    \end{itemize}
    \item Which gender do you identify with? (Spiel et al. \cite{spiel19})
    \begin{itemize}
        \item Woman
        \item Man
        \item Non-binary
        \item Prefer not to disclose
        \item Prefer to self-describe: \textit{(text box)}
    \end{itemize}
    \item Are you color-blind?
    \begin{itemize}
        \item Yes
        \item No
        \item I do not know
        \item Prefer not to disclose
    \end{itemize}
    \item Please perform this short visualization literacy test: 
\href{https://tools.visualdata.wustl.edu/experiment/}{MiniVLAT} (12 questions, 25 seconds to answer each) \cite{pandey23}
\end{enumerate}

\subsection{Tasks}
In this phase, we will ask you to perform three tasks with two interactive systems. Please answer the questions as best as you can. There are no time constraints. If you are not sure, choose the ``I am not sure'' option. While solving the tasks, please think aloud, i.e., say out loud what you observe, plan to do, and what you do.

\textit{The system order changed across participants.}

\subsubsection{Storytelling System: The SDG Atlas}
\begin{quote}
    In 2015, the United Nations agreed on 17 Sustainable Development Goals (SDGs) that aim at achieving “peace and prosperity for people and the planet”. Where do we stand in our efforts to achieve these goals? This SDG Atlas from 2023 explores this question through 17 immersive and interactive visual stories showing a plethora of state-of-the-art data visualizations (adapted quote from \href{https://www.alicethudt.de/projects/atlas-of-sustainable-development-goals}{atlas designer}).
\end{quote}

Please answer the following questions by interacting with the atlas.

\begin{enumerate}
    \item Which of these countries protects the highest percentage of its national waters?
    \begin{itemize}
        \item Greenland.
        \item Brazil.
        \item Singapore.
        \item Jordan.
        \item I am not sure.
    \end{itemize}
    \item How did tuberculosis cases evolve in the Philippines from 2010 to 2012? The number of tuberculosis cases…
    \begin{itemize}
        \item …decreased each year.
        \item …decreased once and then increased.
        \item …increased each year.
        \item …increased once and then decreased.
        \item I am not sure.
    \end{itemize}
    \item Report interesting patterns in the partnerships involving donations from some countries to others for development. Please mention numerical evidence.
    \begin{itemize}
        \item Example: \textit{Europe and Central Asia are the regions donating the most worldwide (339,148 million USD), mostly to countries in the Middle East and North Africa (92,374 million USD) and Sub-Saharan Africa (90,447 million USD).}
    \end{itemize}
    \item Can you explain your thought process? What elements guided your decision? Did you find any part confusing? (Nobre et al. \cite{nobre24})
\end{enumerate}

\subsubsection{Visual Analytics System: EgoLines}

\begin{quote}
    EgoLines is an interactive visualization that supports the egocentric analysis of dynamic networks. The egocentric analysis of dynamic networks focuses on discovering the temporal patterns of a subnetwork around a specific central actor (i.e., an ego-network). These types of analyses are useful to provide insights about how the central actor interacts with other actors. The system shows a network of academic authorship (adapted from the paper's abstract \cite{zhao16}).
\end{quote}

Please answer the following questions by interacting with EgoLines.

\begin{enumerate}
    \item In 1996, how many co-authors did Sara Kiesler have?
    \begin{itemize}
        \item 5 co-authors
        \item 4 co-authors
        \item 3 co-authors
        \item I am not sure
    \end{itemize}
    \item Who had the longest relationship with J. Mackinlay?
    \begin{itemize}
        \item S. K. Card
        \item C. Stolte
        \item R. Rao
        \item I am not sure
    \end{itemize}
    \item Describe the temporal evolution of two clusters that merged into one at any given year.
    \begin{itemize}
        \item Example: \textit{Among the collaborators of M. Chalmers, there were three clusters (green, purple, orange) between 1996 and 2013. In 2008, two clusters disappeared, with an author from each one (Brown \& Reeves) becoming part of the remaining (orange) cluster.}
    \end{itemize}
    \item Can you explain your thought process? What elements guided your decision? Did you find any part confusing? (Nobre et al. \cite{nobre24})
\end{enumerate}

\subsection{Semi-structured Interview}
\begin{enumerate}
    \item What was your strategy for solving the tasks on the atlas?
    \item What was your strategy for solving the tasks on EgoLines?
    \item What skills did you make use of for interacting with these systems?
    \item Is there anything else you would like to share?
\end{enumerate}

\end{document}